\documentclass[a4paper, traditabstract,usenames,dvipsnames]{aa}  
\usepackage{graphicx}
\usepackage[rightcaption]{sidecap}
\usepackage{txfonts}
\usepackage{amssymb}
\usepackage{bbm}
\usepackage{microtype}
\usepackage{url}
\usepackage{booktabs}
\usepackage{arydshln}
\usepackage{enumitem}
\usepackage{footnote}
\usepackage{soul}
\usepackage{commath}
\usepackage{pifont}
\usepackage[bookmarks, colorlinks, breaklinks, ]{hyperref}  
\usepackage{xcolor}
\hypersetup{linkcolor=blue,citecolor=MidnightBlue,filecolor=black,urlcolor=MidnightBlue} 
\def \KL {K_{\rm L}}
\def\Mv {M_{500}}
\def \Rv {R_{500}}
\def \Lv {L_{500}}
\def \Tv {\theta_{500}}
\def \dLv {\Delta_{\rm log}(\Lv)}
\def \dLvrel {\Delta_{\rm log}(\Lv)/\sigma_{log}(\Lv)}

\def \Lvo {L_{500, \rm ovlp}}
\def \LX {L_{\rm X}}
\def \FX {F_{\rm X}}
\def \YX {Y_{\rm X}}
\def \Rap{R_{\rm ap}}
\def \Fap{F_{\rm ap}}
\def \Lap{L_{\rm ap}}
\def\errLv{\sigma_{\Lv}/\Lv}

\def\fgv {f_{\rm g,500}}
\def\RL {R_{\rm L,500}}

\def\LM {$\Lv$--$\Mv$}
\def\LT {$\Lv$--$T$}
\def\keV {\rm keV}
\def\ergs {\rm erg\ s^{-1}}
\def\msun{{M_{\odot}}}
\def\bando {[0.1$--$2.4]\ \keV}
\def\bandi {[0.5$--$2.]\ \keV}

\newfont{\gwpfont}{cmssq8 scaled 1000}
\newcommand{\rexcess}{{\gwpfont REXCESS}}
\def \xmm {\hbox{\it XMM-Newton}}
\def \chandra {\hbox{\it Chandra}}
\def \rosat {\hbox{\it ROSAT}}
\def \planck {\hbox{\it Planck}}

\def\lesssim{\mathrel{\hbox{\rlap{\hbox{\lower4pt\hbox{$\sim$}}}\hbox{$<$}}}}
\def\gtrsim{\mathrel{\hbox{\rlap{\hbox{\lower4pt\hbox{$\sim$}}}\hbox{$>$}}}}

\def \zs {z_{\rm spec}}
\def \zp {z_{\rm phot}}

\begin{document} 

   \title{MCXC-II: Second release of the Meta-Catalogue of X-Ray detected Clusters of galaxies}

    \author{T. Sadibekova\inst{1}, M. Arnaud\inst{1}, G.W. Pratt\inst{1}, P. Tarr\'io\inst{2}, and J.-B. Melin\inst{3} 
    }

    \institute{Universit\'e Paris-Saclay, Universit\'e Paris Cit\'e, CEA, CNRS, AIM, 91191, Gif-sur-Yvette, France\\
              \email{tatyana.sadibekova@cea.fr}
        \and
    Observatorio Astron\'omico Nacional, IGN, Calle Alfonso XII 3, E-28014 Madrid, Spain	
        \and 
       Université Paris-Saclay, CEA, Département de Physique des Particules, 91191, Gif-sur-Yvette, France
    }

\titlerunning{MCXC-II}
\authorrunning{Sadibekova et al.}
 
  \abstract
   {We present the second release of the Meta-catalogue of X-Ray detected Clusters of galaxies (hereafter MCXC-II). MCXC-II has been compiled from publicly available ROSAT All Sky Survey-based (NORAS, REFLEX, BCS, SGP, NEP, MACS, CIZA, and RXGCC) and serendipitous (160SD, 400SD, SHARC, WARPS, and EMSS) X-ray cluster catalogues. Redshifts were systematically checked and updated when necessary, with additional redshift information (type and origin) added. The X-ray data were standardised to an overdensity of 500,  using a new procedure based on the use of the original flux and aperture measurements available in the input catalogues. MCXC-II contains 2221 entries, now including objects from the REFLEX-II and RXGCC surveys, in addition to providing a complete and fully-homogenised sub-catalogue of all published MACS clusters. Duplicate entries from overlaps between the survey areas of the individual input catalogues were carefully handled. For each cluster, the MCXC-II provides three identifiers: redshift, coordinates, and membership in the original catalogue, along with standardised [0.1-2.4] keV band luminosity, $\Lv$, total mass, $\Mv$, and radius, $\Rv$. Uncertainties on $\Lv$ were computed from the flux errors in the original catalogues. MCXC-II additionally provides information on overlaps between the input catalogues, reporting the luminosity and its uncertainty when measurements from different surveys are available, along with notes on individual objects. 
   }

   \maketitle

\section{Introduction}
As the nodes of the cosmic web, clusters of galaxies provide information on the assembly and evolution of large-scale structure and on the underlying cosmology hosting these processes. Emitting in X-rays through bremsstrahlung emission of the hot, gaseous intra-cluster medium (ICM), cluster sources have been apparent in the early all-sky X-ray surveys by {\it Uhuru}, HEAO-1, and {\it Ariel-V}. The launch of  {\it Einstein}, the first X-ray satellite with imaging capability, allowed for the detection of previously unknown clusters from their X-ray emission \citep{1994ApJS...94..583G}. The advent of the R\"ontgen SATelleit (ROSAT) and its associated all-sky survey (RASS) in 1990-1991 led to the discovery of hundreds of clusters, with several hundred more found serendipitously in the later pointed observation phase (see references in Sect.~\ref{sec:firstcat}).

The cluster catalogues extracted from the RASS and the ROSAT pointed observation phase have provided the targets for deeper follow-up observations with the subsequent generation of satellites \xmm,  \chandra\, and {\it Suzaku}, and for identification of existing clusters in surveys in other wavelengths. However, the different conventions employed in the construction of these catalogues (e.g., cosmology, integration radius etc.), have made it difficult to standardise the information contained within. 

In this context, the first Meta-Catalogue of X-ray detected Clusters of galaxies (hereafter MCXC-I) was published by \citet{2011A&A...534A.109P}. As a meta-catalogue, the purpose of the MCXC-I was to assemble the information on X-ray clusters from published samples based on ROSAT and {\it Einstein} cluster detections and (critically) to homogenise the primary catalogue measurement: the X-ray luminosity, $L_{\rm X}$. Measurements from all major publicly-available X-ray catalogues were standardised to the same cosmology and integration radius, using an approach based on use of the [0.1-2.4] keV X-ray luminosity as a mass proxy. Using an empirically-measured average profile and a scaling law relating the X-ray luminosity to the underlying mass, all luminosity measurements were standardised to an integration radius of $\Rv$\footnote{$\Rv$ is the radius in which the mean cluster density is 500 times the critical density of the universe at the cluster redshift, $\rho_{\rm c}(z)$. $\Tv$ and $\Mv$ are the corresponding angular radius and mass}. MCXC-I additionally furnished information on overlaps between the input catalogues, gave the luminosity ratios when measurements from different surveys were available, and provided notes on individual objects.

The majority of the catalogues in MCXC-I were based on X-ray detections of extended sources, either in the original RASS or in deeper observations from the subsequent pointed phase. The advantage of such an approach is that the selection function is purely based on the X-ray characteristics of the sources in the catalogue. The exceptions were the {\it Einstein} Extended Medium Sensitivity Survey \citep[][EMSS]{1990ApJS...72..567G}, for which optical spectroscopic follow-up of all 835 sources was tractable, and the Massive Cluster Survey \citep[][MACS]{2001ApJ...553..668E}, for which flux cuts and hardness ratio criteria were applied to the  ROSAT Bright Source Catalogue before systematic optical spectroscopic follow-up was undertaken. Such approaches are obviously very costly in terms of the observing time needed to complete the optical spectroscopic follow-up and are intractable once the number of sources needing follow-up crosses a given threshold.

Since the release of MCXC-I, the advent of large-scale multi-band optical surveys and the subsequent widespread adoption of photometric redshifts has changed the X-ray cluster survey landscape. Examples of catalogues derived from such approaches include MARD-Y3 (\citealt{2019MNRAS.488..739K}), CODEX (\citealt{2020A&A...638A.114F}), and RASS-MCMF (\citealt{2023MNRAS.526.3757K}). Moreover, cluster detection using joint information from X-rays and SZ surveys has also been developed and applied to ROSAT and {\it Planck} data, yielding the COMPRASS catalogue \citep{tar19}. While such approaches have yielded large numbers of new cluster detections, the resulting selection functions depend not only on the X-ray properties of the sources but also on their optical characteristics. 

This second release\footnote{A first revised version, including REFLEX-II and new MACS clusters published in 2012, was used in the validation of the \planck\ catalogue \citep{psz1rev}. An intermediate revised version, with updated content and revised redshifts, was put online in 2021 in the framework of the ERC M2C project: \url{https://www.galaxyclusterdb.eu/m2c/}.} of the MCXC (hereafter MCXC-II) now includes objects from the REFLEX II  \citep{2012A&A...538A..35C} and RXGCC \citep{2022A&A...658A..59X} surveys, where the cluster detection is based on measurements of X-ray source flux and extent. MCXC-II also provides a complete and fully homogenised sub-catalogue of all published Massive Cluster Survey (MACS) clusters.  Furthermore, MCXC-II revisits and substantially revises the original approach to $\Lv$ homogenisation. This is achieved by recalculating the $\Lv$ values from the original X-ray flux or count rate measurements where available in the original source catalogues. This approach notably allows us to include uncertainties in the luminosity measurements, which were not available for the MCXC-I release. As a further addition to the catalogue, we provide a new flag on the provenance and nature of the redshift measurement for each system. We leave the issue of the homogenisation of catalogues that make use of large-scale multi-band optical surveys to future work.

The paper is organised as follows. Section \ref{sec:cat} describes the new entries, including the search for new redshifts, updates on the overlap and revised merging strategy and Section \ref{sec:zrev} presents 
the redshift revision. The estimate of  $\Lv$ and $\Mv$ and  statistical errors  is detailed  in Section \ref{sec:lm}. Section \ref{sec:mcxcII} present the  properties of   MCXC-II.  In Section ~\ref{sec:discussion}, we discuss the differences between MCXC-II and MCXC-I, as well as systematic errors and the general issue of homogenisation of catalogues. Section~\ref{sec:conclusion} presents our conclusions. 

\begin{table*}[!t]
    \centering
\caption{\label{tab:summary_mcxc} Summary of the MCXC-II catalogues and sub-catalogues, the revised version of Table 1 from \cite{2011A&A...534A.109P}.  }

\begin{tabular}{lllrrr}
\toprule
\toprule
  \multicolumn{1}{c}{{\bf Catalogue}} &
  \multicolumn{1}{c}{{\bf Reference}} &
  \multicolumn{1}{c}{{\bf {\it VizieR}}} &
  \multicolumn{3}{c}{{\bf Number of clusters}}\\  
  
  \multicolumn{1}{c}{} &
  \multicolumn{1}{c}{} &
  \multicolumn{1}{c}{} &
  \multicolumn{1}{c}{$N_{\rm cat}$} & \multicolumn{1}{c}{$N_{\rm in}$} & \multicolumn{1}{c}{$N_{\rm MCXC}$}\\
  \midrule 
 
{\bf NORAS/REFLEX} &   &  \\
~~~~REFLEX & {\citet{2004A&A...425..367B}} &\href{https://cdsarc.cds.unistra.fr/viz-bin/cat/J/A+A/425/367}{J/A+A/425/367} &447& \tablefootmark{a} 444 & 440\\
~~~~REFLEXII & {\citet{2012A&A...538A..35C}} & & 22& 22 & 22 \\
\smallskip
~~~~NORAS & {\citet{2000ApJS..129..435B}}& \href{https://cdsarc.cds.unistra.fr/viz-bin/cat/J/ApJS/129/435}{J/ApJS/129/435} &484 &\tablefootmark{b} 469  & 457\\
{\bf 400SD}   &   {\citet{2007ApJS..172..561B}}& \href{https://cdsarc.cds.unistra.fr/viz-bin/cat/J/ApJS/172/561}{J/ApJS/172/561} &  &  \\
~~~~400SD\_SER &&& 242 & 242 & 236 \\
\smallskip
~~~~400SD\_NONSER &&& 24 & 24 & 20 \\
\smallskip
{\bf 160SD}  & {\citet{2003ApJ...594..154M}}&\href{https://cdsarc.cds.unistra.fr/viz-bin/cat/J/ApJ/594/154}{J/ApJ/594/154} & 223 & \tablefootmark{c} 204& 94 \\
{\bf BCS} &  & &   \\
~~~~BCS & {\citet{1998MNRAS.301..881E}} & \href{https://cdsarc.cds.unistra.fr/viz-bin/cat/J/MNRAS/301/881}{J/MNRAS/301/881}&206& \tablefootmark{d} 205& 45 \\
\smallskip
~~~~eBCS & {\citet{2000MNRAS.318..333E}}&\href{https://cdsarc.cds.unistra.fr/viz-bin/cat/J/MNRAS/318/333}{J/MNRAS/318/333} & 107 & 107 & 31  \\
\smallskip
{\bf SGP}  & {\citet{2002ApJS..140..239C,2003ApJS..144..299C}}&\href{https://cdsarc.cds.unistra.fr/viz-bin/cat/J/ApJS/140/239}{J/ApJS/140/239} & 186 &\tablefootmark{e} 176   &73 \\
 {\bf SHARC}&&&\\
~~~~SHARC\_SOUTH& {\citet{2003MNRAS.341.1093B}}& \href{https://cdsarc.cds.unistra.fr/viz-bin/cat/J/MNRAS/341/1093}{J/MNRAS/341/1093} & 32 & 32 & 15\\
\smallskip
~~~~SHARC\_BRIGHT& {\citet{2000ApJS..126..209R}}& \href{https://cdsarc.cds.unistra.fr/viz-bin/cat/J/ApJS/126/209}{J/ApJS/126/209} & 37 &37 & 14\\
{\bf WARPS} \\
~~~~WARPS I& {\citet{2002ApJS..140..265P}}& \href{https://cdsarc.cds.unistra.fr/viz-bin/cat/J/ApJS/140/265}{J/ApJS/140/265} &39 & \tablefootmark{f} 38 & 13 \\
\smallskip
~~~~WARPS II& {\citet{2008ApJS..176..374H}}& \href{https://cdsarc.cds.unistra.fr/viz-bin/cat/J/ApJS/176/374}{J/ApJS/176/374} &125&125& 68 \\
\smallskip 
{\bf NEP} & {\citet{2006ApJS..162..304H}}&\href{https://cdsarc.cds.unistra.fr/viz-bin/cat/J/ApJS/162/304}{J/ApJS/162/304} & 63 & 63 & 48 \\
{\bf MACS}  & ~~~~~see Sect.\ref{sec:newcat} \\
 ~~~~MACS\_DR1&{\citet{2007ApJ...661L..33E}}&& & 12 & 12\\
 ~~~~MACS\_DR2&{\citet{2010MNRAS.407...83E}} && &35 & 23 \\
 ~~~~MACS\_DR3&{\citet{2012MNRAS.420.2120M}}&& &22 & 19\\
\smallskip
 ~~~~MACS\_MISC&{\citet{2018MNRAS.479..844R}}&& & 16 & 9 \\
{\bf CIZA} & \\
~~~~CIZA I & {\citet{2002ApJ...580..774E}}&\href{https://cdsarc.cds.unistra.fr/viz-bin/cat/J/ApJ/580/774}{J/ApJ/580/774} & 73 & 73 & 72  \\
\smallskip
~~~~CIZA II & {\citet{2007ApJ...662..224K}}& \href{https://cdsarc.cds.unistra.fr/viz-bin/cat/J/ApJ/662/224}{J/ApJ/662/224} & 57 & 57 & 56 \\
{\bf EMSS} & \\
~~~~EMSS\_1994& {\citet{1994ApJS...94..583G}}& \href{https://cdsarc.cds.unistra.fr/viz-bin/cat/IX/15}{IX/15} &106 &  \tablefootmark{g}  81  & 47\\
\smallskip
~~~~EMSS\_2004& {\citet{2004ApJ...609..603H}} & & 23&   \tablefootmark{g} 21 & 14\\
{\bf RXGCC} & {\citet{2022A&A...658A..59X}}& \href{https://cdsarc.cds.unistra.fr/viz-bin/cat/J/A+A/658/A59}{J/A+A/658/A59} & 944 & \tablefootmark{h}  943  & 393 \\

\midrule
{\bf Total}  &  &  && 3448  & 2221  
\\
\bottomrule
\end{tabular}
\tablefoot{
Col.~1: Catalogue and sub-catalogue name.  
These are listed in priority order, essentially following that defined for MCXC. Col.~2: reference and corresponding  {\it VizieR} table. Cols.~3-5  Number of clusters at the different steps of the meta-catalogue construction. Col.~3:  number of clusters in the published (sub-)catalogue table. Col.~4: number of clusters after merging of sub-catalogues and removal of false clusters or those without redshift. The origin of the difference between $N_{\rm cat}$ and $N_{\rm in}$ is given in the footnote. $N_{\rm in}$ may differ from  the values given by \cite{2011A&A...534A.109P} following the revision  described Sect.~ \ref{sec:newmcxc}. Col.~5: $N_{\rm MCXC}$ is the number of clusters in MCXC-II after handling multiple entries between catalogues, i.e. whose parameters are derived from this source. \\
(a) Three duplicates of NORAS clusters were removed by P11 when merging NORAS and REFLEX catalogues. 
(b) Ten REFLEX cluster duplicates were removed by P11 when merging the NORAS and REFLEX catalogues, in addition to two Virgo galaxies. Two clusters identified as false and one still without redshift were removed after the present revision (see Sect.~\ref{sec:noras}).
(c)  The 160SD table includes 22 objects flagged as likely false detections, and one for which the optical follow-up was not possible. The present work confirms and provides redshift for five objects (see Sect.\ref{sec:160SD}). The other 18 objects were removed, in addition to one further cluster later invalidated by \citet{2007ApJS..172..561B}.
(d) \object{MCXC~J1332.7+5032} appears both in BCS and eBCS and was removed by P11.
(e) Ten clusters without redshift were removed after the present revision (Sect.~\ref{sec:sgp}). 
(f) One cluster without redshift was removed after the present revision (Sect.~\ref{sec:warps}). 
(g) EMSS\_2004 is the ASCA follow-up of 23 clusters in the original EMSS\_1994 catalogue. \object{MS1208.7} and \object{MS0147.8}, which do not have luminosity measurements, were removed by P11. The remaining  21 EMSS\_2004 clusters were removed by P11 from the EMSS\_1994 catalogue, in addition to three further clusters, following \citealt{2004ApJ...609..603H}.
(h) \object{RXGCC~908}, a point source in the XMM-{\it Newton} archive observation, was removed.
  }
\end{table*}


We adopt a $\Lambda$CDM cosmology with $H_0 = 70~{\rm km}~{\rm s}^{-1}~{\rm Mpc}^{-1}$, $\Omega_{\rm m} = 0.3$ and $\Omega_{\Lambda} = 0.7$ throughout the paper. The quantity $E(z)$ is the ratio of the Hubble constant at redshift $z$ to its present value, $H_0$, namely, $E(z)^2 = \Omega_{\rm m}(1 + z)^3 + \Omega_{\Lambda}$.

\section{MCXC-II content}\label{sec:cat}

In this section, we describe the addition of new clusters to MCXC-II and how the information on the cross-matching between catalogues was updated. Before describing the second release, we give a brief overview of the content of the MCXC-I that stands as our point of departure.

\subsection{Overview of first release}\label{sec:firstcat}

The MCXC-I catalogue of \citet{2011A&A...534A.109P} consists of  1743 entries. These data were compiled from 16 publicly available X-ray catalogues derived primarily from ROSAT observations. The exceptions are the {\it Einstein} based EMSS catalogue, for which the X-ray quantities for part of the sample were obtained from ASCA follow-up observations, and the RASS-derived MACS catalogue, where the X-ray quantities were obtained from {\it Chandra} follow-up observations.

The simplest classification scheme for these input catalogues is into contiguous area (i.e. RASS-survey) and serendipitous (i.e. pointed) sub-catalogues, as follows:

\begin{itemize}[]
    \item 
    \textbf{RASS-based catalogues}:
     the ROSAT-ESO Flux Limited X-ray galaxy cluster survey \citep[REFLEX;][]{2004A&A...425..367B}; the Northern ROSAT All-Sky galaxy cluster survey \citep[NORAS;][]{2000ApJS..129..435B}; the Clusters In the Zone of Avoidance survey \citep[CIZA and CIZA II;][]{2002ApJ...580..774E,2007ApJ...662..224K}; the ROSAT Brightest Cluster Sample and extended Brightest Cluster Sample \citep[BCS and eBCS;][]{1998MNRAS.301..881E,2000MNRAS.318..333E}; the South Galactic Pole survey \citep[SGP;][]{2002ApJS..140..239C}; the Massive Cluster Survey \citep[MACS;][]{2007ApJ...661L..33E, 2008ApJS..174..117M,2010MNRAS.407...83E}; and the North Ecliptic Pole survey \citep[NEP;][]{2006ApJS..162..304H}.\\
    \item \textbf{Serendipitous catalogues}:
    the 160 Square Degree survey \citep[160SD][]{2003ApJ...594..154M}; the 400 Square Degree survey \citep[400SD][]{2007ApJS..172..561B}; the Wide-Angle ROSAT Pointed Survey \citep[WARPS and WARPS II;][]{2002ApJS..140..265P,2008ApJS..176..374H}; 
    the Einstein Observatory Extended Medium Sensitivity Survey \citep[EMSS;][]{1994ApJS...94..583G,2004ApJ...609..603H}; the Bright Serendipitous High-Redshift Archival ROSAT Cluster survey \citep[SHARC BRIGHT][]{2000ApJS..126..209R,2003MNRAS.341.1093B}; and Southern SHARC survey \citep[SHARC SOUTH][]{2003MNRAS.341.1093B}.   
\end{itemize}

\citet{2011A&A...534A.109P} first merged surveys with identical data types into a single catalogue, considering each original survey as a sub-catalogue. This concerns  BCS, CIZA, and WARPS, which were merged with their lower-flux extensions eBCS, CIZA-II and  WARPS-II, respectively.  Similarly,  REFLEX and NORAS were merged into a single NORAS/REFLEX catalogue. SHARC SOUTH and SHARC BRIGHT were added to the NORAS/REFLEX catalogue as the respective surveys only differ in their flux limit and sky area. For the very few clusters in common between sub-catalogues (e.g. owing to small region overlaps), the most optimal measurement was kept. 

\citet{2011A&A...534A.109P} then carefully identified clusters appearing in several catalogues and defined a priority order to determine the catalogue to be used as the primary input source. This priority is based principally on catalogue size and the availability of an aperture luminosity (see Sect.~\ref{sec:ovlp} for further details). Unless otherwise stated in very specific cases, in the present work we retain the catalogue priority defined for the MCXC-I. We refer to this primary source catalogue simply as the cluster catalogue (or sub-catalogue).  As in \citet{2011A&A...534A.109P}, we will use the term `overlap' to refer to the catalogues, corresponding sources, and/or derived properties that are not derived from the primary source catalogue.
 
For each cluster, MCXC-I provides a name, coordinates, redshift, standardised [0.1-2.4] keV band luminosity, $\Lv$, total mass, $\Mv$, and radius, $\Rv$, calculated in each case using the luminosity from the source catalogue. MCXC-I additionally furnishes information on overlaps between the input catalogues and the luminosity ratios when measurements from different surveys were available. The MCXC-I release is available online from CDS/VizieR\footnote{\url{http://cdsarc.u-strasbg.fr/viz-bin/qcat?J/A+A/534/A109}} and NED\footnote{\url{https://ned.ipac.caltech.edu/inrefcode?search_type=Search&refcode=2011A\%26A...534A.109P}} servers.

\subsection{New entries from original MCXC-I source catalogues}\label{sec:newmcxc}

When constructing MCXC-I, \citet{2011A&A...534A.109P} only considered objects with a measured redshift and luminosity. They also removed clusters identified as false or likely false in the input source catalogues.  We first revised this information using literature and galaxy surveys, particularly the SDSS DR17\footnote{\url{https://www.sdss.org/}},
2dF\footnote{\url{http://www.2dfgrs.net/}}, and 6dF\footnote{\url{http://www.6dfgrs.net/}} surveys. This revision concerns four catalogues: SGP, NORAS, WARPS, and 160SD. NORAS, SGP, and WARPS were the three MCXC-I input catalogues that do not include full validation and/or redshift information, derived either from literature searches or dedicated optical follow-up studies. The 160SD catalogue includes some candidates with probable false status, which can now be changed to bona fide clusters. 

\subsubsection{NORAS clusters}\label{sec:noras}

The NORAS catalogue   \citep[][Table 7]{2000ApJS..129..435B} includes 27 sources without redshift: nine candidates from the nominal cluster detection in RASS (their Table~1), one being identified as the Abell cluster \object{Abell\,2302} without a redshift at the time of publication, and 18 RASS detections at the positions of known Abell clusters (their Table~7). We searched for new counterparts and redshifts by combining information from several sources.

\begin{enumerate}[]
\item NED database search:
\begin{itemize}[]
\item By {\tt Name} on the  Abell clusters;
\item In {\tt refcode} mode on the NORAS publication. This provides the NED cross-identification of the NORAS clusters with other objects in the literature, and the corresponding redshift measurements when available;
\item  Within 5\arcmin of the NORAS position. This reveals possible counterparts missed by the NED cross-identification algorithm. This proved to be essential because NED missed most of the cross-matches between the NORAS and SDSS-based cluster catalogues.  
\end{itemize}
\item Complementary cross-match with optical and SZ catalogues. Although in principle redundant with the NED positional search, this directly provides useful information on the potential counterpart  (such as on source redshift or mass proxies).
\begin{itemize}
\item SZ detected clusters: we used the meta-catalogue of SZ clusters MCSZ \citep{mcsz} which includes {\it Planck}, ACT and SPT clusters, with a compilation of new redshifts from published Planck cluster optical follow-up, not all of which was included in NED.
\item  The updated versions of the redMaPPer (RM) catalogue based on SDSS DR8 and DES verification data \citep{2016ApJS..224....1R}, as well as the  WHL12 catalogue \citep{2012ApJS..199...34W}, updated by \citet{2015ApJ...807..178W} using SDSS-DR12 data. When a cluster counterpart appeared in both WHL and RM catalogues with a spectroscopic redshift, we adopted the WHL value, which is based on more recent SDSS data and is computed from the mean of spectroscopic members (including the BCG), while the RM spectroscopic redshift value is that of the central galaxy only.  
\end{itemize}
\end{enumerate}

%
\begin{table*}[!h]
\caption{Summary of new MCXC-II clusters from the NORAS and SGP catalogues, with newly identified  redshifts.}   
\centering
\resizebox{\textwidth}{!} {
\begin{tabular}{ lllcllllll }
\toprule
\toprule
        \multicolumn{1}{c}{{\bf Name MCXC}} &
        \multicolumn{1}{c}{} &
    \multicolumn{2}{c}{{  Redshift  }} & 
       \multicolumn{1}{c}{} &
     \multicolumn{4}{c}{{  Redshift origin}} &
     \multicolumn{1}{c}{{ XSZ match }} \\
\cmidrule{3-4} 
 \cmidrule{6-9} 
        \multicolumn{1}{c}{{  }} & 
        \multicolumn{1}{c}{{  }} & 
   \multicolumn{1}{c}{{$z$}} &  
   \multicolumn{1}{c}{{type}} &  
        \multicolumn{1}{c}{{  }} & 
    \multicolumn{1}{c}{{Counterpart}} &
    \multicolumn{1}{c}{{Dist[~$\arcmin$\,;~{$\Rv$]}}} &
         \multicolumn{1}{l}{{Ref.}} &
   \multicolumn{1}{l}{{ Nz }}  \\
\midrule
        \multicolumn{1}{c}{{ \bf NORAS }}  & \multicolumn{1}{c}{ {ID CAT} } & \\
\cmidrule{1-2} 
J0909.3+5133 & A746 & 0.215 & S && \object{PSZ2 G166.62+42.13} & 2.06; 0.43 \tablefootmark{(b)}&1 & & PSZRX \\ 
J0922.2+1225 & A791 & 0.1866 & S && \object{WHL J092216.1+122554} \tablefootmark{(a)}  &0.35; 0.08 &  2   \\
J0957.5+1938 & A903 & 0.1709& S &&  \object{WHL J095734.1+193814} \tablefootmark{(a)} &0.56; 0.11 &  2   \\
J1009.0+1400 & A937 & 0.167 & S && \object{WHL J100904.7+140130}  & 1.67; 0.34&3,pw &  3   \\
J1012.2+0625 & A949 & 0.0749 & S && \object{WHL J101223.3+062544}  & 1.39; 0.17& 2 &     \\
J1013.7+1946 & A952 & 0.1121 & S && \object{WHL~J101346.4+194550} \tablefootmark{(a)} & 1.03; 0.14& 2    \\
J1016.2+4108 & A958 & 0.2755 & S && \object{WHL J101620.6+410545} \tablefootmark{(a)}    &2.63; 0.55  \tablefootmark{(b)} &2,pw &13   \\
J1125.2+4229 & A1253 & 0.1891 & S && \object{WHL J112516.4+422912} \tablefootmark{(a)}    & 0.53; 0.11 &2,pw &17 \\
J1210.0+0630  &  - & 0.1355 & S && \object{RM J121007.8+063128.3}   & 1.58; 0.30&4,pw&  7  \\
J1213.0+3411 & A1492 & 0.2292 & S && \object{WHL J121307.5+341034} \tablefootmark{(a)}    &1.04; 0.24 &2 \\
J1233.9+1511 & A1560 & 0.285 & S && \object{ACT-CL J1233.9+1511}  & 1.00; 0.21&5& &  PSZ2, PSZRX \\  
J1310.4+2151 & A1686 & 0.2734 & S && \object{RMJ131022.2+215005.6} & 1.97; 0.50  \tablefootmark{(b)} &4,pw &  5,M &    PSZ2 \\
J1320.6+3746 & A1715 & 0.2385& S && \object{RMJ132028.2+374623.3} & 2.39; 0.56 \tablefootmark{(b)} & 4,pw &  11  \\
J1327.3+0337 & A1743 & 0.225 & S &&  \object{ACT-CL J1327.3+0337} &0.29; 0.06 & 5 &   \\
J1331.5+0451 & A1753 & 0.1723 & S &&  \object{ACT-CL J1331.6+0452} &2.35; 0.46 \tablefootmark{(b)} &5  \\
J1644.9+0140& -& 0.336 & P & & \object{MACS J1644.9+0139} &1.03; 0.31& 6 & &     ACT \\
J1647.4+0441 &  - & 0.2845 & P && \object{WHL J164727.5+044048}  &0.49;0.13 &7&  &    ACT \\
J1738.1+6006 & -  & 0.372 & P && \object{MACS J1738.1+6006} &0.23; 0.07 &6 & & PSZRX  \\  
J1819.9+5710 & A2302 & 0.179 & E && \object{A2302} & 1.62; 0.31  &8& &PSZ1  \\  
J1920.1+6318  & - & 0.0752 & P &&  \object{RXGCC 802} & 1.41; 0.18 &9 & & \\
J1921.3+7433 & -  & 0.101 &   S && \object{PSZ2 G106.11+2411} & 0.59; 0.08&10 & &  PSZRX\\  
\midrule
        \multicolumn{1}{c}{{ \bf SGP }} & \multicolumn{1}{c}{ {ID CAT} } & \\
\cmidrule{1-2} 
J0007.4-2809 & A2726 & 0.0609 & S 	&& \object{A2726}   &2.28; 0.27& 11  \\
J0012.9-0853 &  & 0.338 & S & & \object{ACT-CL J0012.8-0855} \tablefootmark{(c)}& 4.20; 1.17 \tablefootmark{(b)} & 5 &  &  PSZ2    \\
J0033.8-0750 & A56 & 0.304 & S 	&&  \object{MACS J0033.8-0751} &0.73; 0.19 &6,pw  & 6  & PSZ2, PSZRX \\ 
J0035.6+0138 &  & 0.0799 & S && redMaPPer MEM\_MATCH 07496 ?& 1.10; 0.15&12,pw \tablefootmark{(d)} & 16 &  \\ 
J0122.2-2131 & A185 & 0.1663 & P 		&& \object{WHL J012214.9-213105}  & 0.59; 0.12& 7 \\
J0139.9-0555  &  & 0.4505&  S 	&&  \object{ACT-CL J0140.0-0554} & 0.77; 0.25 & 5 &&    MACS-DR3 \\ 
J0143.4-4614 & A2937 & 0.0776 & S 	&& \object{A2937} & 1.46; 0.20& 13 &   \\  
J0148.2-3155 & A2943 & 0.1489 & S 	&&  \object{A2943} & 0.05; 0.01& 14 &  &   PSZRX   \\
J0148.3-0406 &        & 0.0885 & S  && \object{WHL J014827.6-040747} &2.06; 0.29 &2 \\
J0213.9-0253 &   & 0.1720 & S 			&& \object{WHL J021356.4-025325} &0.47; 0.09 & 7,pw &  10  &   \\
J0218.3-3141 &   & 0.2755 & S 	&&  \object{SPT-CLJ0218-3142}  & 0.77; 0.19& 15,pw \tablefootmark{(e)}&  1  &     ACT/PSZ2/PSZRX\\
J0229.9-1316 &  & 0.05647 & S 	&&  \object{[DZ2015] 408} group &3.97; 0.42  \tablefootmark{(b)}  & 16 &   \\
J0303.2-2735 	& A3082 & 0.268 & P &&  \object{SPT-CL J0303-2736} &0.84; 0.20 & 15    \\ 
J0317.7-4848 & A3113 & 0.164 & S 		&&  \object{SPT-CLJ0317-4849} & 0.98; 0.19&15 & &   PSZ2,PSZRX \\  
J2231.2-3802 &   & 0.0756 & S 		&& \object{SJD J2231.1-3801}  & 1.87; 0.25& 17 &  \\
J2241.9-4235 & A3902 & 0.200 & P 		&& \object{SPT-CLJ2241-4236} &0.53; 0.12& 15   \\  
J2355.1-2834 & A4054 & 0.1866 & S 	&& \object{A4054} &2.20; 0.48 \tablefootmark{(b)} & 14& \\
J2356.0-0129 &   & 0.038 & S 			&& \object{[DZ2015] 540} group incl. AGN & 8.82; 0.72 \tablefootmark{(b)} & 16 &     \\
\hline
\end{tabular}}
\tablebib{
(1) \citet{str18}; 
(2)  \citet{2015ApJ...807..178W};
(3) \citet{2010ApJS..187..272W}; 
 (4) \citet{2016ApJS..224....1R}; 
(5) \citet{2021ApJS..253....3H};
(6)  \citet{2018MNRAS.479..844R}; 
(7) \citet{2012ApJS..199...34W};
 (8)  \citet{1999ApJ...519..533D}; 
(9) \citet{2022A&A...658A..59X}; 
(10) \citet{2019AstL...45...49Z};
(11) \citet{2018ApJ...869..145C};
(12) \citet{wet22}
(13) \citet{2009AJ....137.4795C};
(14) \citet{2002MNRAS.329...87D}; 
(15)  \citet{2015ApJS..216...27B};
(16) \citet{dia15}; 
(17) \citet{2016MNRAS.458.3083S};
}
\tablefoot{  
Col. 1-2: MCXC name and identification in source catalogue; Col.  3-4: Redshift and  redshift type (S: spectroscopic; P: photometric, E:estimated); Col. 5-8: details on the redshift estimate;  Counterpart identified whose redshift is estimated, its distance to the X--ray position, and corresponding reference. Present work (pw) is added if we refined the published redshift estimate, using $N_{\rm z}$ (Col. 8) galaxies with $z_{spec}$  taken from SDSS-DR18, except J0218.3-3141 based on  2dFGRS;   the BCG is always included with M meaning possible merging system, with more than one BCG. Col. 9: Match with  clusters  in other X-ray or SZ catalogues: ACT \citep{2021ApJS..253....3H},  PSZ1 \citep{psz1rev}, PSZ2 \citep{psz2}; PSZRX refers to COMPRASS, the  combined X-ray-SZ detection  catalogue of \citet{tar19}.
\\

(a)  also matches a Redmapper cluster \citep{2016ApJS..224....1R} 
(b) The large distance and cross-identification is discussed in Appendix~\ref{app:new_znorasssgp}
(c): also \object{WHL J001248.9-085535} ;  (d) The ID is uncertain. The published spectroscopic redshift of redMaPPer MEM\_MATCH  07496 of the DES Redmapper catalogue \citep{wet22} is $z=0.104$ is different with our SDSS redshift. There is a clear X-ray emission in the SWIFT image. (e) NED preferred redshift is 0.275, equal to SPT zphot. However, ACT gives zspec=0.270 from DeCals. Our analysis from 2dFGRS confirms z=0.275  }
  \label{tab:newznorassgp}
\end{table*}

We were able to revise the validation or provide a redshift for 26 of the 27 sources, largely owing to the broad overlap between the NORAS and SDSS (and BOSS) footprints and the existence of SZ surveys:
\begin{itemize}[]
\item 
Twenty-one clusters have new redshifts. All but three of these come from cross-identification with optical clusters from SDSS-based catalogues, redMaPPer \citep{2016ApJS..224....1R}  and/or WHL \citep{2015ApJ...807..178W},  and/or with SZ clusters with redshifts from the PSZ or ACT catalogues  \citep{psz2,2021ApJS..253....3H}. Except for five clusters, the distance between the NORAS source and its counterpart is smaller than both $2\arcmin$ and $0.3\Tv$, with a median distance of $1.03\arcmin$ and $0.17\Tv$.  Taking into account the RASS resolution ($1.5\arcmin$ half power radius) and the possibility of an offset between the X-ray and the optical/SZ centre, this confirms the cross-identification, as well as the high richness of the optical clusters (e.g. median richness  $\RL=65$ for the optical clusters from \citet{2015ApJ...807..178W}, where $\RL$ is defined from the total r-band luminosity within $\Rv$). The five clusters with the largest counterpart distances are discussed individually in Appendix~\ref{app:noras}.

\item When the redshift of the identified SDSS cluster counterpart was photometric or based only on a single galaxy spectrum, we used SDSS-DR17 to refine the redshift estimate. We looked for galaxies with $z_{\rm spec}$ around the estimated $z$ and within $\Tv$ of the optical centre. The refined redshift, derived from the biweight mean estimate \citep{1990AJ....100...32B},  and the number of galaxies used for redshift estimate, $N_{\rm z}$ is  given in  Table~\ref{tab:newznorassgp}. For RM clusters with $z_{\rm spec}$ from the central galaxy, a new redshift is only given if $N_{\rm z}\ge 5$. Detailed information, including notes on some clusters, are given in Table~\ref{tab:newznorassgp}.

\item 
Three NORAS clusters are in fact eBCS clusters, which are included in  MCXC-I (\object{MCXC\,J0906.4+1020}, \object{MCXC\,J0448.2+0953}  and \object{MCXC\,J1323.5+1117}). \cite{2011A&A...534A.109P} performed the cross-identification after removing objects without redshift in the source catalogues, which could not be entered in the overlap information owing to the lack of luminosity measurement.

\item
Two NORAS sources, \object{RX\,J0910.0+3533} and \object{RXC\,J0913.5+8133},  are identified in NED with a star and a QSO, respectively. These are targeted  NORAS detections towards the Abell clusters \object{Abell\,0752} and \object{Abell\,0723}. The separation between the X-ray and Abell position is large, at $7.8\arcmin$ and $6.7\arcmin$ respectively, while the objects in question have a very low extent probability parameter. Thus further supports our conclusion that the counterparts are in fact false cluster detections. 
\end{itemize}

In summary, 24 sources were found to be bona fide clusters with redshifts and 2 turned out to be false candidates. We failed to find a clear counterpart for the last object, RXC J2035.7+0046. The nearest NED counterpart, $3.4\arcmin$ away, is a poor WHL group at $z=0.22$.

%
\begin{table*}[!h]
\caption{\footnotesize Summary of new MCXC-II  clusters from the 160SD catalogue \citep{1998ApJ...502..558V,2003ApJ...594..154M}, previously classified as likely false.  }   
\centering

\begin{tabular}{lrllllcl}
\toprule
\toprule
        \multicolumn{1}{c}{{\bf Name MCXC}} &
        \multicolumn{1}{c}{[{VMF98]}} &
    \multicolumn{2}{c}{{  Redshift  }} & &
     \multicolumn{3}{c}{{  Redshift origin}} \\
\cmidrule{3-4} 
 \cmidrule{6-8} 
        \multicolumn{1}{c}{{  }} & 
        \multicolumn{1}{c}{{Number }} & 
   \multicolumn{1}{c}{{$z$}} &  
   \multicolumn{1}{c}{{type}} & &
     \multicolumn{1}{c}{{Counterpart }} &
 \multicolumn{1}{c}{{ Dist[~$\arcmin$\,;~$\Tv$]}} &
  \multicolumn{1}{c}{{ Reference }} \\
\midrule
J0236.0-5225 & 27 & 0.59& P && \object{X-CLASS 0418} & 0.21; 0.13 & \citealt{2017MNRAS.468..662R}  \\
J0831.2+4905 & 54 & 0.500 & S && \object{RM J083112.0+490455.2} & 0.68, 0.34 &\citealt{2016ApJS..224....1R}, pw$^{(a)}$ \\
J0953.5+4758&  78 & 0.2028 & S && \object{PDCS 040} & 0.94; 0.31 & \citealt{1999AJ....118.2002H}   \\
J1418.7+0644 & 160 &  0.673 & P &&  \object{ACT-CL J1418.7+0644}  & 0.15; 0.08 & \citealt{2021ApJS..253....3H}   \\
J2004.8-5603 & 197 &  0.761 & P && \object{ACT-CL J2004.8-5603} & 0.50;  0.31 &  \citealt{2021ApJS..253....3H} \\
\hline
\end{tabular}
\tablefoot{Cols. 1-2: MCXC name and index  in \citet{1998ApJ...502..558V} catalogue [VMF98]. Cols. 2-3: redshift and  redshift type (S: spectroscopic; P: photometric); Cols. 4-6: redshift estimate details; redshift of identified counterpart; separation distance from 160SD position (in arcmin and in units of $\Tv$); reference. $^{\rm (a)}$ \citealt{2016ApJS..224....1R} estimated a photometric redshift of $z_{\rm}=0.52$. From SDSS DR16 data of four galaxies, we derived a spectroscopic value of $z_{\rm spec}=0.500$.}
 \label{tab:newz160SD}
\end{table*}


%
\subsubsection{SGP clusters}\label{sec:sgp}

The  SGP catalogues, obtained from RASS data in the South Galactic Pole region, include 29 sources without redshift, of which 11 objects were identified with Abell clusters.  One  SGP object, \object{RXC\,J0152.5-2853}, is in fact a MACS cluster, which was included in MCXC-I. This was missed by \cite{2011A&A...534A.109P}, because the cross-identification was performed after removing this object without redshift from the SGP catalogue. 

For the other objects, we performed a counterpart and redshift search identical to that for the NORAS clusters described above. The main difference with the SGP catalogue is that the overlap with SDSS is much smaller than for NORAS; however, this is partially compensated for by coverage with the 2dF and 6dF southern optical surveys and the SPT SZ survey. However, the SGP survey is deeper than NORAS and the identification is more difficult, in particular for low-mass objects at low redshift.

As detailed in Table~\ref{tab:newznorassgp}, 18 clusters now have a redshift derived from the literature. These include: four Abell clusters with redshift available from 2dF data; six objects cross-identified with SZ clusters; one MACS cluster (not included as such in MCXC-II as no X-ray flux was published); four objects cross-identified with WHL or RM clusters; and three with optical local groups. 

For most of the objects, the association is obvious in view of the small separation distance ($D<2\arcmin$ or $<0.3\,\Tv$) and the estimated mass of the counterpart. However, the cross-identification of four SGP objects required deeper individual studies, in view of their larger separation distances and/or the low mass of the potential counterpart. The new counterparts include the association of \object{MCXC\,J0012.9-0853} with the ACT cluster \object{ACT-CL\,J0012.8-0855} at $D=4\arcmin$, the confirmation of the identification of \object{MCXC\,J2355.1-2834} with \object{Abell\,4054} ($D=0.5\,\Tv$), and 
the cross-match of \object{MCXC\,J0229.9-1316} and \object{MCXC\,J2356.0-0129} with local [DZ2015] compact groups. In the first and last cases, there is clear contamination from point-like emission. These four clusters are discussed individually in Appendix~\ref{app:sgp}.

\subsubsection{WARPS clusters}\label{sec:warps}

Five WARPS-I sources were not included in MCXC-I, owing to a lack of redshift and/or luminosity information. These are additional clusters and candidates which do not  belong to the statistically complete, flux-limited, WARPS sample \citep[][Table 5]{2002ApJS..140..265P}: 

\begin{itemize}[]
\item Two low-flux detections, \object{WARP\,J0236.0-5224} and \object{WARP\,J0255.3+0004}, were without redshift, the optical follow-up having not been completed.  \object{WARP\,J0236.0-5224} was associated  with the 160SD cluster \object{[VMF98]\,027} by \citet[][Table 5]{2002ApJS..140..265P}, but was flagged as a false candidate by  \citet{1998ApJ...502..558V} and \citet{2003ApJ...594..154M}. We found that this source is in fact the distant cluster \object{X-CLASS\,0418}, at $z=0.59$, re-discovered in the XMM-{\it Newton} serendipitous survey by  \citet{2017MNRAS.468..662R}. \object{WARP\,J0255.3+0004} is in the SDSS footprint, but we were unable to find any obvious counterpart.
\item \object{WARP\,J1407.6+3415}  and \object{WARP\,J2320.7+1659} do not have a published luminosity. \object{WARP\,J2320.7+1659} is a cluster at $z=0.499$ contaminated by an AGN point source at $z=1.8$. Concerning \object{WARP\,J1407.6+3415}, \citet{2002ApJS..140..265P} discussed the low quality of its redshift estimate ($z=0.577$), and the possibility of AGN contamination. The {\it Chandra} archive image clearly shows extended emission; however,  there is no obvious AGN contamination, while SDSS-DR17 spectroscopic data confirm the redshift measurement. A galaxy with $z_{\rm spec}=0.5805$, is located at the {\it Chandra} X-ray peak, likely the BCG.  We found another galaxy with $z_{\rm spec}=0.5748$  at  $\sim 0.5\,\Tv$. Both clusters are now included in MCXC-II, with a luminosity estimated from the flux (see Sect.~\ref{sec:method}). The redshift of  \object{WARP\,J1407.6+3415}  is set to $z_{\rm spec}=0.5805$, and a note on the point source contamination was added for \object{WARP\,J2320.7+1659}. 
\item  \object{WARP\,J1515.5+4346} has neither a published luminosity nor a flux measurement,  the identification of the source being uncertain. \citet{2002ApJS..140..265P} discuss two possible redshifts for the counterpart, $z=0.136$ and $z=0.237$. However, they mis-identified the source with the 160SD cluster \object{[VMF98]\,169} (\object{MCXC\,J1515.6+4350}), 4\arcmin\ to the North, at $z=0.243$ \citep{2003ApJ...594..154M}.  \object{WARP\,J1515.5+4346} in fact coincides with \object{[VMF98]\,168} (\object{MCXC\,1515.5+4346}, $z=0.137$), located 20\arcsec\ away. \citet{2018ApJ...855..100S} also detected a clear overdensity at that location and redshift in the HectoMAP Cluster Survey (HMxcl151550.0+434556) )\footnote{There is also some confusion between \object{[VMF98]\,168} and \object{[VMF]\,169}, in the discussion of the corresponding MCXC-I clusters by \citet{2018ApJ...855..100S}. This is likely due to an inversion of the redshift of the two 160SD clusters in the original catalogue \citep{1998ApJ...502..558V}, which was subsequently corrected by \citet{2003ApJ...594..154M}. The MCXC-I positions and redshift are correct.}.  We could not compute a luminosity for this WARP detection, since its flux was not published, but the object is now included in the overlap information for \object{MCXC\,1515.5+4346}. 
\end{itemize}

In summary, all WARPS-I are included in MCXC-II, with the exception of \object{WARP\,J0255.3+0004}.  

Finally, there is some ambiguity associated with the  WARP-II double-peaked source, \object{WARPS\,J1419.9+0634}.  \citet{2008ApJS..176..374H} cross-identified the source  with the 160SD cluster \object{RX\,J1419.9+0634} (\object{[VMF98]\,162}),  at $z=0.549$ \citep{2003ApJ...594..154M}. From their optical follow-up,  they split the source into two clusters,  \object{WARPS\,J1419.9+0634\,W} at $z=0.5641$ and  \object{WARPS\,J1419.9+0634\,E} at $z=0.5740$.  \object{MCXC\,J1419.8+0634} in the MCXC catalogue (\citep{2011A&A...534A.109P}) corresponds to \object{WARPS\,J1419.9+0634\,W}, with \object{[VMF98]\,162} as overlap. It is unclear whether there are two separate clusters (likely in the process of merging, given their similar redshift), or whether \object{WARPS\,J1419.9+0634} is a single bimodal (post-merger) cluster. For completeness, we added \object{WARPS\,J1419.9+0634\,E} as \object{MCXC\,J1419.9+0634} to the MCXC-II catalogue. 

%
 \begin{table*}[t]
\caption{ \label{tab:dupl}  MCXC-I cluster pairs at a separation distance of less than $\Tv$, that are considered as the same object in MCXC-II.  } 

\resizebox{\textwidth}{!}{\centering
\begin{tabular}{llllllllllllllll}
\toprule
\toprule
  \multicolumn{2}{c}{MCXC} &&
  \multicolumn{2}{c}{Sub-catalogue} &&
  \multicolumn{2}{c}{Distance} &&
 \multicolumn{3}{c}{Redshift} &&
    \multicolumn{2}{c}{$\Lv$} &
   \multicolumn{1}{c}{Note} \\

\cmidrule{1-2} 
\cmidrule{4-5} 
\cmidrule{10-12} 
\cmidrule{14-15} 
 \multicolumn{1}{c}{ID1} & 
 \multicolumn{1}{c}{ID2} && 
 \multicolumn{1}{c}{ID1} &  
 \multicolumn{1}{c}{ID2} &&
 \multicolumn{1}{c}{(\,\arcmin\,)} &
\multicolumn{1}{c}{($\Tv$)} &&
 \multicolumn{1}{c}{ID1} &
 \multicolumn{1}{c}{ID2} &
 \multicolumn{1}{c}{MCXC} &&
 \multicolumn{1}{c}{ID1} &
 \multicolumn{1}{c}{ID2} \\ 
\midrule
{\bf J0034.2-0204} 	&J0034.6-0208   && SGP & REFLEX && 6.75  & 0.70  && 0.0822 & 0.0812  & 0.0822  && 1.24  & 1.28  &Simbad;    XMM ,  C\\
{\bf J0125.4+0145} 	& J0125.4+0144      	&& NORAS & BCS                	&& 1.8 &  0.09	&& 0.0183 & 0.0181      &  0.0183	&& 0.042 & 0.061    	& XMM \\
{\bf J1010.2+5430}	& J1010.2+5429	&& 400SD\_SER  & NORAS   	&& 0.9 & 0.10  	&& 0.047    & 0.045     & 0.047 	&& 0.013 &0.076   	& NED,PSPC pointed; A, \\
{\bf J1058.1+0135} 	& J1058.2+0136  	&& REFLEX & 400SD\_SER 	&& 1.9 &  0.17     	&& 0.0398 & 0.0385    & 0.0398    	&& 0.084 &  0.068       & NED;  XMM \\
{\bf J1311.7+2201} 	& J1311.5+2200       	&& NORAS & eBCS     && 2.8 &  	0.60        &&    0.1716  & 0.266    &  0.1716	&& 1.00   &   2.48     & XMM; B,C  \\
{\bf J1652.9+4009}	& J1652.6+4011	&&  NORAS  & eBCS	         && 3.8 & 0.66     	&& 0.1492 &  0.1481	   & 0.1492   	&& 1.67  & 1.57     	& NED; Chandra; C  \\
{\bf J2350.5+2929}	&J2350.5+2931	&&  NORAS &BCS      	   	&& 2.0  & 	  0.33      	&&0.1498  & 0.095	   & 0.1498  	&&  2.03 &   2.66   & NED; B \\
\bottomrule
\end{tabular}}
\tablefoot{Cols. 1-2: Name of each cluster. Cols. 3-4: source catalogue. Cols. 5-7: redshift in source catalogue; final MCXC-II redshift.  Cols. 8-9: $\Lv$ luminosity computed from the flux at the final redshift. Col. 10: Notes, where we indicate whether the duplicate was identified by NED or Simbad, respectively, and the existence of \xmm,  \chandra, or ROSAT-PSPC  pointed observations. Cases of differences in $\Lv$ (A), in redshift (B), or of large positional offset as compared to $\Tv$ (C) are flagged and discussed in Appendix~\ref{app:dupl}.}
\end{table*}

\subsubsection{Update of 160SD cluster status}\label{sec:160SD}

The 160SD catalogue includes 22 objects flagged as likely false detections, and one source with no follow-up as it is obscured by Arcturus. Some of these sources could be bona fide, likely distant, clusters. A NED search on the source name \footnote{We note that NED  lists the source \object{RX\,J0857.7+2747} ((\object{[VMF98]\,065}) as a  $z=0.5$ cluster. This source was invalidated by  \citet{2003ApJ...594..154M} and the \xmm\  image confirms that it is indeed a superposition of point sources.} indicates that four sources are indeed $z>0.5$ clusters:
\begin{itemize}[]
\item \object{RX\,J0236.0-5225} (\object{[VMF98]\,027}) is an X-CLASS cluster at $z=0.59$, re-discovered serendipitiously in \xmm\ observations by \citet{2017MNRAS.468..662R}.
\item Two objects, \object{RX\,J1418.7+0644} (\object{[VMF98]\,160}) and \object{RX\,J2004.8-5603} (\object{[VMF98]\,197}) coincide with  SZ clusters at $z=0.673$ and $z=0.761$, respectively, from the ACT survey \citep{2021ApJS..253....3H}. The latter was originally considered by \citet{1998ApJ...502..558V} to be a cluster at $z_{\rm phot} = 0.7$.
\item \object{RX\,J0831.2+4905}  (\object{[VMF98]\,054}) matches with the rich ($\lambda=47$) 
 redMaPPer cluster \object{RM\,J083112.0+490455.2} at $z_{\rm phot} = 0.52$ \citep{2016ApJS..224....1R}. We used DSS-DR17 spectroscopic data to refine the redshift of this object. 
\end{itemize}

The small distance between the ROSAT position and the SZ or optical counterpart, $[0.1$--$0.3]\ \Tv$  (Table \ref{tab:newz160SD}), and the consistency of the mass proxy estimates support the above associations.

The case of a fifth source, \object{RX\,J0953.5+4758} (\object{[VMF98]\,078}) is less clear. It is cross-identified by NED with \object{PDCS\,040}, a poor (richness class 0 according to the classification proposed by \citealt{1958ApJS....3..211A}) cluster at $z=0.203$ from the Palomar Distant Cluster Survey \citep{1999AJ....118.2002H}.  It may be surprising that a counterpart at such a redshift was not identified in the 160SD follow-up.  However,  the derived X-ray mass is consistently low:  $\Mv = 7 \times 10^{13}\ \msun$ for  $z=0.203$, that is, a group-scale object.  The distance between the optical and X-ray centres is $0.9\arcmin$.  Although larger than for the other clusters, this distance is still much less than $\sim0.3\ \Tv$, with all five spectroscopic galaxy members lying within $\Tv$.  We therefore added this cluster to MCXC-II, adding a note on a possible chance association. A  search for possible counterparts around the X-ray position for 160SD sources flagged as likely false did not reveal any other firm confirmation. 

\citet{2007ApJS..172..561B} further flagged four 160SD clusters from their optical follow-up of the 400SD survey  as likely false. This was further supported by high-resolution {\it Chandra} observations of one of the sources, \object{[VMF98]\,167} or \object{MCXC\,J1500.8+2244},  which shows a superposition of point sources without extended emission. The source is removed from MCXC-II. 

 In summary, five new  160SD clusters are included in MCXC-II (see Table \ref{tab:newz160SD}), and one is removed.  

\subsection{Duplicate objects}\label{sec:dupl}

\citet{2011A&A...534A.109P}  identified objects that appeared in different catalogues based on angular separation distance, redshift, and examination of RASS and pointed ROSAT observations (see their Sect.~4). Their approach was conservative, meaning that they did not remove any cluster if the association was not certain. 

Defining a duplicate as the cross-identification of two MCXC-I clusters as the same object, NED identifies four cases, and Simbad identifies one duplicate more. We reviewed these cases and systematically searched for other duplicates. The angular separation distance of the same object detected in two surveys may be not only depend on the angular resolution of the survey, but also on object size (the centre determination depends on the detection method, particularly for complex morphologies). Therefore, we used the distance, $D$, of each cluster to its closest neighbour relative to  $\Tv$, as the criterion. We found 15 pairs with a separation distance less than $\Tv$.
\begin{itemize}[label=$-$]
\item Four are simply well-resolved pairs of different clusters detected in the same survey. In three cases, the two objects are at the same redshift, and therefore possibly in a pre-merging state. 
\item Three other pairs correspond to close-by clusters lying at the same redshift, detected in two different surveys. These are  \object{MCXC\,J2318.5+1842} (BCS, $z=0.039$, \object{Abell\,2572B}) at $1.6\arcmin$ from  \object{MCXC\,J2318.4+1843} (NORAS, $z=0.040$, \object{Abell\,2572}); \object{MCXC\,J2218.2-0350} (EMSS, $z=0.09$) at $6.7\arcmin\ (0.8\, \Tv$) from  \object{MCXC\,J2218.6-0346} (REFLEX, $z=0.09$)  both also detected in SGP; and \object{MCXC\,J2306.5-1319} (REFLEX, revised $z=0.11$) at $6.9\arcmin\ (\sim 0.8\,\Tv$) from \object{MCXC\,J2306.8$-$1324} (SGP,  $z=0.066$). In all cases the \xmm\ image shows that these are different objects.  
\item One case, \object{MCXC\,J1329.5+1147} (NORAS, $z=0.022$) at  $4.1\arcmin\ (\sim 0.4\,\Tv$) from \object{MCXC\,J1329.4+1143} (400SD, $z=0.023$) is ambiguous.  The \xmm\ image shows a very diffuse object with a complex morphology and the NORAS   luminosity is about three times greater than the 400SD value. Several bright point sources are also superimposed on the diffuse emission. The 400SD and NORAS objects may be different components of the same structure, and we therefore kept them both. 
%
%
\begin{table*}[t]
     \caption{Summary of the construction of the MACS catalogue and its inclusion in MCXC-II.   }
   \centering
 \resizebox{\textwidth}{!} {
    \begin{tabular}{lcllclcccl}
    \toprule
    \toprule
 Data Release &  $N_{\rm DR}$ & Reference 	&  {\tt SUB\_CAT}     	& $N_{\rm subcat}$ & OVLP$_{\rm prime}$  & $N_{\rm OVLP,prime}$ & $N_{\rm MCXC-II}$ & OVLP$_{\rm sec.}$  & $N_{\rm OVLP, secondary}$ \\
    \midrule
        DR1 	& 12 &\citet{2007ApJ...661L..33E}        	&  MACS\_DR1 	& 12 	 	&      -                &  0	& 12	& 2 &EMSS   \\
           	&    	&                                                      	&                          	&     		&                       &	&	&  1 &NORAS \\
        DR2 	& 35 & \citet{2010MNRAS.407...83E} 	&  MACS\_DR2 	&  35	    	& NORAS 	& 6	& 23	&  1 &EMSS  \\
            	&    	&                                                       	&                           	&        	& REFLEX        &5	&	& 2 & REFLEX    \\
            	&    	&                                                        &                           	&        	& BCS        	& 1	&	&    1&  SGP       \\
        DR3 & 54 & \citet{2012MNRAS.420.2120M} 	& MACS\_DR3   	&  22   	& NORAS 	& 2	& 19    \\  
             	&    	&                             	            		&                               &            	&SGP               	& 1         	&                        &	&      \\     
       MISC & 45 & \citet{2018MNRAS.479..844R} 	& MACS\_MISC       	& 16     	&  NORAS 	& 5	& 9\\
           	&     &                           	            		&                               &            	&  REFLEX      & 2 \\
    \midrule
    TOTAL  &    &                    			                 &                           	& 85       	&                        & 22	& 63 \\
    \bottomrule 
       \end{tabular} }
       \tablefoot{Cols. 1-3: Data release with sample size and reference. Col. 4: Name of the sub-catalogue including the clusters first appearing in the data release (number given in  Col. 5). Cols. 6-7:  Overlapping  MCXC-II primary input catalogues and corresponding number of clusters.  Col. 8: Number of  MACS clusters included in MCXC-II after removing these multiple entries.  Cols. 9-10 Number of secondary detections of MACS MCXC-II clusters and corresponding catalogues. }
    \label{tab:macscat}
\end{table*}

\item Finally we found seven duplicates, including those already identified by NED or Simbad. These are listed in Table~\ref{tab:dupl}.  Two cases are trivial: a single object is detected in the \xmm\ archive image, the distance is small ($<2\arcmin$ or $0.17\,\Tv$), and the redshifts are consistent. The other cases required deeper investigation because of luminosity or redshift discrepancies (cases A and B in Tab.~\ref{tab:dupl}) or separation distances larger than $0.5\,\Tv$ (case C). The association is based on higher resolution observations with \xmm, \chandra, or ROSAT pointings, and ancillary optical data (primarily SDSS). Appendix~\ref{app:dupl} contains details for each pair. 
\end{itemize}

\subsection{New clusters from MACS and new RASS-based catalogues} 
\label{sec:newcat}

In addition to the input catalogues listed above, the MCXC-II was augmented with new clusters from the REFLEX II, MACS, and RXGCC catalogues. 

\subsubsection{REFLEX-II }\label{sec:reflexII}
REFLEX-II is the extension of the REFLEX catalogue to a lower-flux limit corresponding to 0.6 times the REFLEX limit, as described by \citep{2013A&A...555A..30B}. The 22 clusters at redshift above $z>0.2$ have been published by \citet{ 2012A&A...538A..35C} and are included in MCXC-II. All the objects are new, and there is no overlap with other MCXC-II source catalogues.

\subsubsection{MACS}\label{sec:newmacs}

First described in \citet{2001ApJ...553..668E}, MACS has not been released or published as a single homogeneous catalogue. In the first release of the MCXC, \citet{2011A&A...534A.109P} assembled a consolidated catalogue of 39 MACS clusters, plus 11 objects common to other catalogues, by collecting and merging data from various MACS sub-samples that had been published up to that date \citep{2007ApJ...661L..33E, 2010MNRAS.407...83E, 2008ApJS..174..117M, 2010MNRAS.406.1773M}. Since then, two further MACS sub-samples have been published \citep{2012MNRAS.420.2120M, 2018MNRAS.479..844R}. 

For MCXC-II, we revised the MACS merging strategy to simplify the integration of the new entries from more recent publications and we re-adapted the sub-catalogue naming scheme for consistency. This is summarised in Table~\ref{tab:macscat}.
We assigned a data release (DR) number to each MACS sub-sample in order of its publication date, and constructed the corresponding sub-catalogue with the clusters first appearing in each release.
    \begin{itemize}[]
        \item[-] MACS\_DR1 is a complete sample of the most distant MACS clusters with $z>0.5$, published in \citet{2007ApJ...661L..33E}. It contains 12 objects, with \chandra\ luminosities. 
        \item[-] MACS\_DR2 is a flux-limited sample of 34 clusters in the redshift range  $0.3<z<0.5$, above the flux limit of $2\times10^{-12}$ $\ergs$~cm$^2$ in the $[0.1$--$2.4]\ \keV$ energy band, published in \citet{2010MNRAS.407...83E}. Physical parameters from \chandra\ follow-up were published by \citet{2010MNRAS.406.1773M}. Three further sources were listed by \citet{2010MNRAS.407...83E}: a QSO (\object{MACS\,J1542.0-2915}) and two sources contaminated by AGN emission, \object{MACS\,J0047.3-0810} and \object{MACS\,J1824.3+4309}. The \chandra\ image of the latter shows clear extended emission, and the corresponding cluster properties are published by \citet{2008ApJS..174..117M}. This object is therefore included in the MACS\_DR2 sample, which contains 35 objects in total. 
        \item[-] MACS\_DR3 was built from the third release of MACS clusters,  published by \citet{2012MNRAS.420.2120M}. Based on a complete sample with $\LX>5 \times 10^{44} \ergs$, the publication includes a total of 54 clusters with \chandra\ follow-up. However, only RASS luminosities are provided. DR3 adds 22 clusters to the previous DR1 and DR2 samples. 
        \item[-] Additional MACS clusters recovered from non-catalogue studies were assembled into the supplementary MACS$\_$MISC sub-catalogue.  Here, we retained only MACS objects with available redshift and luminosity information. MACS$\_$MISC is currently based on the study of \citet{2018MNRAS.479..844R}, an HST follow-up of 86 clusters at $0.3<z<0.5$. It also includes  some clusters for the as-yet unpublished extension of MACS to the southern hemisphere (the SMACS catalogue). We consider the 45 clusters with \chandra\  follow-up, which adds 16 clusters to the previous sub-catalogues.
        \item[-] All clusters studied by \citet{2008ApJS..174..117M} are now included in one of the above sub-samples. 
    \end{itemize}

The final concatenated MACS catalogue consists of 85 clusters. The information was integrated into MCXC-II, after handling of multiple entries (see Table~\ref{tab:macscat}). A total of 63 MACS clusters are included in MCXC-II as primary objects. A further 22 clusters that appear in other MCXC-II input catalogues (NORAS, REFLEX, SGP, or BCS) are included in the overlap information, following the standard catalogue input priority scheme. Four further MACS DR1 or DR2 objects coincide with clusters from the NORAS, REFLEX, or SGP catalogues. One is the SGP cluster \object{RXC\,J0152.5-2853} (mentioned in Sect.\ref{sec:sgp}) and three were put as primary objects by \citet{2011A&A...534A.109P} because the MACS redshift was more precise.  In principle, since we computed the luminosity from the flux, we could have updated the redshift and used the standard priority for these systems. However, for continuity of the catalogue, we kept the original MCXC-I primary catalogue designation. Finally, two MACS objects also appear in the EMSS catalogues.

The MCXC-II position of the 12 DR1 clusters is the \chandra\ X-ray centroid determined by \citet{2007ApJ...661L..33E}. For the other clusters, we adopted the latest \chandra\ X-ray peak position published by \citet{2018MNRAS.479..844R}, when available (31 clusters).  The positions for the remaining nine DR2 and eight DR3 clusters were taken from the original catalogues, given by \citet[][the \chandra\ X-ray peak]{2010MNRAS.407...83E} and \citet{2012MNRAS.420.2120M}, respectively. The position of \object{MACS\,J1824.3+4309} is its \chandra\ X-ray centroid as published by \citet{2008ApJS..174..117M}.  

The MACS\_DR1, MACS\_DR2, and MACS$\_$MISC cluster redshifts are  spectroscopic redshifts from the original publications \citep[][respectively]{2007ApJ...661L..33E, 2010MNRAS.407...83E, 2008ApJS..174..117M, 2018MNRAS.479..844R}, except for five DR2 clusters (of which three are prime objects in MCXC-II). For these, the redshift given by \citet{2018MNRAS.479..844R} is slightly different, and we adopted this more recent spectroscopic value. The redshift of the 19 DR3 clusters given by \citet{2012MNRAS.420.2120M} were based on photometric data. For 14 of these, we consolidated the redshift with spectroscopic values from the literature: 11 redshifts from a cross-match with the optical catalogue of \citet{2015ApJ...807..178W}, three redshifts from a cross-match with the ACT catalogue \citep{2021ApJS..253....3H}, and one CLASH cluster \citep{2016ApJS..224...33B}. The difference with the original MACS values is small, $\delta(z)<0.008$, with a median value of $\delta(z)=0.003$. For the remaining clusters,  we adopted either the catalogue value \citep{2012MNRAS.420.2120M}, or the values given by \citep{2018MNRAS.479..844R}, which were published with more significant digits, when available. The MCXC-II position and redshift may therefore differ slightly from those published by \citet{2011A&A...534A.109P} for common clusters, since they used a single reference for all parameters (the primary catalogue from which the luminosity was extracted).

The standardisation of the X-ray luminosity $\Lv$ and mass $\Mv$ measurements for the MACS catalogue is different to that applied to the other {\it ROSAT}-based catalogues. This is described in Sect.~\ref{sec:lm500_macs}.  

\subsubsection{RXGCC}\label{sec:rxgcc}
RXGCC \citep{2022A&A...658A..59X} is a catalogue of galaxy clusters constructed using a dedicated source detection and characterisation algorithm optimised for extended sources in the RASS. It contains 944 groups and clusters above the Galactic plane ($|b|>20^\circ$). The redshifts were estimated from the distribution of spectroscopic and photometric redshifts of galaxies surrounding each detection.

\citet{2022A&A...658A..59X} provide a cross-identification with previously-known clusters from several catalogues, including with MCXC-I \citep{2011A&A...534A.109P}. This cross-identification was based on the positional separation distance $d$ between the objects (using $d<15\arcmin$ and $d<0.5$ Mpc as an association criterion) and their redshift difference ($\Delta z < 0.01$). However, to include the RXGCC catalogue into MCXC-II, we have decided to undertake our own cross-identification between RXGCC and MCXC-II. This allows us to take into account the updates of MCXC-II with respect to the MCXC-I release (including the additional clusters described above, and redshift updates described in Sect. \ref{sec:zrev}), and to use cross-identification criteria that are better adapted to the cluster physical sizes. In particular, our cross-match relies on the angular separation distance between the clusters ($d$) and on their relative distance in terms of the angular size of the MCXC-II cluster ($d/\Tv^{\rm MCXC}$).

%
\begin{figure}[!t]
	\begin{centering}
		\includegraphics[width=\columnwidth]{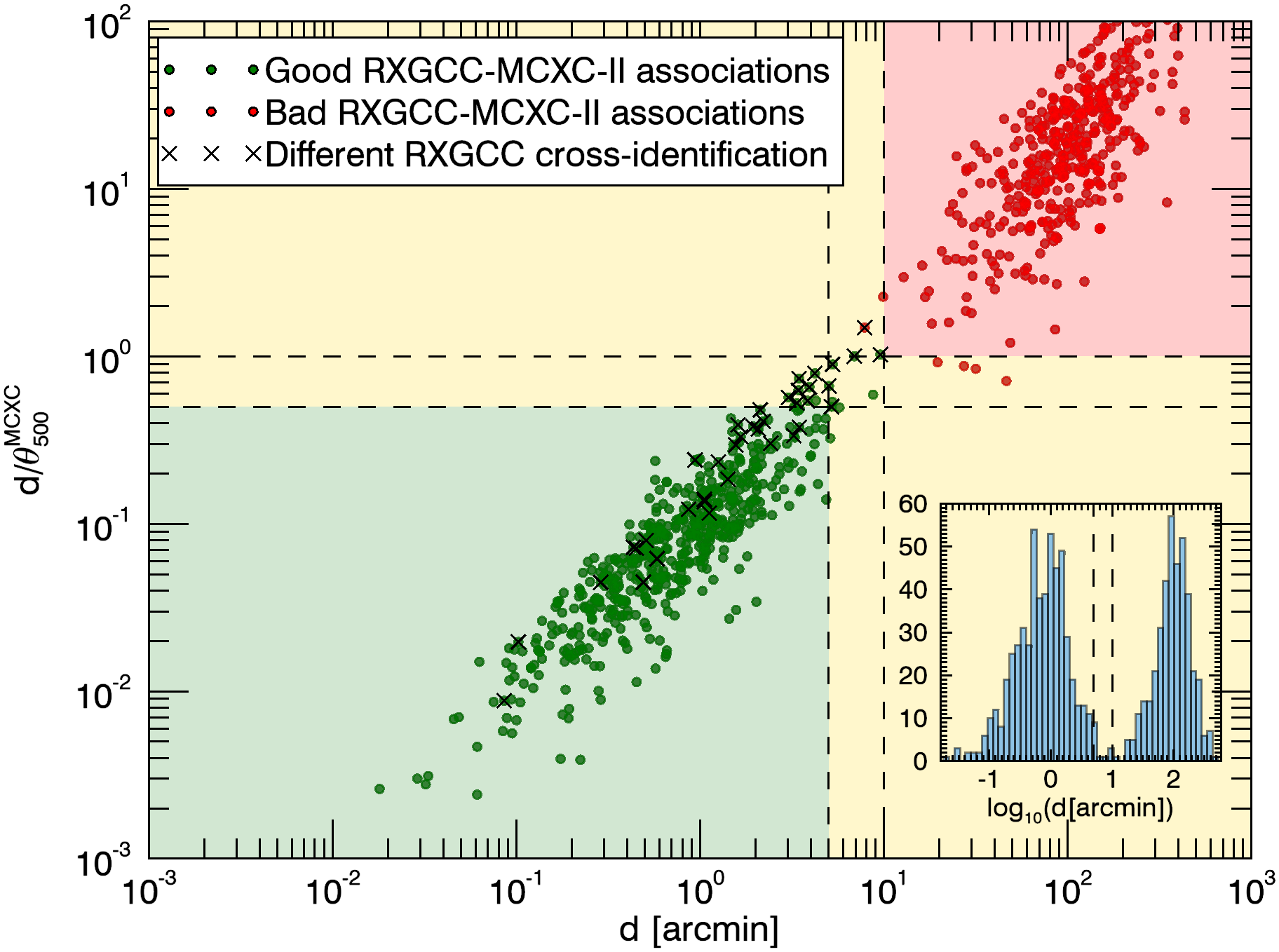}
		\caption{\footnotesize Separation distance $d$ between each RXGCC candidate and its closest MCXC-II cluster is plotted against their relative separation distance in terms of the angular size of the MCXC-II cluster $\Tv^{\rm MCXC}$. Associations falling in the green/red coloured regions are considered to be good/bad, respectively. Associations falling in the intermediate yellow region were analysed individually. Green and red dots represent the final good and bad associations. The black crosses mark the associations for which there is a disagreement between our cross-identification and that undertaken by  \citet[][; GCXSZ column]{2022A&A...658A..59X}. The subpanel on the lower-right corner shows the histogram of separation distances. }  
		\label{fig:RXGCC_MCXC_crossmatch}   
	\end{centering}
\end{figure}

To identify the RXGCC clusters that are already included in one (or more) of the catalogues used to construct MCXC-II, we determined for each RXGCC cluster its nearest MCXC-II cluster. Figure~\ref{fig:RXGCC_MCXC_crossmatch} shows these possible associations in the $d$ versus $d/\,\Tv^{\rm MCXC}$ plane. We distinguish two clouds of points: those with a small separation in absolute and in relative terms (the green-shaded area defined as $d<5\arcmin$ and $d<0.5\, \Tv^{\rm MCXC}$), which correspond to correct associations, and those with a large separation in absolute and in relative terms (the red-shaded area defined as $d>10\arcmin$ and $d>\,\Tv^{\rm MCXC}$), which correspond to incorrect associations. We further individually analysed all the potential  associations with an intermediate separation distance (the yellow-shaded area defined as $d<10\arcmin$ and $d>0.5\,\Tv^{\rm MCXC}$, or $d>5\arcmin$ and $d<\Tv^{\rm MCXC}$) using \xmm, \chandra, SWIFT, ROSAT PSPC, and SDSS images (when available). We found that all the cluster pairs with $ d>10\arcmin$ were bad associations, and that all the pairs with $d<10\arcmin$ except two seemed to be good associations. These two exceptions are the two pairs with the largest relative separation $d/\ \Tv^{\rm MCXC}$ in the intermediate region. 

\begin{itemize}
	\item \object{RXGCC\,767} and \object{MCXC\,J1807.4+6946} ($d=2.27\ \Tv^{\rm MCXC}$), which are indeed two different objects that can be distinguished in the ROSAT PSPC image.
	\item \object{RXGCC\,908} and \object{MCXC\,J2256.9+0532} ($d=1.55\ \Tv^{\rm MCXC}$): SWIFT, \chandra\ and \xmm\ images show extended emission from two nearby sources and one bright point source. The MCXC position lies between the two extended sources, but the RXGCC position is centred on the point source emission. We consider that this RXGCC cluster is a false detection, since the point source is driving the detection, but it is not identified as a point source due to the nearby cluster that adds some extension. 
\end{itemize}

Based on this analysis, we decided to use the following criteria for the RXGCC-MCXC-II association: if the separation distance $d$ between the RXGCC cluster and its closest MCXC-II cluster is less than 10\,arcmin and less than  $1.5\ \Tv^{\rm MCXC}$, we associate the RXGCC cluster to the existing MCXC-II cluster. Otherwise, the RXGCC cluster is considered to be a new entry for the MCXC-II catalogue, except for the identified RXGCC false detection mentioned above. This results in 550 RXGCC clusters associated with already existing MCXC-II clusters, 393 new clusters for MCXC-II, and one RXGCC cluster that is not included in MCXC-II. 

There are some differences between our RXGCC-MCXC-II cross-identification and the RXGCC-MCXC cross-identification included in RXGCC, owing to the use of different cross-identification criteria. In particular, there are 37 pairs of RXGCC-MCXC-II clusters that were not cross-identified in the RXGCC catalogue (marked with a cross in Fig.~\ref{fig:RXGCC_MCXC_crossmatch}), 25 in the green-shaded area, and 12 in the yellow-shaded area.


\begin{SCfigure*}
 \includegraphics[width=0.7\textwidth]{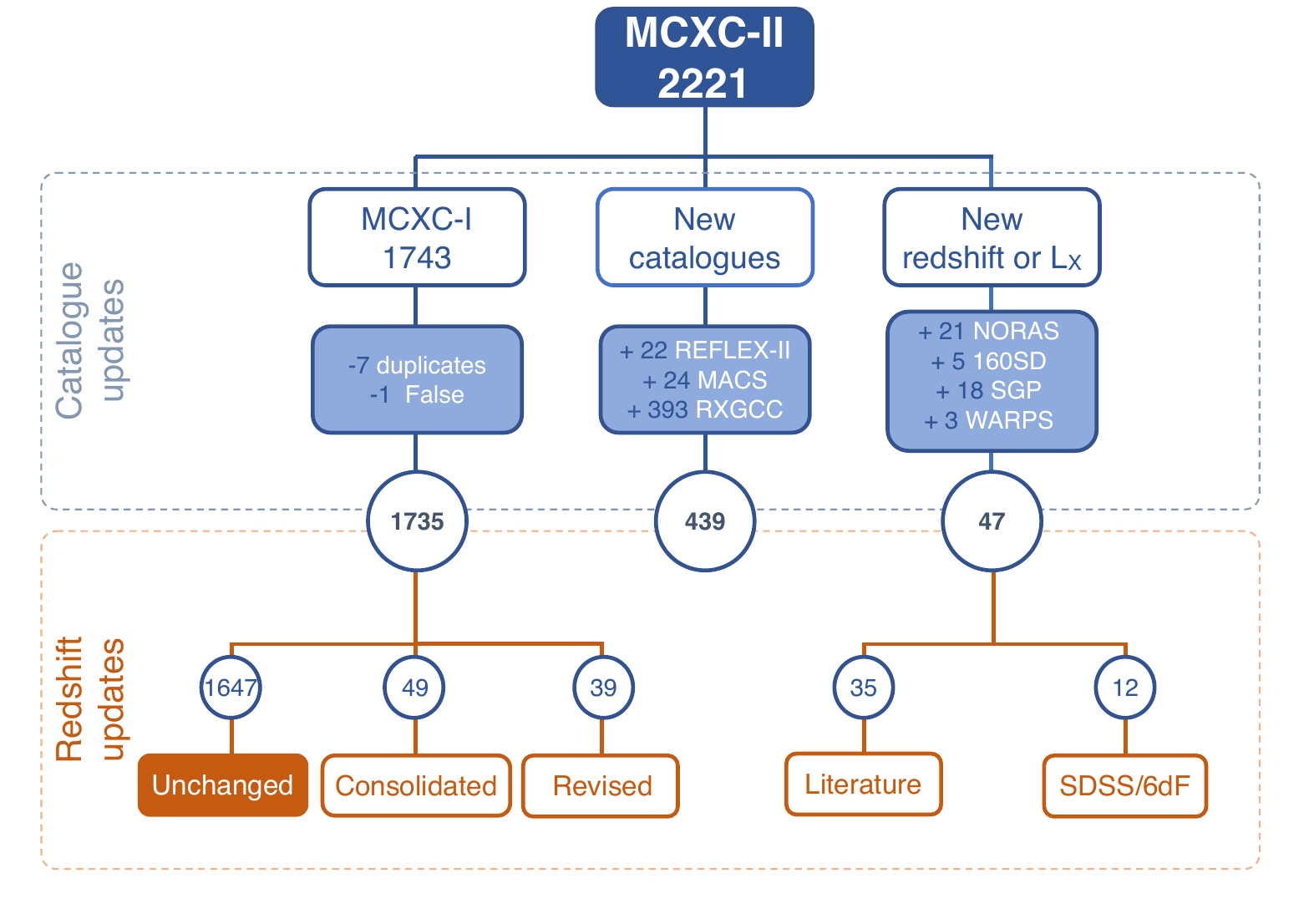}
  \caption{MCXC-II catalogue update summary diagram. 
    Top panel: Revision of catalogue content, from left to right: 1)  removal of one false cluster and seven duplicates (Sect.~\ref{sec:dupl}) in the original MCXC-I catalogue \citep{2011A&A...534A.109P}; 2) entries from new ROSAT-based catalogues (Sect.~\ref{sec:newcat}); 3) new entries from source catalogues not previously considered owing to a lack of redshift or luminosity information (Sect.~\ref{sec:newmcxc}).  Bottom panel:  redshift revision of MCXC-II redshift (Sect.~\ref{sec:zrev}) and new redshift sources (Sect.~\ref{sec:newmcxc}).
    \vspace{2cm}
    }
  \label{fig:sum_diagram}
\end{SCfigure*}


We analysed all of these pairs individually. The disagreement in 15 of the 25 pairs in the green area is due to a new cluster or a redshift update in MCXC-II with respect to MCXC-I. Using the new redshift, the clusters would have also been cross-matched with the RXGCC criteria. Another case of disagreement in the green area is the pair \object{MCXC\,J1314.4-2515} - \object{RXGCC\,493}, which is not associated in RXGCC because the distance ($d$=0.51 Mpc) is just above their threshold (this fixed distance criterion becomes more restrictive as the redshift increases). However, the \xmm\ image shows a unique cluster and so, taking into account the angular extent of the source ($d=0.41\ \Tv^{\rm MCXC}$), we consider this to be a correct association. In the nine remaining pairs of disagreement in the green area, there is a difference between the MCXC and RXGCC redshifts, in some cases owing to a superposition of different objects in the line-of-sight (as in 
\object{MCXC\,J1601.3+5354}, detailed in Appendix~\ref{app:rxgcc}), or the detection of two or more nearby clusters as a single object (as in \object{MCXC\,J0034.2-0204}, \object{MCXC\,J0909.1+1059}, and \object{MCXC\,J1032.2+4015}, also described in Appendix~\ref{app:rxgcc}). We considered the association to be correct in these cases since the detected X-ray signal is the same for MCXC and RXGCC, but the redshift assignment is undertaken differently. In our case, we chose the component that contributes the majority of the X-ray signal, whereas in RXGCC the redshift was assigned to the component with the most galaxies at that redshift. In other cases, the redshift difference is likely explained by the combination of spectroscopic and photometric redshifts included in the RXGCC, which produces a less accurate redshift estimation than the spectroscopic redshift preferred by the  MCXC. This was the case for \object{MCXC\,J1700.7+6412} (detailed in Appendix~\ref{app:rxgcc}), \object{MCXC\,J1751.6+6719}, \object{MCXC\,J1755.7+6752}, and \object{MCXC\,J1921.3+7433}. The last case with a redshift difference is the pair \object{MCXCJ1659.6+6826} - \object{RXGCC 685}, which is in fact a collection of point sources and not a real cluster (as detailed in Appendix~\ref{app:rxgcc}).  

Regarding the yellow-shaded area, there are 12 pairs that we have associated, but RXGCC does not associate, either due to a large redshift difference, to a large separation distance, or both. Except for two cases, the redshift disagreement disappears when considering the updated MCXC-II redshifts (described below in Sect.~\ref{sec:zrev}). One of the exceptions is \object{MCXC\,J0056.0-3732}, for which the RXGCC redshift is incorrectly selected (see details in Appendix~\ref{app:rxgcc}). The second exception is \object{MCXC\,J1414.2+7115}, for which the redshift difference is just borderline ($z_{\rm MCXC}=0.225$, $z_{\rm RXGCC}=0.215$). The \xmm\ and \chandra\ images of this cluster show several extended emission sources, but both the MCXC and RXGCC aperture radii cover all of these sources (although the central position is chosen differently). Therefore, we consider that this is a correct association. On the other hand, the cross-identifications that are missing in RXGCC due to a large separation distance ($d>0.5$ Mpc) can be explained by the different positions chosen by MCXC and RXGCC. Except for one case, MCXC generally chooses the peak of the X-ray emission whereas RXGCC chooses the centroid. This creates a separation between the MCXC and the RXGCC positions when the X-ray emission is elongated, in such a way that it exceeds the RXGCC distance threshold. The exception is the triple cluster \object{MCXC\,J0956.4-1004}, where the MCXC position is centred in between the three components, and the RXGCC position is centred in one of the components (see details in Appendix~\ref{app:rxgcc}).

Finally, we verified the coherence between the redshifts of all associated RXGCC-MCXC-II clusters, and individually checked all the cases with $|z_{\rm RXGCC}-z_{\rm MCXC}|/(1+z_{\rm MCXC}) > 0.01$ or $|z_{\rm RXGCC}-z_{\rm MCXC}|/z_{\rm MCXC} > 0.1$. A total of nine pairs were found for the first criterion.
For eight of these, there was a disagreement in the cross-identification as mentioned above. The final pair concerns \object{MCXC\,J2306.5-1319}, for which we have updated the redshift (see details in Appendix~\ref{app:zrev}).
For the second criterion, we found six additional pairs: one of which has a cross-identification disagreement mentioned above (\object{MCXC\,J1755.7+6752}), as well as five low-redshift pairs for which the redshift difference can be explained by the use of photometric redshifts in the RXGCC catalogue (\object{MCXC\,J0152.9-1345}, \object{MCXC\,J1506.4+0136}, \object{MCXC\,J1654.7+5854}, \object{MCXC\,J1705.1-8210}, and \object{MCXC\,J1836.5+6344}). The final MCXC-II catalogue includes a note for all the cases with a significant redshift difference, which is usually due to the superposition of different clusters.

\subsection{Update of multiple-entry information and final MCXC-II content}\label{sec:ovlp}

Updating the MCXC-II content as described in Sect.~\ref{sec:cat} has an impact on the overlap information provided in the catalogue. New multiple entries primarily originate from the overlap between the MCXC-II and the new catalogues, but new duplicates also occur between objects in the MCXC-II input source catalogues. 

When a cluster appears in several catalogues, \citet{2011A&A...534A.109P} had to choose the source catalogue (hereafter, 'primary catalogue') from which they extracted all the relevant physical properties (position, redshift, and luminosity). They, therefore, defined a priority hierarchy between input catalogues, based principally on catalogue size and the availability of an aperture luminosity (rather than a total value), in order to maximise the homogeneity of the final meta-catalogue. There were two exceptions to the general priority rule. When the redshift difference was larger than 10\%,  they chose the catalogue with the best redshift.  In case of multiple systems, they adopted the measurement that best separated the components or was better centred on the main component. We maintained this strategy for complex systems. However, the first exception is no longer necessary: since we recompute the luminosity from the flux, we can simply update the redshift in the primary catalogue.

In the treatment of new multiple entries, we generally follow the philosophy of \citet{2011A&A...534A.109P}. In particular, we maintained the same ranking of MCXC-I input catalogues, as given in Table \ref{tab:summary_mcxc}. A further concern was the continuity of the MCXC catalogue itself and how it is used. We, therefore, resolved to change the cluster reference catalogue and physical parameters only when necessary, and we did not change the primary catalogue of objects already present in MCXC-I, unless physically justified. In practice, the following treatments were used for the different cases of  multiple detection:

\begin{itemize}[]
\item For duplicates already identified in MCXC-I, we kept the primary catalogue.  In particular, this concerns the five BCS-NORAS  and three MACS-(NORAS or REFLEX) pairs whose primary choice does not follow the catalogue priority but is based on redshift considerations \citep[see][Table~B.1]{2011A&A...534A.109P}.
\item For the seven new MCXC-II cluster duplicates identified Sect.~\ref{sec:dupl}), the choice of object to be retained was based on catalogue priority, except for \object{MCXC\,J1010.2+5430} (400SD) and \object{MCXC\,J0034.2-0204} (SGP). The former is preferred to \object{MCXC\,J1010.2+5429} (NORAS) because the NORAS detection is contaminated by a point source. The latter is preferred to  \object{MCXC\,J0034.6-0208} (REFLEX)  because it is well centred on the main component of the multiple system.  Both cases are discussed in detail in Appendix~\ref{app:dupl}.
\item The new clusters from source catalogues considered in MCXC-II because of newly available redshift measurements (Sect.~\ref{sec:newmcxc}) are essentially new single entries, as expected. The only exception is \object{MCXC\,J0236.0-5225}, which is listed both in 160SD (where it is formally flagged as false) and WARPS (Sect.~\ref{sec:warps}). 
However, there are also four new pairs introduced by our use of input catalogue flux measurements, rather than luminosities. These pairs comprise the three NORAS objects (Sect.~\ref{sec:noras}) and the SGP  cluster (Sect.~\ref{sec:sgp}), which were excluded by \citet{2011A&A...534A.109P} owing to the lack of a redshift and therefore luminosity measurement. These were matched with three eBCS clusters and one MACS\_DR2 cluster. We computed the luminosity from the flux at the cluster redshift for these objects. For catalogue continuity, we did not change the primary source, in spite of the higher NORAS and SGP catalogue priority, but the overlap information for these objects is complete.
\item During the revision, we further realised that six clusters with redshift are in fact missing in the overlap information. These include two EMSS clusters (\object{MS\,0906.5+1110} and \object{MS\,2318.7-2328}), the BCS detection of \object{Abell\,0168}, one 160SD cluster (\object{[VMF98]\,196}), one SHARC\_BRIGHT cluster (\object{RX\,J1142.2+1026}) and one MACS\_DR2 cluster (\object{MACS J2228.5+2036}). Their corresponding input sub-catalogue and luminosity are now included in the overlap information of their counterpart in higher priority catalogues:  \object{MCXC\,J0909.1+1059} (NORAS), \object{MCXC\,J2321.4-2312} (REFLEX), \object{MCXC\,J0115.2+0019}  (REFLEX), \object{MCXC\,J2003.4-5556} (400SD), \object{MCXC\,J1142.2+1027} (400SD), and  \object{MCXC\,J2228.6+2036} (NORAS), respectively.
\item The cross-identification of clusters from new catalogues (REFLEX-II,  the MACS extension, and RXGCC) with other MCXC-II clusters,  is described in the corresponding Sections (Sects.~\ref{sec:reflexII}, ~\ref{sec:newmacs}, and~\ref{sec:rxgcc}, respectively). 
 REFLEX-II has no overlap with other input catalogues, and only one cluster in common with RXGCC, and was therefore simply added as a sub-catalogue of the NORAS/REFLEX catalogue. 
The 24 new MACS clusters not matching previously known MCXC-II clusters from higher priority catalogues  (NORAS, REFLEX, or SGP) were introduced as primary objects. 
There is a large overlap between the RXGCC catalogue and the RASS based catalogues, and its inclusion was undertaken in such a way as to maximise the MCXC continuity.  Although the RXGCC catalogue size is slightly larger than that of NORAS+REFLEX (the highest priority catalogue) and does include aperture flux measurements, we simply added the  RXGCC catalogue at the end of the list. This choice avoids an artificial change of the parameters of the 550 objects in common with other MCXC-II clusters, while the  RXGCC and MCXC-I luminosities are in good agreement (Sect. ~\ref{sec:overlap}). We found that none of the complex association cases discussed in Sects.~\ref{sec:rxgcc} and~\ref{app:rxgcc}) required a change of primary choice.
\end{itemize}

In all cases,  the information on additional detections is provided in the overlap fields of each MCXC-II cluster. We give the names of overlapping catalogues and the corresponding $\Lv$ values,  computed from the respective flux at the adopted redshift. These may differ from the MCXC-I values, which were based on source catalogue luminosities, and thus subject to additional systematic differences in cases of redshift discrepancy between catalogues. 

The final MCXC-II catalogue content is summarised in Table~\ref{tab:summary_mcxc}. For each input catalogue,  we give the number of clusters included in MCXC-II  (excluding false objects, or those still without redshift), and the number of clusters after handling of multiple entries between catalogues.
Figure~\ref{fig:sum_diagram} summarises the catalogue update steps. MCXC-II comprises a total of 2221 clusters, as compared to the 1743 MCXC-I clusters. This 27\% increase is primarily due to the inclusion of the RXGCC. A detailed comparison between MCXC-I and MCXC-II is presented in Sect.~\ref{sec:comp_mcxc}.

\section{Cluster redshift revision}\label{sec:zrev}

\subsection{General method }

In the first release of the MCXC, the redshift $z$ of each object was taken to be that given in the parent catalogue. A cross-check was undertaken in the case of catalogue overlaps, and objects with redshifts differing by more than 10\% were subject to manual verification as described in Appendix~B of \citet{2011A&A...534A.109P}. 
The redshift values in the parent catalogues come from literature searches for known objects, dedicated optical follow up, or a mixture of both. Although spectroscopic redshifts are the 'gold standard', the value is often based on measurements for one galaxy, which may be a foreground object. \citet{2014ApJ...783...80R} compared the MCXC-I redshift and  SDSS-based redMaPPer photometric redshift of 323 MCXC-I clusters at $z>0.1$, and found excellent agreement, albeit with eight prominent outliers (their Fig.~1). For six of these, the photometric measurements indicated that the MCXC-I redshift was incorrect, while the two remaining objects were affected by $\zp$ systematics. 

For MCXC-II, we performed a systematic revision of the MCXC-I redshift. In this undertaking, our goal was not to identify the best available redshift for each object, a task which is far beyond the scope of the paper (and which is arguably somewhat subjective). Our aim was simply to identify and update 'problematic' redshifts as broadly as possible; in particular, where there is a significant impact on the luminosity and (thus) mass estimate. For this, we compared the MCXC-II redshifts with the preferred values of the NED and Simbad database as of August 2023, as well as the spectroscopic redshift of optical counterparts from large catalogues from large optical/NIR surveys (SDSS, WISE, etc.) \footnote{A first revision was performed in 2017-2019, which was used for the cross-identification of ComPRASS objects \citep{tar19} and is the version available in the M2C Cluster Data Base opened in 2021. The new catalogues and/or the NED/Simbad information essentially confirm the redshifts in this previous revision, with the addition of new revision cases. A table of differences will be put on the M2C database page.}.

The criteria for further redshift inspection were: i) a difference from the MCXC value larger than $10\%$ and/or ii) a $\Delta(z)/(1+z)>0.01$. Via the luminosity-distance factor, the first criterion corresponds to systematic uncertainty of $\gtrsim 20\%$ on the luminosity,  similar to its typical statistical precision. The second criterion corresponds to $\sim 3000$ km ~s$^{-1}$, or three times the typical velocity dispersion of a massive cluster, and is an indicator of a possible redshift measurement based on a  foreground/background object. This dominates the $\Delta(z)/z < 10\%$ requirement at $z>0.1$, above the median MCXC- redshift.  
The decision to update the redshift was based on a manual inspection of each case. Depending on data availability and case complexity, the information we used may include the following.   
\begin{itemize}[label=$-$]
\item Information on the quality of the original redshift, particularly the number of galaxies used for its estimation. This was retrieved from the original publication or follow-up information.
\item Complementary information from NED and Simbad,  such as a search around the position, and the consideration of various $z$ estimates. The main cluster identified as the source of the redshift measurement is also important in complex cases of multiple (or close-by) systems at the X-ray position. 
\item Comparison with redshift of possible counterparts. This includes  clusters from the Abell catalogue \citep{1989ApJS...70....1A} and recent optical catalogues based on large galaxy surveys:
\begin{itemize}
\item 
 redMaPPer catalogues (hereafter RM) published by   \citet{2016ApJS..224....1R}, including  clusters from 
 DES Science Verification data and the updated version of the catalogue of  \citet{2015MNRAS.453...38R} from SDSS-DR8 data.  The spectroscopic redshift of the BCG is available for $60\%$ of the SDSS clusters.
\item
The catalogues  of  \citet[][hereafter WH]{2015ApJ...807..178W} and  \citet[][herafter WHY] {2018MNRAS.475..343W} based on SDSS-DR12 and 2MASS, WISE, and SuperCOSMOS data, respectively.  The WH catalogue constitutes an update of the WHL12 catalogue \citep{2012ApJS..199...34W}, with new clusters at high $z$. Spectroscopic redshifts, estimated from all spectroscopic members, are available for $\sim 75\%$ of the WH clusters.
\item 
The catalogue of $ 0.01<z<0.2$ clusters identified from over-densities in redshift phase space from SDSS-DR13 data by \citet[][herafter GalWCat]{2020ApJS..246....2A}.
\end{itemize}
We also considered the  \xmm\  serendipitous catalogue of \citet[][herafter  X-CLASS]{kou21}, using the redshift information (position and redshift of spectroscopic galaxies) available from their database\footnote{\url{https://xmm-xclass.in2p3.fr}}. We also make use of the RXGCC database\footnote{\url{https://github.com/wwxu/rxgcc.github.io/tree/master}}, which contains redshift histograms and overlays of  X-ray images and galaxy distributions at various redshift peaks. 
\item Inspection of X-ray images from RASS, \xmm, \chandra, and SWIFT (available in the M2C database), and/or pointed ROSAT PSPC observations, and SDSS optical images, with the positions of optically-detected clusters and galaxies with spectroscopically-measured redshifts overlaid.
\end{itemize}

\subsection{From NED and Simbad} 
\label{sec:zrevnedsb}

\begin{table*}
\caption{Clusters with revised redshifts.}
\resizebox{\textwidth}{!} {
\begin{tabular}{lrllclrrlllll}
\toprule
NAME MCXC     & SUB\_CAT   &$z_{\rm cat}$ &$z$& Type &Ref.			&$dz/z(\%)$ &       $dz/(1+z)$ & Source & Counterpart & Rationale\\
\midrule
     J0016.3-3121 &             SGP &   0.0805 &   0.1063 &  S 	&1	&   {\bf 32} &     {\bf 0.024} 	& NEDSB/Ocat &   WHY, RXGCC  		& 1, 2  (Nz=36)\\
     J0019.6+2517 &           NORAS &   0.1353 &   0.3657 &  S 	& 2	&  {\bf 170} &       {\bf 0.203} 	& Ocat         	& WH      			& 3  \bf{los struct.}     \\
     J0043.8+2424 &             BCS &   0.0830 &   0.0837 &  S 	& 2	&    1 &    0.001                    	& NED/Ocat 	&   WH    			& 2  (Nz=23)    \\
     J0152.9-1345 &          REFLEX &   0.0050 &   0.0057 &  S 	& 3	&   14 &    0.001                     	& NED          	&   RXGCC 			 & 3 \object{NGC720} dominated    \\
     J0210.4-3929 &           160SD &   0.1650 &   0.3058 &  S 	& 4	&   {\bf 85} &     {\bf 0.121} 	& NED         	 & XCLASS 			&  1, 3   \\
     J0507.7-0915 &          REFLEX &   0.0398 &   0.1483 &  S 	&5	& {\bf  273} &     {\bf 0.104} 	& Ocat          	&     WHY, RXGCC  		& 1   \\
     J0748.1+1832 &           NORAS &   0.0400 &   0.0460 &  S 	&6	&   15 &    0.006                	 	& SB/Ocat 	&   GalWCat, RXGCC 	&  2 (Nz=76)  \\
     J0935.4+0729 &           NORAS &   0.2610 &   0.2160 &  S 	& 7	&   17 &   {\bf -0.036} 		& NEDSB/Ocat &   WH ,RM  			& typo   \\
     J1008.7+1147 &           NORAS &   0.2245 &   0.2599 &  S 	&    8            	&   16 &    0.029                     	& SB          	&                               		&    2   \\
     J1011.4+5450 &  400SD\_SER &   0.2940 &   0.3798 &  S 	& 2 	&   29 &    {\bf 0.066} 		& Ocat          	&  WH                    		& 3   \\
     J1016.6+2448 &           NORAS &   0.0811 &   0.1732 &  S 	& 2 	&  {\bf 114} &     {\bf 0.085} 	& NED/Ocat      &    WH, RM     		& 1, 2 (Nz=21)\\
     J1017.5+5934 &           NORAS &   0.3530 &   0.2880 &  S 	& 2 	&   18 &  {\bf  -0.048 }              	& NED/Ocat 	&   WH  				& 1, 2 (Nz=17)    \\
     J1022.0+3830 &           NORAS &   0.0491 &   0.0543 &  S 	&6 	&   11 &    0.005                        	& SB/Ocat 	&    GalWCat, RXGCC 	&  2,  (Nz=45) \\
     J1025.0+4750 &           NORAS &   0.0520 &   0.0626 &  S 	&6 	&   20 &    0.010                     	& SB/Ocat  	&   GalWCat, WH, RXGCC  & 2  (Nz=24)    \\
     J1036.6-2731 &          REFLEX &   0.0126 &   0.0133 &  S &	 9	&    6 &    0.001                   	& SB          	&  RXGCC   			& 2   \\
     J1159.2+4947 &           NORAS &   0.2110 &   0.3486 &  S & 2 	&   {\bf 65}&     {\bf 0.114} 	         & NEDSB/Ocat & WH, RM, RXGCC 	&1   \\
     J1236.4+1631 &           NORAS &   0.0780 &   0.0700 &  S &6 	&   10 &   -0.007                      	& SB/Ocat  	&   GalWCat  			& 2 (Nz=33) \\
     J1339.0+2745 &         WARPSII &   0.1300 &   0.1631 &  S & 2 	&   25 &    0.029                      	& Ocat          	&    WH  				& 2   \\
     J1340.9+3958 &  400SD\_SER &   0.1690 &   0.2772 &  S & 2 	&   {\bf 64} &     {\bf 0.093} 	& Ocat          	&  WH, RM  			& 1, 3  confusion   \\
     J1343.4+4053 &           160SD &   0.1400 &   0.2560 &  S & 10 	&   {\bf 83} &     {\bf 0.102 }	& NEDSB/Ocat	&    WH, RM   			& 1 \\
     J1343.7+5538 & SHARC\_BRIGHT &   0.0766 &   0.0685 &  S & 2 &   11 &   -0.008                     	& Ocat          	&     WH 				&  2 (Nz=16)\\
     J1359.2+2758 &           NORAS &   0.0612 &   0.0751 &  S &6 	&   23 &    0.013                      	& SB/Ocat  	&   GalWCat, WH ,RXGCC & 3 \bf{los struct} (Nz=56)  \\
     J1415.1+3612 &           WARPS &   0.7000 &   1.0260 &  S & 11 	&   {\bf 47} &     {\bf 0.192}     	& NEDSB  	&   XCLASS 			& 2 (Nz=  25)   \\
     J1421.6+3717 &           NORAS &   0.1813 &   0.1623 &  S & 12 	&   10 &   -0.016                      	& NEDSB/Ocat &  WH,  RM , RXGCC   	& 2 (Nz=119)   \\
     J1447.4+0827 &           NORAS &   0.1954 &   0.3760 &  S & 2 	&   {\bf 92} &     {\bf 0.151} 	& NEDSB/Ocat &  WH,  RM 			& 1, 2 \\ 
     J1501.1+0141 &          REFLEX &   0.0050 &   0.0066 &  S & 13 &   {\bf 32} &      0.002		& NED          	&  RXGCC    			& 3 \object{NGC5813} dominated   \\
     J1520.9+4840 &           NORAS &   0.1076 &   0.0740 &  S & 14 &   {\bf 31} &   {\bf -0.030}         	& NEDSB/Ocat &    GalWCat, WH, RXGCC & 1, 2 (Nz=38)     \\
     J1532.9+3021 &             BCS &   0.3450 &   0.3625 &  S & 15 	&    5 &    0.013                      	& SB/Ocat   	&   WH  				& 2  \\
     J1544.0+5346 &           160SD &   0.1120 &   0.5000 &  S &   16        	 &  {\bf 346} &      {\bf 0.349 }                     	& SB/Ocat    	&   WH, XCLASS 		& 1, 3 \\ 
     J1605.5+1626 &            eBCS &   0.0370 &   0.0423 &  S &17 	&   14 &    0.005                      	& NED          	& RXGCC     			& 2 (Nz=16)   \\
     J1606.8+1746 &             BCS &   0.0321 &   0.0391 &  S & 18 	&   22 &    0.007                      	& NED          	&        				& 3  \object{A2151E}\\
     J1621.0+2546 &       EMSS\_1994 &   0.1610 &   0.1909 &  S & 2&   19 &    0.026                      	& Ocat          	&   WH, RM 			& 1, 2  (Nz=5)\\
     J1730.4+7422 &           NORAS &   0.1100 &   0.0470 & SP &    19   	&   {\bf 57} &    {\bf -0.057} 	& NED          	&   RXGCC  			 & 1   \\
     J2032.1-5627 &          REFLEX &   0.1380 &   0.2840 &  S & 20	&  {\bf 106} &     {\bf 0.128} 	& NEDSB     	&   XCLASS    			& 3 (Nz=32) los struct  \\
     J2135.2+0125 &          REFLEX &   0.1244 &   0.2290 &  S & 10, pw &   {\bf 84} &     {\bf 0.093}	& NEDSB/Ocat &    WH, RM, RXGCC	 & 1 (Nz=10)    \\
     J2306.5-1319 &          REFLEX &   0.0659 &   0.1095 &  S & 5 	&   {\bf 66} &  {\bf  0.041}  	& Ocat       	&   WHY, RXGCC 		 & 3 confusion \\
     J2326.2-2406 &             SGP &   0.0880 &   0.1116 &  S & 21     	 &   27 	&    0.022 	            	& NEDSB/Ocat	&   WHY         			& 2 (Nz=21)\\
     J2334.0+0704 &           NORAS &   0.0990 &   0.2950 &  S &     21, pw        	&  {\bf 198 }	&   {\bf  0.178} 	& SB          	&    					& 3  (Nz=20), mis-ID with \object{A2620}     \\
     J2341.1+0018 &           NORAS &   0.1100 &   0.2768 &  S & 2, pw     &  {\bf 151} &     {\bf 0.150}     	& Ocat          	&  WH  			& 1, 2  (Nz=18)  \\
\bottomrule
   
\end{tabular}}
\tablebib{
(1)\citet{2002MNRAS.329...87D}; 
(2)  \citet{2015ApJ...807..178W};
(3)  \citet{2002ApJS..140..239C};
(4) \citet{2006ApJ...646..133M};
(5) \citet{2018MNRAS.475..343W};
(6)  \citet{2020ApJS..246....2A};
(7) \citet{1995ApJS...96..343Q};
(8)       \citet{orl21};
(9) \citet{2004AJ....128.1558S};
(10) \citet{2016ApJS..224....1R};
(11) \citet{2009ApJ...707L..12H};
(12) \citet{2013ApJ...767...15R};
(13) \citet{2016MNRAS.460.1758H};
(14) \citet{2007MNRAS.379..867V};
(15) \citet{2010MNRAS.407...83E};
(16) \citet{kou21}; 
(17)  \citet{1999MNRAS.305..259W};
(18) \citet{2022MNRAS.509.3470M};
 (19)            \citet{2022A&A...658A..59X};  
(20) \citet{2012ApJ...761...22S};
(21) \citet{2002AJ....123.1200C}
}

\tablefoot{Cols. 1-2: Cluster MCXC Name and input catalogue. Col. 3: MCXC redshift value from input catalogue $z_{\rm cat}$. Cols. 4-6: Revised redshift,  $z$, type and and reference. Cols. 7-8 Difference, 
$\Delta(z)=(z- z_{\rm cat})$, between the two values,  in percentage terms $\bar{\Delta(z)}/z_{\rm cat}$ or  divided by $(1+z_{\rm cat})$.  Col. 9. Reason for the redshift revision: discrepancy with NED (NED) or SIMBAD (SB) preferred value, or discrepancy with both (NEDSB); discrepancy with the spectroscopic redshift of the associated cluster from optical catalogues (Ocat) based on large galaxy surveys (principally SDSS). Col 10. Matching  cluster with $\zs$  in the  catalogues of \citet[][WH]{2015ApJ...807..178W},  \citet[][RM]{2016ApJS..224....1R}, \citet[][WHY]{2018MNRAS.475..343W}, \citet[][GalWCat]{2020ApJS..246....2A},   \citet[][XCLASS]{kou21},  and  \citet[][RXGCC]{2022A&A...658A..59X}. Col 11. Rationale for change: (1) original value based on one or two likely foreground galaxies; (2) more precise value based on a larger number of galaxies, or more robust value; (3) Complex case (e.g. multi-component systems). Values with large differences, as well as cases of line-of-sight structures,  are indicated in bold and discussed in the Appendix. }

\label{tab:zrev}
\end{table*}
For 98 MCXC-II clusters, the MCXC-I redshift difference with respect to the preferred NED and/or Simbad value is larger than $10\%$. This includes 18 discrepant redshifts with both NED and Simbad, 37 with NED only and 43 with Simbad only. The $\Delta(z)/(1+z)<0.01$ criterion adds 45 additional discrepant redshift cases, of which only two are in common between NED and Simbad. 

The majority of the cases (68\%) were not retained for revision. There are a few complex cases of confusion, or even incorrect cross-identification of the MCXC cluster.  However, the majority of the redshift discrepancies concern older references (and generally less precise redshifts) in NED or Simbad, or the choice of a photometric preferred redshift rather than spectroscopic values. Examples of 'old' measurements include: Abell cluster redshifts based on one or two galaxies where better estimates are available in the X-ray catalogues; and the redshifts for the 16SD catalogue in Simbad, which were mostly taken from the original publication   \cite{1998ApJ...502..558V} rather than the values derived from dedicated latter follow-up by \citet{2003ApJ...594..154M}. Concerning photometric redshift measurements, NED favours redshifts with uncertainties, which is the case for the photometric redshifts in the RM catalogue. Several cases of redshift mismatch correspond to photometric redshifts taken from the RM catalogue \citep{ 2016ApJS..224....1R}, with differences consistent with the spectroscopic value.  

In  30 cases, we decided to revise the MCXC-II redshift. The original MCXC-I redshift, $z_{\rm cat}$, and the revised value are listed in Table~\ref{tab:zrev}, together with the redshift type and reference. Note that the final redshift is based on examination of the available information, and is not necessarily the NED or Simbad preferred value. The 23 largest revisions, with $\Delta(z)/z > 30\%$ or $\Delta(z)/(1+z)>0.03$ are marked in bold in the table and are discussed individually in Appendix~\ref{app:zrev}.

At low redshift ($z<0.1$) the revision greatly benefitted from the GalWCat catalogue, and large-scale spectroscopic follow-up of cluster samples \citep[][e.g.]{1999MNRAS.305..259W, 2004AJ....128.1558S} or complex fields \citep[e.g.][]{2022MNRAS.509.3470M}, which allows us to refine the catalogue values. Large outliers, with $\Delta(z)/(1+z)>0.01$,  were observed above $z>0.1$. Several cases simply correspond to a catalogue redshift value based on one or two likely foreground galaxies, for example,  RASS clusters matched with Abell systems that have a redshift estimated from only one or two galaxies. Other cases mostly included complex fields, with several close-by objects or even structures at different redshifts along the line of sight. These important revisions are usually supported by converging evidence from the cross-match with optical clusters (GalWCat, RM and WH/WHY) and/or X-CLASS/RXGCC data, with the revised redshifts usually based on a large number of spectroscopic galaxies (see columns 11 and 12 of Table~\ref{tab:zrev}).

\begin{figure*}[t]
\begin{centering}
    \includegraphics[width=0.9\columnwidth]{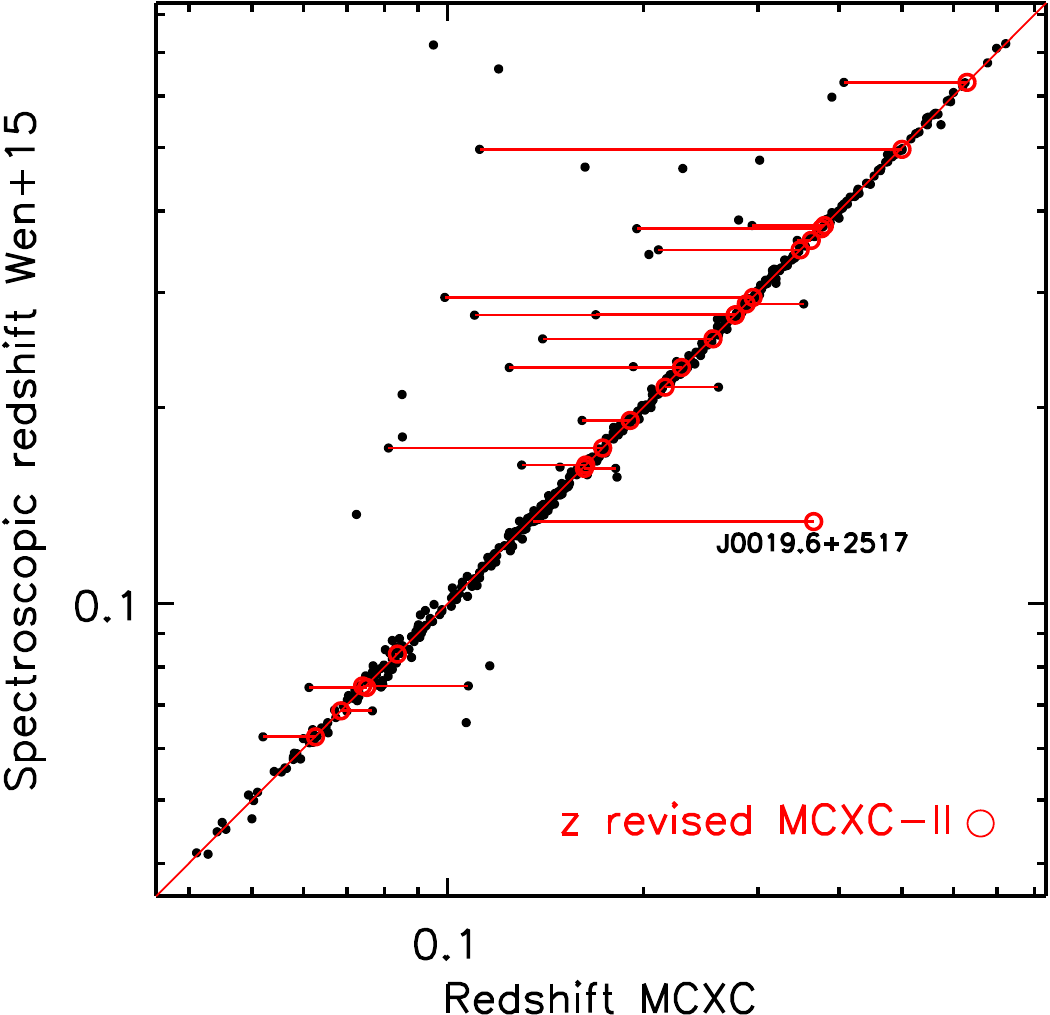}
    \hspace{0.4cm}
    \includegraphics[width=0.9\columnwidth]{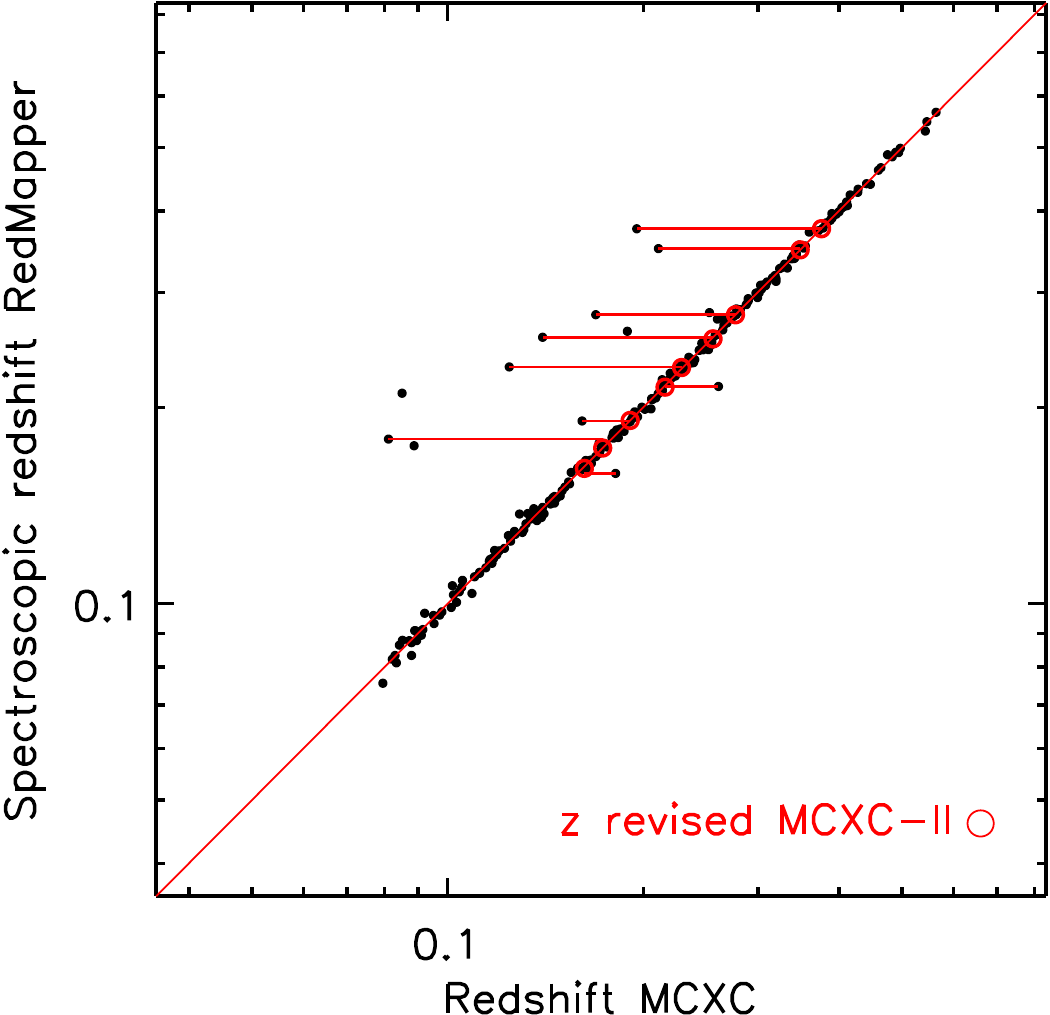}
    \caption{ Comparison between MCXC-II redshifts and spectroscopic redshift values from SDSS-based catalogues.  The revised values discussed in Sect.\ref{sec:zrev} are marked in red, and connected to the  MCXC-I values by a red line. Left panel: Comparison with the closest potential counterpart  from the \citet[][WH]{2015ApJ...807..178W} catalogue  (see text). \object{MCXC\,J0019.6+2517}, labelled in the figure, is a line-of-sight structure where the main component is identified with the richest second-closest WH cluster.  Right panel:  Same for redMaPPer clusters from the updated version  of  the catalogue of \citet{2015MNRAS.453...38R}, published by \citet{2016ApJS..224....1R}}  
 \label{fig:zzopt}   
\end{centering}
\end{figure*}

\subsection{Additional revision from  cross-match with optical catalogues} \label{sec:zrevopt}

In a second step, we cross-identified the remaining clusters with matching NED/Simbad redshift with large optical RM, WH/WHY, and GalWcat catalogues. The difference criteria remained the same, only considering optical clusters with spectroscopic redshits. We started with the WH catalogue, which is based on more recent (and thus complete) SDSS spectroscopic data. 

Cross-identification with optical catalogues is not trivial, owing to physical offsets between X-ray and optical centres and superpositions along the line of sight. In particular, the probability of chance association is expected to increase with decreasing redshift as the source angular extent becomes larger. Furthermore, SDSS-based catalogues are incomplete at low redshift, and may miss the optical counterpart even of nearby massive objects, while distant low-mass optical groups can appear in projection against the large angular extent of the X-ray cluster. We cannot exclude these cases using $\Delta(z)$  constraints,  as those would artificially exclude cases of false redshifts.  For each MCXC-II object in the original MCXC-I catalogue we identified the closest optical cluster, or the second-closest optical cluster if the latter was richer than the former.

To put a constraint on the separation distance $D$, we first looked at the cluster position in the $D$--$D/\Tv$ plane. We considered both the  $\Tv$ estimated from the X--ray measurements ($\theta_{\rm 500, MCXC}$), and the optical value derived from the mass--richness relation calibrated in the respective optical catalogues ($\theta_{\rm 500, opt})$. Contrary to the case of the cross-match between MCXC and RXGCC (Fig.~\ref{fig:RXGCC_MCXC_crossmatch}), we did not observe two well-separated clouds. We therefore further examined the location of clusters with matching SDSS redshifts in the $D$--$D/\Tv$  plane, taking into account their redshift difference.  This allowed us to define the following criteria for possible optical counterpart:   $z_{\rm MCXC}>0.04$ and $D/ \theta_{\rm 500, opt} <1 $   (in practice this also corresponds to  $D/ \theta_{\rm 500, MCXC}  < 1$).

 We identified 533 potential WH counterparts, whose optical redshifts are compared to the  MCXC-I redshift in the left panel of Fig. \ref{fig:zzopt}. The agreement is excellent for 498 objects with a median ratio of  $0.9995\pm0.0005$ (that is, zero bias) and a 68\% dispersion of only $\Delta(z)= 0.0024$.  There are  35 outliers. 
Thirteen cases correspond to MCXC-II redshifts that are discrepant with NED or Simbad values,  whose redshifts have been revised to a new redshift in agreement with the SDSS value (Sect.~\ref{sec:zrevnedsb}). After individual examination of the 22 remaining cases, we further revised the redshift of seven objects (Table~\ref{tab:zrev} ). The largest revisions are discussed in Appendix~\ref{app:zrev}, and correspond to the same typical cases as discussed in Sect.~\ref{sec:zrevnedsb}.  The remaining 15 WH outliers, but for two entries, are background objects. They are at much larger redshift than the MCXC-II cluster ($\Delta(z)=[0.04$--$0.6]$, see Fig.~\ref{fig:zzopt}), while the choice of MCXC-II redshift is further supported by redMaPPer and/or GalWCat counterpart, not present in WH catalogues.

We repeated the procedure for the catalogue of \citet[][WHY]{2018MNRAS.475..343W}. Here we only examined the cases of clusters not already in the \citet{2015ApJ...807..178W} catalogue, as indicated in their table. Of eight new discrepant cases, we further retained two revisions, discussed in detail in Appendix~\ref{app:zrev}. In the case of two REFLEX cluster, the procedure was inconclusive owing to a lack of additional data, so we conservatively kept the MCXC value. 

We then considered possible discrepancies with redMaPPer spectroscopic redshifts. The pre-selection of potential counterparts was defined following the same method as for the WH clusters. We used a slightly higher redshift threshold, $z_{\rm MCXC}>0.075$, and $D/ \theta_{\rm 500, opt}$. The right-hand panel of Fig.~\ref{fig:zzopt} compares the MCXC-I and RM redshifts for the 250 objects in common.  There is an excellent agreement for the vast majority of cases, with a median ratio of  $1.0006\pm  0.0008$ and 
 a 68\% dispersion of only $\Delta(z)= 0.0022$, for the 237 objects with $\Delta(z)/z < 0.1$.  Of the 13 outliers, nine had already been revised in the previous steps, in agreement with the RM value.  The remaining four outliers are clearly background objects that were not retained for revision.  

Finally, we cross-matched with the GalWCat catalogue.   No further redshift revision was required. This was expected as Simbad includes the  GalWCat cross-match, and discrepant redshifts were mostly identified with the NED/Simbad test. 

\subsection{Redshift type and reference and consolidation}\label{sec:ztyperef}

For each cluster, we provide additional information beyond the redshift value: the bibliographic reference {\tt Z\_REF}, given as far as possible in the form of {\tt BIBCODE},  the redshift type {\tt Z\_TYPE}, and a flag, {\tt Z\_FLAG}.   

For {\tt Z\_TYPE}, we distinguish firstly between photometric ({\tt P}) and spectroscopic ({\tt S}) redshift. A few redshifts are only estimated (e.g. from the magnitude of the BCG) and these have a redshift type designated '{\tt E}'. We added the category '{\tt SP}'  for RXGCC redshifts (see below).  When we could not determine the provenance of the redshift, it was set to '{\tt U}' for unknown. 

{\tt Z\_FLAG} refers to the redshift revision. We set {\tt Z\_FLAG} to '{\tt NEW}' for NORAS, SGP, 160SD or WARPS entries based on newly-available redshifts (Tables~\ref{tab:newznorassgp} and ~\ref{tab:newz160SD}).   Z{\tt \_FLAG} is set to '{\tt Revised}'   for clusters with revised redshift presented in Sect.\ref{sec:zrev} (Table~\ref{tab:zrev}). For the other clusters, the redshift is nominally taken from the source catalogue and {\tt Z\_FLAG} is therefore set to '{\tt Catalogue}'. In some cases, we used another reference as described below, and the corresponding clusters  are denoted by {\tt Z\_FLAG} = '{\tt Consolidated}'

For clusters with new or revised redshifts, the redshift type and reference are given in the corresponding Sections (Sects.~\ref{sec:newmcxc} and \ref{sec:zrev}).  The redshift type and reference of MACS clusters is described in Sect.~\ref{sec:newmacs}. For RXGCC,  the redshift was primarily determined from a compilation of galaxy redshifts, combining spectroscopic and photometric values. In that case {\tt Z\_TYPE} is set to '{\tt SP}'. When this was not possible, \citet{2022A&A...658A..59X} gave a redshift taken from the literature, simply specifying whether the cross-identified cluster was from optical or X-ray catalogues. In that case we set {\tt Z\_TYPE} to '{\tt U}'. In all cases, {\tt Z\_REF} is the {\tt BIBCODE} of the RXGCC publication. 

For the other clusters, we started with the information in the catalogue publication, as now described. For five catalogues, NEP, SHARC\_SOUTH, WARPS, WARPSII, and EMSS\_1994, the redshifts were all derived from dedicated follow-up.  We thus put the reference of the catalogue, except for EMSS and NEP, the reference being the previously published follow-up \citep[][respectively]{1991ApJS...76..813S,2003ApJS..149...29G}. 

For the other catalogues, the redshift source is a mixture of dedicated follow-up (published separately or not) and/or literature (including the galaxy catalogue). In that case, the reference for each cluster can be retrieved from the {\it VizieR} table associated with each cluster catalogue. It lists the correspondence between the reference code (number or letter) given in the catalogue and the {\tt BIBCODE} when available and/or a description of the reference. Some homogenisation and manipulation were necessary

We set {\tt Z\_REF} to the {\tt BIBCODE} of the catalogue  publication when:
\begin{itemize}[noitemsep,topsep=0pt]
\item  The redshift comes from dedicated follow-up, if no corresponding published article is listed. Note that more information on part of the SGP and REFLEX observing campaigns, are published in the PhD thesis of K.~Romer (1995) and in \citet{2009A&A...499..357G}, respectively.
This also concerns references quoted as `in preparation', explicitly as a forthcoming follow-up article (that we could not find in some cases) or by the catalogue authors. However, in the specific case of BCS, the BCG redshift for clusters listed with the reference `Crawford et al, in prep.'  have been published by  \citet{1999MNRAS.306..857C}. In that case, we replaced the catalogue value and reference with this publication. 

\item The redshift is a compilation of several sources from the literature (possibly combined with the follow-up). This concerns REFLEX, NORAS, SGP and  CIZA~I catalogues. 

\item The source is a galaxy catalogue, from which the cluster redshift is estimated. In that case we add the catalogue name to the {\tt BIBCODE}.  This concerns the CfA catalogues,	Cat. <VII/193>\footnote{\url{https://cdsarc.cds.unistra.fr/viz-bin/cat/VII/193}},  by \citet{1992ZCAT..C......0H,1990ApJS...72..433H} used for BCS/eBCS (although it is not always clear which version were used), the reference catalogue of Bright galaxies, Cat. <VII/155>\footnote{\url{https://cdsarc.cds.unistra.fr/viz-bin/cat/VII/155}}  by \citet{1991rc3..book.....D}, used for NORAS, BCS and eBCS, and Cat. <VII/115>\footnote{\url{https://cdsarc.cds.unistra.fr/viz-bin/cat/VII/115}} by \citet{1989spce.book.....L} used for NEP and SGP. 
\end{itemize}
In all these case the redshift type (essentially spectroscopic) is unambiguous from the follow-up description and/or the quoted source.  

 When the redshift source was a published article,  we simply set {\tt Z\_REF} to its corresponding {\tt BIBCODE}. For publications based on previous work, for example, the compilation of \citet[][]{1999ApJS..125...35S}, we did not trace back to the original reference,  unless if needed to determine the redshift type or when needed settle the cases of problematic redshifts (Sect.~\ref{sec:zrev}). The only exception is the 400SD clusters for which \citet{2007ApJS..172..561B} refer to the 160SD publication \citep{2003ApJ...594..154M}. In this case, the redshift source is taken from the latter.  The redshift type can generally be found in the publication in question.  

Finally, there are cases where the source given in the catalogue is not sufficient to determine the redshift type, such as `private communication', publications `in preparation' or a few other problematic references. In these cases,  we searched for other redshift sources. If available, we list the corresponding reference (given in {\tt Z\_REF}), as well as the redshift for consistency. The difference with the catalogue value is negligible. These cases are flagged by {\tt Z\_FLAG} set to 'Consolidated'. 

\begin{table*}[!ht]
    \centering
    \caption{\label{tab:Fx_methods} Summary of the flux measurements and standardisation method. }
\begin{tabular}{llcl}
\toprule
\toprule
     \multicolumn{1}{c}{{\bf Catalogue}} &
    \multicolumn{1}{c}{{\bf Input }} &  
    \multicolumn{1}{c}{{\bf Input band}} &
    \multicolumn{1}{c}{{\bf Transformation}} \\
   
    \multicolumn{1}{c}{}&
    \multicolumn{1}{c}{\bf from catalogue}&
    \multicolumn{1}{c}{\bf in keV}&
    \multicolumn{1}{c}{} \\    
\midrule 
 {\bf \normalsize{RASS-based}}&&&\\
 \hdashline[0.1pt/2pt]
REFLEX, &$F_{\rm ap}, eF_{\rm ap}, R_{ap}$       	& $[0.1$--$2.4]$	   & $-$ \\
\hdashline[0.1pt/2pt]
NORAS & $F_{\rm ap}, eF_{\rm ap}, R_{ap}$         	& $[0.1$--$2.4]$	  &$-$\\
 \hdashline[0.1pt/2pt]
BCS, eBCS & $F_{tot}, CR_{tot}, eCR_{tot}$   	& $[0.1$--$2.4]$	 	 &$F_{\rm ap}=F_{tot}\times CR_{\rm VTP}/CR_{tot}$\\
&$R_{\rm VTP}, CR_{\rm VTP}, $                      &                          	&$eF_{\rm ap}/F_{\rm ap} = eCR_{tot}/CR_{tot}$\\
&&&$R_{ap}=R_{\rm VTP}$\\
\hdashline[0.1pt/2pt]
SGP &$F_{tot},$                                              	&$[0.1$--$2.4]$	&      	$eF_{tot}=F_{tot}\times eCR_{[0.5-2.0]}/CR_{[0.5-2.0]}$ \\ 
	&$CR_{[0.5-2.0]},eCR_{[0.5-2.0]}$ 		&                             	&\\
\hdashline[0.1pt/2pt]
NEP &$F_{tot}, R_{circ}, SC,$                        	& $[0.5$--$2.0]$       	&$F_{\rm ap}=F_{tot}/SC$\\
&$ CR_{[0.1-2.4]}, eCR_{[0.1-2.4]}$                   &                         &$eF_{\rm ap}/F_{\rm ap}= eCR_{[0.1-2.4]}/CR_{[0.1-2.4]}$\\
&&&$R_{ap}=R_{circ}$\\
\hdashline[0.1pt/2pt]
CIZA I, II & $F_{tot},CR_{tot}, eCR_{tot}$       	& $[0.1$--$2.4]$	   &$eF_{tot}/F_{tot}= eCR_{tot}/CR_{tot}$\\
\hdashline[0.1pt/2pt]
RXGCC 	& $F_{500}, CR_{500}$                  	&$[0.1$--$2.4]$	       	& $F_{\rm ap}=F_{500} \times CR_{\rm SIG}/CR_{\rm 500}$\\
	      	& $R_{\rm SIG}, CR_{\rm SIG}, eCR_{\rm SIG}$	&		& $eF_{\rm ap}/F_{\rm ap}= eCR_{\rm SIG}/CR_{\rm SIG}$ \\
                 &                                                   	&                          	&$R_{ap}=R_{\rm SIG}$\\

\midrule
{\bf \normalsize{Serendipitous}} &&&\\ 
\hdashline[0.1pt/2pt] 
400SD   & $F_{tot}, eF_{tot}$&$[0.5$--$2.0]$ &$-$\\
\hdashline[0.1pt/2pt]
160SD & $F_{tot}, eF_{tot}$ &$[0.5$--$2.0]$ &$-$\\
\hdashline[0.1pt/2pt]
\hdashline[0.1pt/2pt]
SHARC  & $F_{tot}, R_{80}, CR_{tot}, eCR_{tot}$&$[0.5$--$2.0]$ &$F_{\rm ap} = 0.8 \times F_{tot}$ \\
 &&&$eF_{\rm ap}=F_{\rm ap}\times eCR_{tot}/CR_{tot}$ \\
&&&$R_{ap}=R_{80}$ \\
WARPS I,II& $F_{tot}$ &$[0.5$--$2.0]$ &$eF_{tot}=10\%~F_{tot}$ \\
\hdashline[0.1pt/2pt]
EMSS  & $F_{tot}, eF_{tot}$ {\small(EMSS/ASCA)}& $[0.3$--$3.5]$ & $-$\\
\bottomrule
\end{tabular}
  \tablefoot{Columns: (1) catalogue; (2) Published quantities used as input: $CR_{tot}$, $F_{tot}$ denote total count rates and fluxes, $CR_{ap}$, $F_{ap}$ count rates and fluxes in a given aperture $R_{ap}$, errors on quantities are noted by the prefix $e$; see text for other quantities (3) Energy band of the input flux (4): Equations used to derive the aperture flux and flux errors. Count rates and their errors are used to obtain the fractional flux uncertainties.}
  \end{table*}

\section{Luminosity and mass determination}\label{sec:lm}

\subsection{General method}  \label{sec:method}

The X-ray luminosity information published in each catalogues is not homogeneous, being given in various energy bands and apertures, and for different cosmologies. MCXC-I provided standardised luminosity and mass within $\Rv$, $\Lv$ and $\Mv$.  As detailed in Sect.~3.4.1 of \citet[][]{2011A&A...534A.109P}, the published catalogue luminosities were first converted to $[0.1$--$2.4]\ \keV$ band luminosities for the reference cosmology when necessary and then aperture-corrected. 

In a significant improvement with respect to the original release, MCXC-II has been fully updated with new $\Lv$ calculations. For $96\%$ of MCXC-II, the $\Lv$ was recalculated using the X-ray flux information from the original source catalogues. 

The remaining 4\% correspond to 14 EMSS objects and the clusters in the MACS catalogue, the specific treatment of which is detailed in Sect.~\ref{sec:lm500_emss} and Sect.~\ref{sec:lm500_macs}. This approach was deemed necessary to include newly-available redshifts for which no luminosity had been published. Moreover, this new approach allowed us to homogenise further the luminosity measurements, to calculate statistical errors on $\Lv$ directly from measured quantities, and to rigorously update the luminosity with new redshifts, if applicable.

The starting observable is the flux in a given energy band at Earth in a given aperture, $F_{\rm ap,[E_1-E_2]}$, which must be converted to $L_{\rm ap,[E_1-E_2]}$, the X-ray luminosity in the same energy band at the source. This conversion is undertaken via the standard K-correction factor $K_{[E_1-E_2]}\ (T,z)$, namely: the ratio of the  luminosity in the measured energy band, $[E_1-E_2]$, to the luminosity in the shifted  energy band of the incoming photons, $[E_1-E_2](1+z)$ as follows:
\begin{eqnarray}
     K_{\rm [E_1-E_2]}\ (T,z) & = & \frac{L_{\rm X,[E_1-E_2]}}{L_{\rm X,[E_1(1+z)-E_2(1+z)]}}  \\ 
     L_{\rm ap, [E_1-E_2]} & =  & 4\ \pi D_{\rm l}^{2}(z)\  F_{\rm ap, [E_1-E_2]}\  K_{\rm [E_1-E_2]}\ (T,z)
     \label{eq:KK}
\end{eqnarray}
where $D_{\rm l}$ is the luminosity distance. The quantity  $K_{[E_1-E_2]}\ (T,z)$ depends on the temperature, $T$, and redshift, $z$, and can be computed from plasma emission models. Starting from the flux allows us to propagate the redshift updates in a rigorous way. Indeed, it is because of the $K(z,T)$ correction that the luminosity in the source catalogue cannot simply be rescaled by the change in luminosity distance owing to the change in redshift.

If the input flux is given in an energy band different from the reference $[0.1$--$2.4]$ keV band, we further need to convert luminosities. We define $K_{\rm L}$ as   
the ratio of the luminosity in a given energy band, $[E_1-E_2]$, to the luminosity in another energy band $[E^\prime_1-E^\prime_2]$. This ratio depends on the considered energy band and on the temperature:
\begin{eqnarray}
    \label{eq:KL}
   L_{\rm ap,[0.1-2.4]} &=& K_{\rm L,[E_1-E_2]}(T)\ L_{\rm ap,[E_1-E_2]}
  \end{eqnarray}
The last step is to convert the luminosity measured in a given aperture to $\Lv$. As in \citet{2011A&A...534A.109P}, we assumed a universal density profile such that:
\begin{eqnarray}
\label{eq:lapl500}
 L_{\rm 500, [0.1-2.4]} &= &f_{\rm ap}(R_{\rm ap}/\Rv)\  L_{\rm ap, [0.1-2.4]}
 \end{eqnarray}
where $f_{\rm ap}(R_{ap}/\Rv)$ depends on the shape of the universal density profile, $\rho_{\rm gas}(r/\Rv)$ (see Appendix~\ref{ap:apcor}). 

To estimate the temperature and $\Rv$ appearing in the above equations, we used the \LM\ and \LT\ relations:

\begin{eqnarray}\label{eq:lt}
    A_{\rm T}\left(\frac{T}{5\ \keV}\right)^{\alpha_{\rm T}}  &=& E(z)^{\beta_{\rm T}}\ \left(\frac{L_{\rm 500, [0.1-2.4]}}{10^{44}\  {\rm \ergs}}\right)  \\
  \label{eq:lm}
  A_{\rm M}\ \left(\frac{\Mv}{3\times10^{14}\ \msun} \right)^{\alpha_{\rm M}} &=& E(z)^{\beta_{\rm M}}\  \left(\frac{L_{\rm 500, [0.1-2.4]}}{10^{44}\  {\rm \ergs}}\right),
\end{eqnarray}
together with the relation between $\Rv$ and $\Mv$
\begin{equation}
\label{eq:r500}
\Rv = \left(\frac{\Mv}{(4\pi/3)\ 500\ \rho_{\rm rc}(z)}\right)^{1/3},
\end{equation}
where $\rho_{\rm rc}(z)$ is the critical density of the Universe at the cluster redshift. 

For a given input flux, the unknown parameters $T$, $\Lv$ and $\Mv$, must simultaneously solve the three equations given in Eqs.~\ref{eq:lapl500}, ~\ref{eq:lt}, and ~\ref{eq:lm}. In practice this is achieved through a double iteration: 

\begin{enumerate}
    \item For a given temperature $T$, $L_{\rm ap,[0.1-2.4]}$ is derived from Eq.\ref{eq:KK}, combined  with Eq.~\ref{eq:KL} if necessary. 
    \item The  corresponding luminosity, $\Lv$, and mass $\Mv$ are then derived iteratively using Eq.~\ref{eq:lapl500} and Eq.~\ref{eq:lm}. The first guess $\Mv$ is derived from Eq.~\ref{eq:lm}, setting $\Lv$ to $\Lap$, then a new $\Lv$ is derived from Eq.~\ref{eq:lapl500} with $\Rv$ from Eq.~\ref{eq:r500}. The process continues until it converges. 
    \item A new temperature $T$ can be derived from Eq.~\ref{eq:lt}, and the procedure restarted from step~1, until convergence. We used a starting temperature of 3 $\keV$ for the iteration procedure. The median number of iterations, $n_{\rm iter}$, necessary to reach a convergence of $\Delta(T)/T <10^{-5}$ is three, with $n_{\rm iter}$ between two and four for $95\%$ of the cases. This rapid convergence, which is insensitive to the starting $T$ value, is due to the nature of the corresponding implicit equation on $T$. This is further discussed in Appendix~\ref{ap:iter}. 
\end{enumerate}

 This method allows for a fully consistent computation of the uncertainties on $\Lv$ and $\Mv$ for each cluster directly from the flux errors. The upper and lower limits on $\Lv$  and $\Mv$ are readily obtained by re-running the procedure with the upper and lower limits on the flux.

\begin{figure}[!t]
\begin{centering}
    \includegraphics[width=\columnwidth]{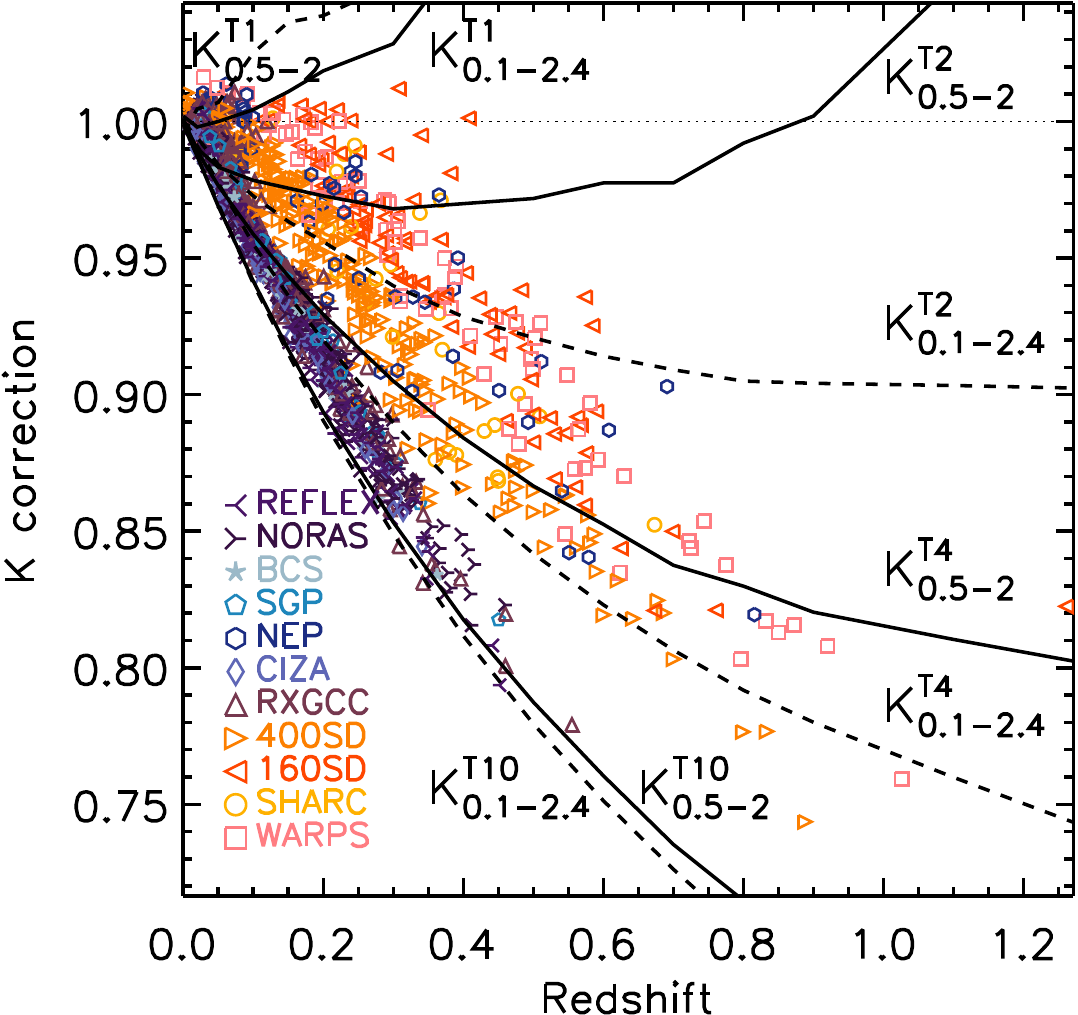}
    \caption{$K$-correction as a function of redshift. Lines: Correction computed with an {\tt apec} model for two energy bands of the ROSAT input flux, $\bandi$ and $\bando$, and four different temperatures (1, 2, 4, and 10~keV).  Each curve is labelled accordingly, with the information in the index and exponent, respectively. For instance: $K^{10}_{0.1-2.4}$ is the $K$-correction for the flux in the energy band  $\bando$ and a cluster temperature of $10\ \keV$. Points: $K$-correction  for each object in the ROSAT catalogues with a measured flux; the temperature being estimated from the iterative procedure described in Sect.\ref{sec:method}. }  
 \label{fig:Kz}   
\end{centering}
\end{figure}

\subsection{Standardised input fluxes}

Table \ref{tab:Fx_methods} gives a synthetic overview of the flux standardisation for each input catalogue.  The aperture flux, $\Fap$,  and corresponding radius, $\Rap$,  are directly available in the REFLEX and NORAS catalogues. For NEP,  \citet{2006ApJS..162..304H} provide $\Rap$ (denoted $R_{\rm circ}$) and $\Fap$ was recovered from the published total fluxes and size correction, $SC$. For SHARC,   \citet{2000ApJS..126..209R} chose an aperture  $R_{80}$ that would contain $80\%$ of the flux for a model cluster. In this case, $\Rap=R_{80}$ and   $\Fap$ is equal to $0.8$ times the published total flux. For BCS, the detection was based on the VTP (Voronoi tesselation and percolation) technique  \citep{1998MNRAS.301..881E}, and the source radius was expressed as $R_{\rm VTP}$ the equivalent radius of the source detected by VTP. We assumed $\Rap = R_{\rm VTP}$ and retrieved the corresponding flux from the total flux multiplied by the ratio between the total and VTP count rates. 
For RXGCC, \citet{2022A&A...658A..59X} provide $\Rap$ (denoted $R_{\rm SIG}$), which was defined from a growth curve analysis, and the corresponding count rate, $CR_{\rm SIG}$.  The aperture flux was retrieved from their published flux within $\Rv$  (estimated iteratively) multiplied by the ratio of count rates within $R_{\rm SIG}$ and $\Rv$. For the other catalogues, only the total flux is available. For these, we used a fixed size correction, as explained in the next Section. 

Errors on the flux were scaled from the relative statistical errors on the observed count rates, when available. Those are therefore as close as possible to purely statistical errors. For SHARC-SOUTH,  we removed quadratically the 5\% systematic uncertainty that \citet{2003MNRAS.341.1093B} added to their count rate errors. No flux errors are available for WARPs clusters. Errors are further discussed in Sect.~\ref{sec:systematic}.

\subsection{Input models} \label{sec:inmod}
\begin{figure}[!t]
\begin{centering}
    \includegraphics[width=0.95\columnwidth]{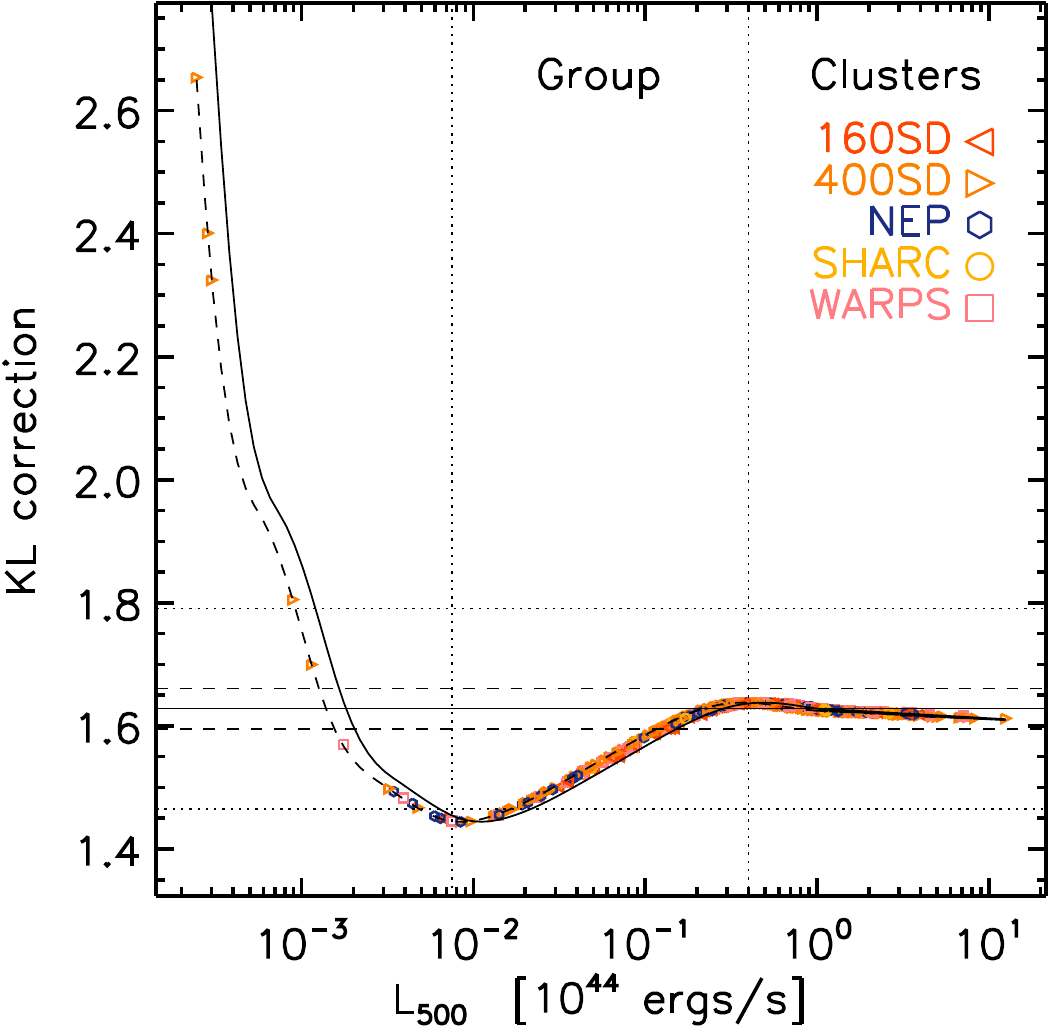}
    \caption{ $\KL$-correction factor,  the ratio of the luminosity in $\bando$ energy band  to that in the $\bandi$ band.  The theoretical variation with luminosity, for $z=0$ and $z=0.5$ are shown with solid and dashed lines, respectively. Here the dependence on temperature $\KL(T)$ has been translated to a variation with luminosity via the \LT\ relation.  The redshift dependence is a small translation in the $\Lv$--$\KL$ plane via the $E(z)$ factor. 
  The horizontal lines show the value for a $10^{44}\ \ergs$ cluster at $z=0$ (solid), $\pm2\%$ (dashed) or $\pm10\%$ (dotted), respectively. 
  The points show the $\KL$-correction (Eq.~\ref{eq:KL}) applied to the  400SD, 160SD, NEP, SHARC, and  WARPS clusters (for which the catalogue flux is given in the $\bandi$  band). Each point corresponds to a cluster from these samples; the temperature being estimated from the iterative procedure described in Sect.\ \ref{sec:method}.  
   } 
      \label{fig:KL}   
\end{centering}
\end{figure}

The $K$ and $\KL$ corrections were computed from an  {\tt apec} model with a fixed abundance of $0.3$ relative to solar.  Figure~\ref{fig:Kz}  shows the variation with redshift of the $K$-correction for the two energy bands of interest for ROSAT catalogues. The correction is always less than $25\%$. 

The $\KL$-correction, plotted in Fig.~\ref{fig:KL} as a function of luminosity, is almost constant in the cluster regime ($\Mv\!>\!10^{14}\ \msun$ or $\Lv \gtrsim 4\times 10^{43}\ \ergs$) with less than $1.6\%$ variation. It decreases in the group regime, by  $10\%$  at $\Lv\!\sim\!7.5\times 10^{42}\ \ergs$ ($\Mv\!\sim\!10^{13}\ \msun$). For lower luminosity/temperature systems, the $\KL$-correction increases again and becomes extremely sensitive to the temperature, due to the increasing contribution of emission lines to the flux. However, there are only six MCXC-II objects with $\Lv\!<\!10^{41}\ \ergs$, most likely galaxy halos, for which the $\KL$-correction is more than $10\%$ larger than the mean.

The $\KL$  differs from that computed by \citet{2011A&A...534A.109P} for low temperature (low luminosity) clusters. \citet{2011A&A...534A.109P} used a correction from a tabulated {\tt MeKaL} model which was extrapolated below $T=1\ \keV$, thus with a weaker temperature dependence. The difference is less than $0.5\%$ in the cluster regime, and does not exceed $\pm 10\%$ between $0.3$~and~$1\ \keV$; however, it reaches a factor of two at $kT=0.2\keV$. 

The aperture correction is the same as that used for constructing MCXC-I, which was based on the \rexcess\ density profile \citep[see][Sect~3.4.1.]{2011A&A...534A.109P} and is detailed in Appendix~\ref{ap:apcor}. If only the total flux was provided in the input catalogue, $\Lv$ was calculated from $L_{tot}$ multiplied by a factor of $0.91385$, corresponding to the ratio between the integral of the profile within $\Rv$ and the integral within a very large radius.  In this case, step~2 is trivial, and the procedure only requires the temperature iteration to estimate $\Lv$, with $\Mv$ computed at the end with Eq.~\ref{eq:lm}.

 We used the non-core-excised \LT\  relation in the $\bando$ band obtained by \cite{Pratt2009} from \rexcess\ data. \citet{2011A&A...534A.109P} used this same relation for the computation of the temperature needed for the $\KL$ conversion factor. 

 We used a re-calibration of \LM\ relation of \citet{2011A&A...534A.109P}, using an evolution factor $E(z)^{-2}$, appropriate for luminosities in the soft energy band, instead of $E(z)^{-7/3}$ assumed by \citet{2010A&A...517A..92A}, whose relation was used for MCXC-I. This local re-calibration was based on \rexcess\  data as in \cite{2010A&A...517A..92A}.
The parameters of Eq.~\ref{eq:lt} and Eq.~\ref{eq:lm}  are then:
\begin{eqnarray}
  	A_{\rm T}=3.46 & \alpha_{\rm T}=3.00 &\beta_{\rm T}=-1 \\
 	\log_{10}(A_{\rm M})=-0.282 & \alpha_{\rm M}=1.675& \beta_{\rm M}=-2.
	\label{eq:ltmparam}
\end{eqnarray}
  
\begin{figure}[!t]
\begin{centering}
    \includegraphics[width=0.9\columnwidth]{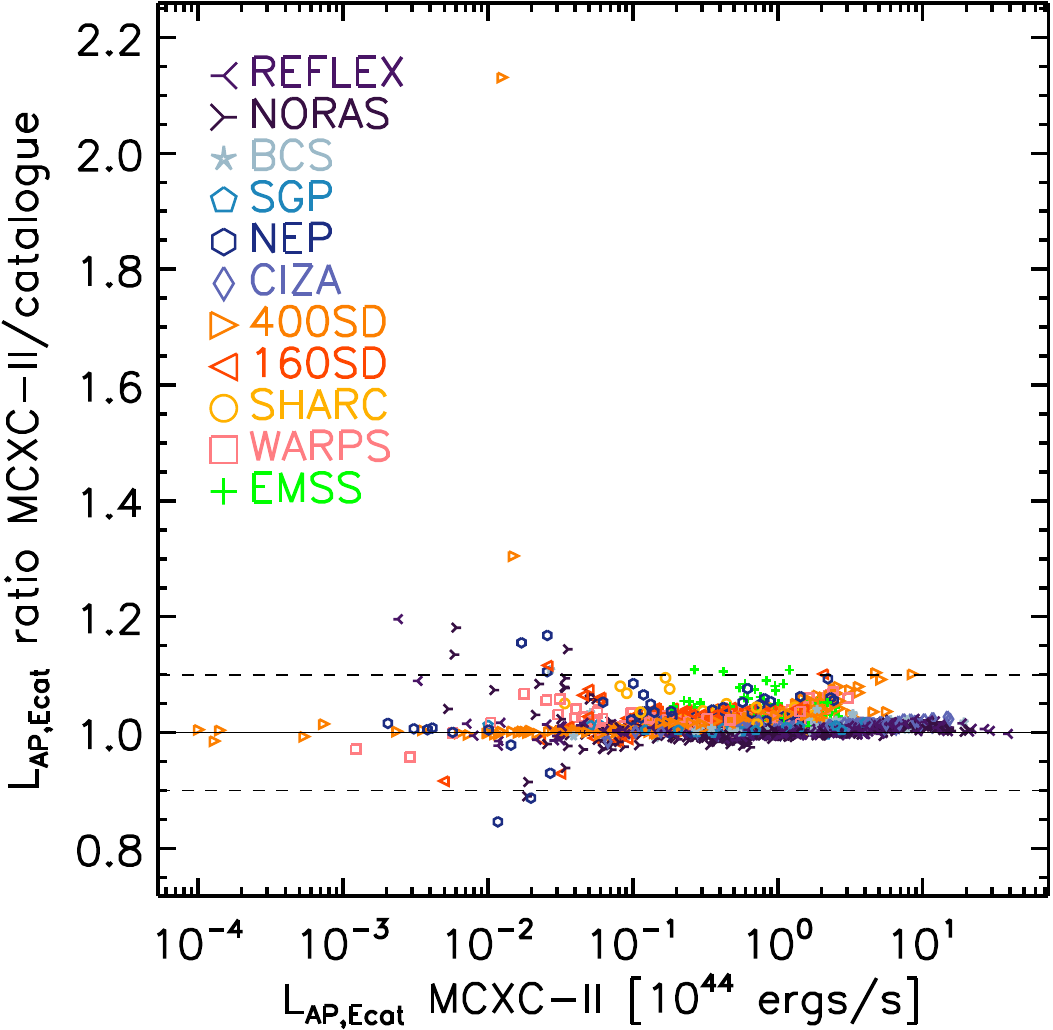}
    \caption{ Aperture luminosity estimated from the flux using the iterative procedure described in Sect.~\ref{sec:method} compared to the luminosity published in the corresponding input catalogue. Only objects in the original MCXC-I input catalogues with unchanged redshift are shown.}    
 \label{fig:comp_laplapcat}
\end{centering}
\end{figure}

\begin{figure*}[!t]
\begin{centering}
	\resizebox{0.9\textwidth}{!} {
       \includegraphics[width=\columnwidth]{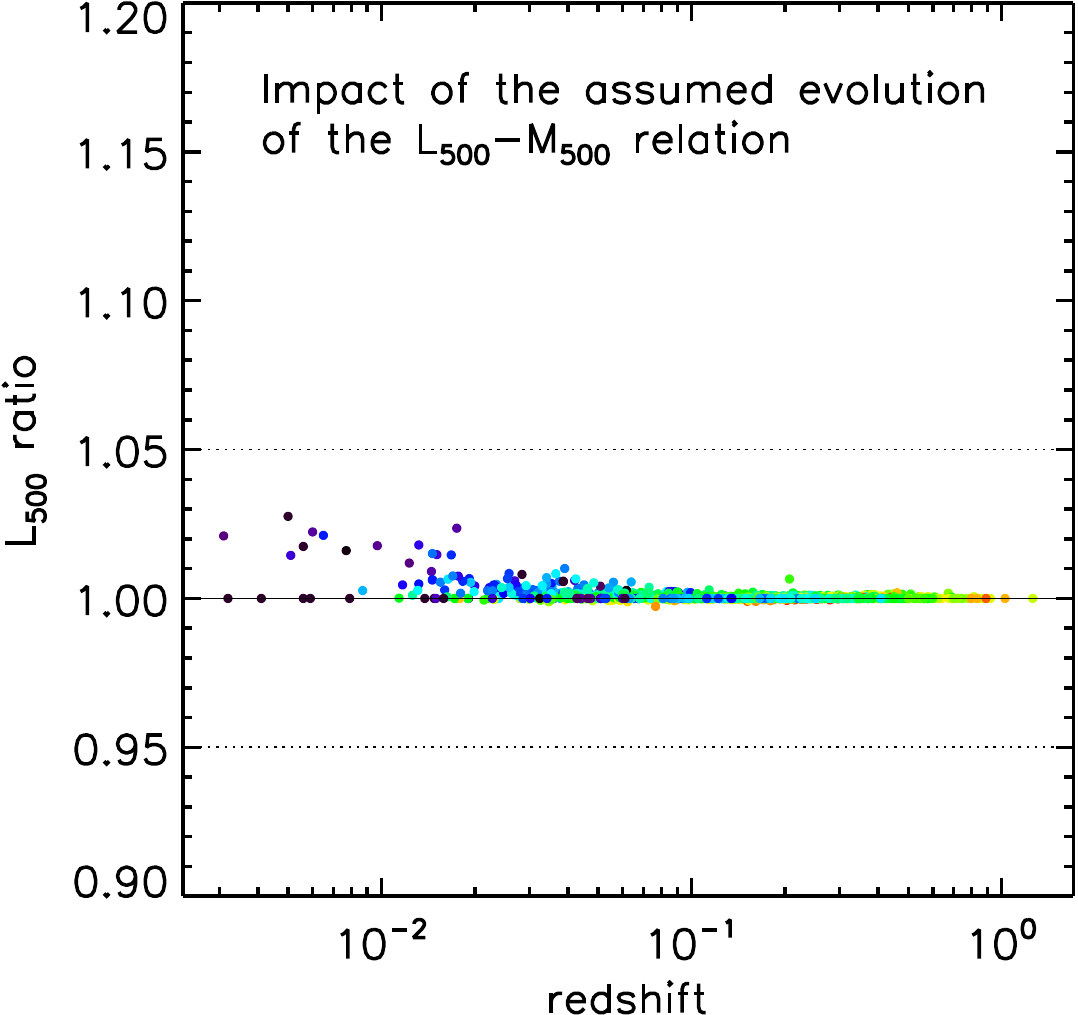}
       \hspace{1.5cm}
        \includegraphics[width=\columnwidth]{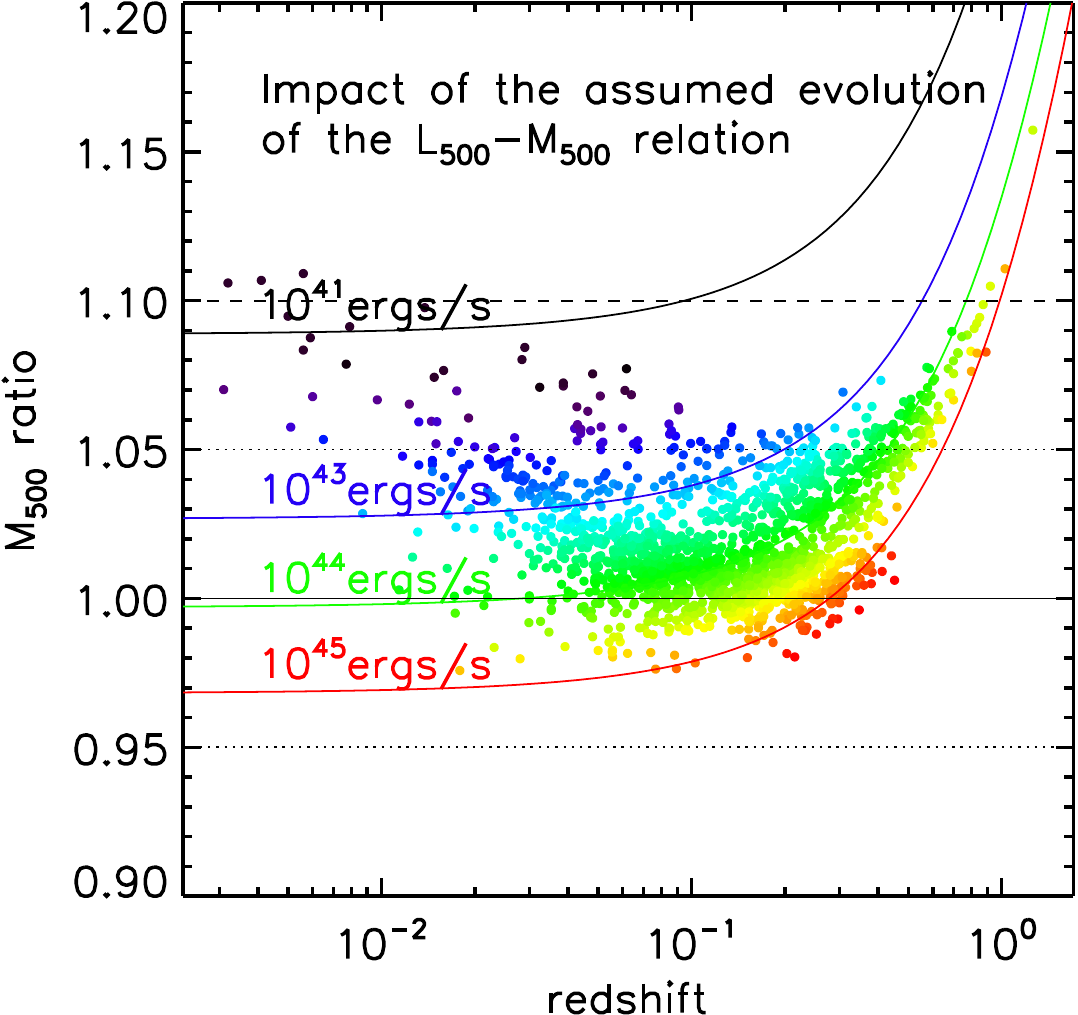}
        }
    \caption{Impact of the change in the evolution factor of the \LM\ relation (Sect.~\ref{sec:lmr}). The $\Lv$ and $\Mv$ values were estimated using the iterative procedure described in Sect.\ref{sec:method}. Left: Ratio of $\Lv$ estimated with the recalibrated \LM\ relation evolving as $E(z)^{-2}$ (Eq.~\ref{eq:ltmparam}) to the value obtained with \rexcess\ relation ($E(z)^{-7/3}$ evolution). Ratios are colour-coded according to cluster luminosity and plotted as a function of redshifts. Right. Same for the mass, $\Mv$. The curves show the theoretical ratio for different luminosity ratios (see Sect.~\ref{sec:lmr}).}
    \label{fig:comp_lmr}
    \end{centering}
\end{figure*}
\subsection{Comparison between original catalogue luminosity and recomputed luminosity}\label{sec:comp_laplapcat}

In Figure~\ref{fig:comp_laplapcat}, we compare the aperture luminosity that we computed from the flux using the iterative process described in Sect.~\ref{sec:method}, to that published in the input catalogues. We obviously considered only clusters with unchanged redshifts, and compared the $\Lap$ value in the energy band of the input flux measurement.  In principle, the two estimates should only differ as a result of the $K$-correction, that is, on the assumed temperature.  

The ratio of the two estimates exhibits a large scatter, which increases at lower luminosities, 
 $\Lap \lesssim 10^{43}\ \ergs$. This is not expected as it concerns low temperature ($T<2\ \keV$), mainly low $z$, objects for which the $K$-correction is very small (Fig.~\ref{fig:Kz}). For the clusters with $\Lap \lesssim 10^{43}\ \ergs$, the $K$-correction we estimated is indeed less than $\pm2\%$.  We found that the difference with the published luminosities is in fact due to truncation errors when the luminosity value has only one significant digit.  
 
For $\Lap\!>\!10^{43}\ \ergs$, the differences between the published and the re-estimated $\Lap$ is less than $0.7\%$ on average, with a standard deviation of $ 1.8\%$, and a maximum difference of   $9\%$. However, for the 400SD catalogue (orange triangles in Fig.~\ref{fig:comp_laplapcat}), the ratio of the two quantities systematically increases with luminosity. This is due to the impact of different $K$-corrections: the value estimated by \citet{{2007ApJS..172..561B}},  recovered from the flux to published luminosity ratio, decreases more steeply with $z$ than our $K$-correction and therefore yields higher  luminosities.  We verified that this different behavior of the $K$-correction is in turn essentially due to a different slope of the assumed \LT\ relation. \citet{{2007ApJS..172..561B}} estimated the temperature from the \LT\ relation of \citet{mark98}, which has a shallower slope than that from \rexcess\ ($\alpha_{\rm T}\sim 2$ compared to $\alpha_{\rm T}\sim 3$), and which therefore yields higher temperatures at high luminosities.  

Finally, there are two prominent unexplained outliers, \object{MCXC\,J0458.9-0029} ($z=0.015$) and  \object{MCXC\,J1329.4+1143} ($z=0.023$), corresponding to two 400SD clusters with $\Lap\!\sim\!10^{43}\ \ergs$. We found that the published luminosity and flux measurements for these objects are inconsistent, and correspond to a $K$-corrections of $K=0.47$ and $K=0.77$, respectively. At these nearby redshifts the $K$-correction should be very close to unity. 

The above (mostly minor) differences contribute to the difference between MCXC-I and MCXC-II $\Lv$ luminosities, which is discussed in detail in Sect.~\ref{sec:comp_mcxc}.

\subsection{Effect of the correction of the evolution factor in the \LM ~relation} \label{sec:lmr} 

Following the full iterative process described in Sect.~\ref{sec:method}, we computed the $\Lv$ and $\Mv$ values from the flux for the new \LM\ relation. For comparison, we reran the process for the  \rexcess\ relation used in the construction of the original MCXC-I:
\begin{eqnarray}
 	\log_{10}(A_{\rm M}^{\rm rex})=-0.274 & \alpha_{\rm M}^{\rm rex}=1.64 & \beta_{\rm M}^{\rm rex}=-7/3.
	\label{eq:lmrexcess}
\end{eqnarray}
These results are denoted $\Lv^{\rm REX}$ and $\Mv^{\rm REX}$, respectively. 

We expect a only small effect on $\Lv$ as the change in evolution factor only affects the aperture correction, via the value of $\Rv$ which scales as $\Mv^{1/3}$. Indeed the effect is less than $2\%$ on $\Lv$, and negligible above $z>0.1$. The counter-intuitive increase of the difference at low redshift (Fig.~\ref{fig:comp_lmr}, left panel) is due to the larger aperture correction at these redshifts, where objects have smaller $\Rap/\Rv$ on average. 

The effect on $\Mv$ is therefore expected to be a direct consequence of the change in the parameters of the \LM\ relation. The right-hand panel of Fig.~\ref{fig:comp_lmr} shows the ratio of the $\Mv$ values as a function of redshift. Cluster points are colour-coded by luminosity.  The Figure  also shows the theoretical variation of the ratio for different values of the luminosity. From Eq.~\ref{eq:lm}, with parameters from Eq.~\ref{eq:ltmparam} and Eq.~\ref{eq:lmrexcess},   the effect is expected to be small:
\begin{equation}
\frac{\Mv}{\Mv^{\rm rex}} = 0.998\ E(z)^{0.23}  \left( \frac{\Lv}{10^{44} \ergs}\right)^{-0.013}.
\end{equation}
At high z, the mass is affected by the weaker evolution factor.  However, the mass is also affected by the corresponding re-calibration of the local mass slope (the normalisation at the pivot is changed by less than $0.2\%$). This is illustrated Fig.~\ref{fig:comp_lmr}.  Below $z\sim0.1$ the mass evolution is negligible, but higher/lower luminosities correspond to lower/higher masses due to the (slightly) shallower mass slope.  With decreasing $z$ the median mass of the MCXC-II catalogue decreases (towards the group regime), and the slightly shallower mass slope yields a {\it higher} $\Mv$ on average. Above $z\sim0.1$, a given luminosity corresponds to higher masses, an effect that increases with $z$ due to the weaker evolution. 
 
In summary, the effect of the change in the evolution factor is small: the median change in $\Mv$ is $1.4\%$. In catalogue terms, 58\% of the cluster masses are affected by less than $2\%$, 31\%  between $2\%$ and $5\%$, and the maximum effect is only $15\%$ ($11\%$ at $z<1$).

\begin{table}[]
\label{tab:macslx}
    \centering
    \caption{References for the input luminosity data for the different MACS sub-catalogues. }
 \resizebox{\columnwidth}{!} {
    \begin{tabular}{llllllll}
    \toprule
    \toprule
       SUB\_CAT  & $L_{\rm X }$ reference & $N_{cl}$ & Data\\
    \midrule
       MACS\_DR1   & \citet{2008ApJS..174..117M} & 6 & \chandra \\                   
                             & \citet{2007ApJ...661L..33E} & 6  & \chandra  \\
        MACS\_DR2    & \citet{2010MNRAS.406.1773M} & 22  &\chandra\\
                     & \citet{2008ApJS..174..117M} & 1 & \chandra \\ 
       MACS\_DR3    & \citet{2008ApJS..174..117M} & 4  & \chandra \\
                     & \citet{2018MNRAS.479..844R} &  8  & \chandra  \\
                     &  \citet{2012MNRAS.420.2120M} & 10 & \rosat  \\
       MACS\_MISC   & \citet{2018MNRAS.479..844R} & 16 &  \chandra  \\
    \bottomrule
 \end{tabular}}
\end{table}

\subsection{Specific treatment of MACS clusters}\label{sec:lm500_macs}


Measurement of the X-ray luminosity of MACS clusters from RASS data alone is more uncertain than for other catalogues, because most of the objects are not resolved by construction.  Following \citet{2011A&A...534A.109P}, we used the \chandra\  information, when available, as published in the catalogues \citep{2007ApJ...661L..33E,2010MNRAS.406.1773M} or in follow-up work \citep{2008ApJS..174..117M,2010MNRAS.406.1773M}. However, for ten MACS\_DR3 clusters, only RASS luminosities are available. 

The published \chandra\ luminosities were derived from the flux and a $K$-correction estimated from measured temperatures, and can directly be used as input. After performing the energy band conversion when necessary, \citet{2011A&A...534A.109P} relied directly on the published \chandra\ $\Lv$ values, and also used the corresponding $\Mv$ published in  \citet{2008ApJS..174..117M} and \citet{2010MNRAS.406.1773M}. The latter were based on mass proxies measured with \chandra\  $\YX$ and the gas mass, respectively. They may therefore differ from the $\Rv$ and $\Mv$ estimated from the \LM\ relation we used in this work.

For consistency, we chose to derive  $\Lv$ and $\Mv$ following the procedure used for other catalogues, as described in Sect.~\ref{sec:method}.

\begin{enumerate}[]
\item As in \citet{2011A&A...534A.109P}, we converted the bolometric luminosities from \citet{2008ApJS..174..117M} into $[0.1$--$2.4] \ \keV$ band luminosities using the \chandra\ temperature and an {\tt apec} plasme emission model. All other published MACS luminosities are in the  $[0.1$--$2.4]\ \keV$ band.
\item We then performed the following aperture correction:
\begin{itemize}[noitemsep,topsep=0pt]
\item[-]  We considered the  $\Lv$ and $\Rv$ values published by \citet{2008ApJS..174..117M} and \citet{2010MNRAS.406.1773M} as $\Lap$ and $\Rap$ values, and re-estimated $\Lv$ with our standard aperture correction procedure when the aperture was known (Eq.~\ref{eq:lapl500}). This correction is in practice very small,  with a ratio of new to original $\Lv$ values of $1.00\pm0.02$ for \citet{2008ApJS..174..117M} and $0.99\pm0.02$ for \citet{2010MNRAS.406.1773M},  respectively.
\item[-] When the aperture was not known, we applied a constant conversion factor based on the \rexcess\ profile shape, as described in Sect.\ref{sec:method}. The RASS-selected total luminosities from \citet{2012MNRAS.420.2120M} were converted to $\Lv$ by applying a conversion factor of 0.96. Similarly, the \chandra\ luminosity  within  ${\rm R_{200}}$ from  \citet{2007ApJ...661L..33E}  was converted to $\Lv$ using a fixed conversion factor of 0.914, as the ${\rm R_{200}}$ value was not published. No correction was applied to the $\Lv$ published by \citet{2018MNRAS.479..844R}, as $\Rv$ was not published. 
\end{itemize}
\item Finally, the $\Mv$ values were estimated from the \LM\ relation (Eq.~\ref{eq:lm}). 
\end{enumerate}
Statistical errors were propagated from the errors on the input luminosities, except for the RASS luminosities \citep{2012MNRAS.420.2120M} for which no uncertainties were published. 

The resulting \chandra-based $\Lv$ for clusters in common between the various publications are in excellent agreement. Not surprisingly, the ratio between the $\Lv$ derived by  \citet{2018MNRAS.479..844R} and that estimated by   \citet{2010MNRAS.406.1773M} is $1.01\pm0.08$, if we exclude the \object{MACS\,J2243.3-0935} outlier. The independent study of \cite{2008ApJS..174..117M} also agrees well with that of \citet{2010MNRAS.406.1773M}, with a derived $\Lv$ ratio of $0.97$ and a standard deviation of $0.07$. There is a slight offset with the values derived from \citet{2007ApJ...661L..33E} for the six distant MACS\_DR1 clusters in common, with a mean ratio of $0.85$ and a standard deviation of 0.06. 

%
\begin{figure*}[!h]
 \begin{minipage}[t]{0.65\textwidth}
 \vspace{0pt}
   \includegraphics[width=\textwidth]{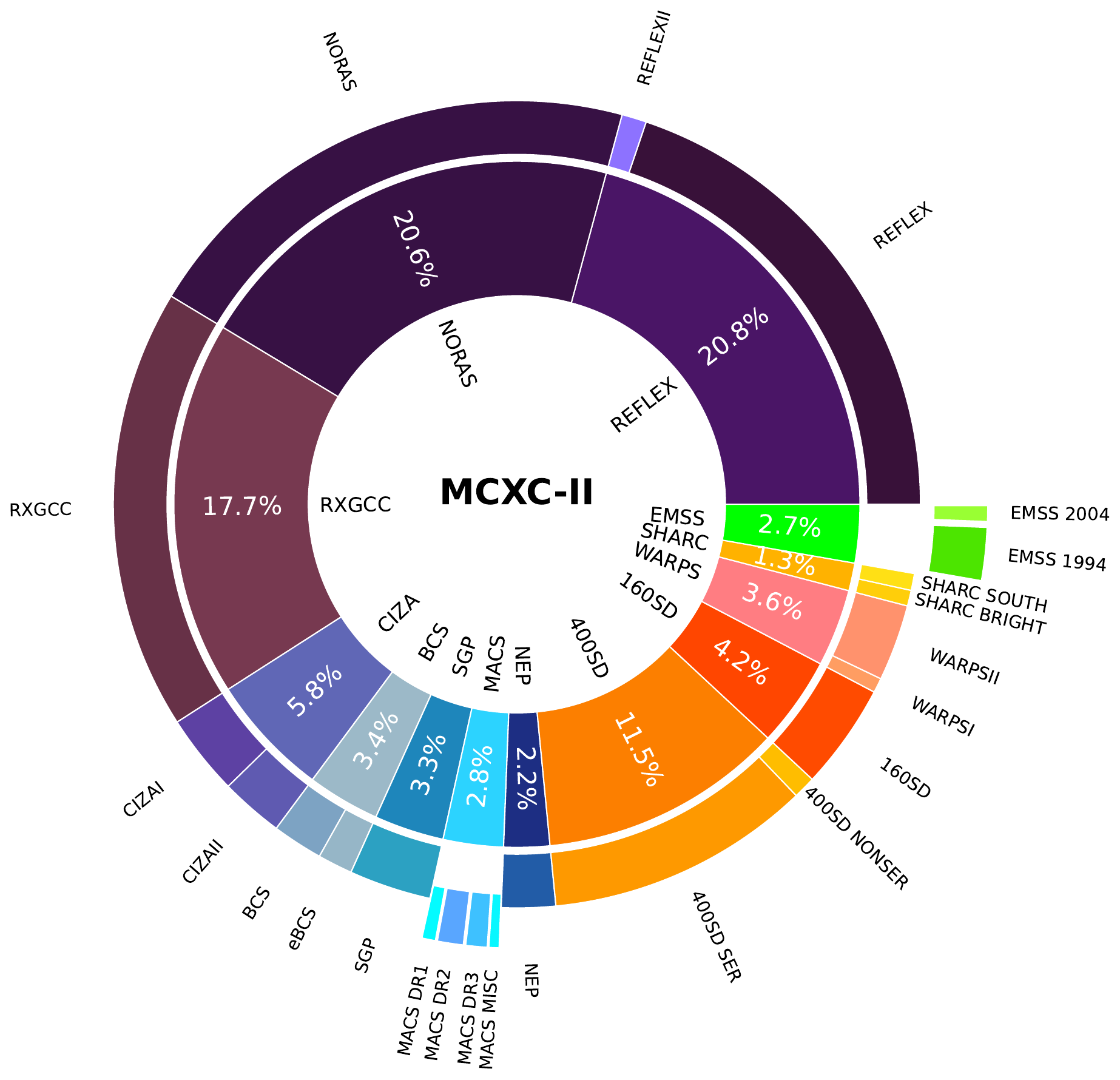}
 \end{minipage}\hfill
 \begin{minipage}[t]{0.3\textwidth}
 \vspace{0pt}
   \caption{
\footnotesize Pie chart representation of the catalogues and sub-catalogues contained in MCXC-II. RASS-based catalogues are depicted with 'cold' colours, while serendipitous catalogues are shown with 'warm' colours. The overall contribution of each (sub-)catalogue to MCXC-II is given by the percentage in the inner annulus. The exact number of objects represented by each percentage is given in Table~\ref{tab:summary_mcxc}.  } \label{fig:piechart}
 \end{minipage}
\end{figure*}

As described in Sect.~\ref{sec:newcat}, some objects appear in several of the MACS studies, notably for the MACS\_DR3 sample. For the choice of the final $\Lv$ and $\Mv$ to be included in MCXC-II, we gave higher priority to \chandra\ measurements than those derived from RASS, then to \chandra\ data with published aperture values and finally considered the size of the sample. This yields to the following sources of information for the MACS\_DR2, DR3 or MISC sub-samples, in priority order: \citet{2010MNRAS.406.1773M}, \citet{2008ApJS..174..117M},  \citealt{2007ApJ...661L..33E} , \citet{2018MNRAS.479..844R}  and \citet{2012MNRAS.420.2120M}. 
This is summarised in Table~\ref{tab:macslx}, which also gives the number of clusters retained from each study. 

\begin{figure*}[!h]
\begin{centering}
      \includegraphics[width=0.925\textwidth]{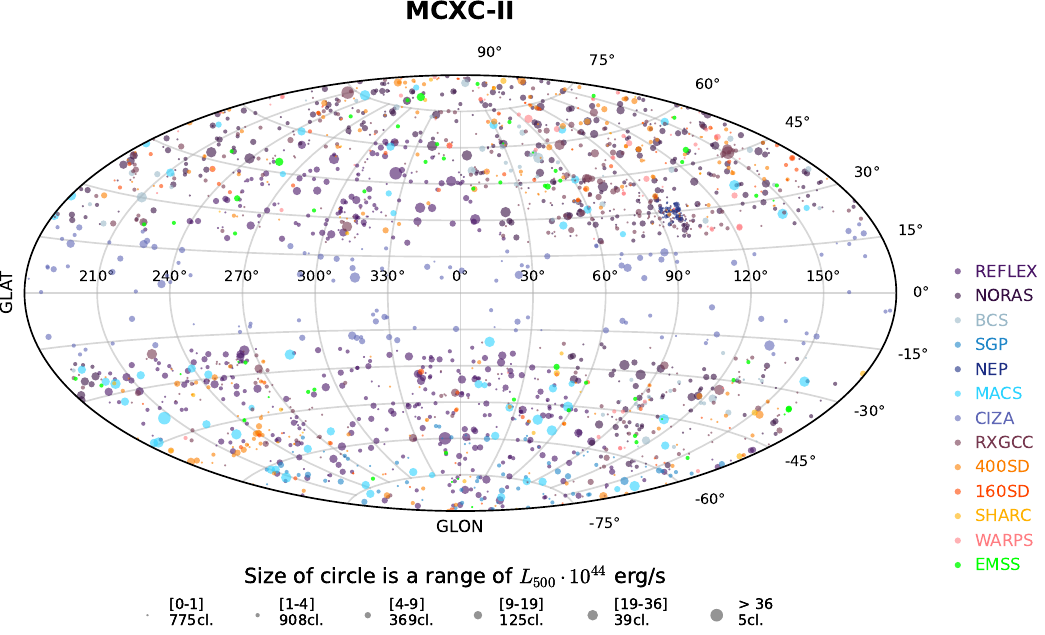}
    \caption{Sky distribution of the MCXC-II clusters. }  
    \label{fig:skymap} 
\end{centering}
\end{figure*}

The luminosity $ \Lv$  of MACS clusters in  MCXC-I and MCXC-II are in excellent agreement, with an error-weighted mean ratio of $1.017\pm0.005$ and a standard deviation consistent with the errors. This follows the agreement between various publications and the present standardisation. However, the MCXC-II  $\Mv$ values differ significantly from those in MCXC-I, which were based on the published \chandra\ mass estimates. The sub-catalogue with the largest difference is that of \citet{2010MNRAS.406.1773M}, where the published mass estimates were obtained with the gas mass $M_{\rm g}$ as a mass proxy, assuming a constant gas mass fraction $f_{\rm g}=0.115$. In contrast, the \rexcess\ data used to determine the $\Mv -\YX$ relation used in the present work show a mass-dependent gas mass fraction with $f_{\rm g} \propto (\Mv / 2.5\times10^{14}\ {\rm M}_{\odot})^{0.21}$ \citep[][see also \citealt{pra22}]{Pratt2009}. Taking into account that the gas mass integrated to $\Rv$ depends on the mass as $M_{\rm g}(<\Rv) \propto \Mv^{2/5}$, one expects that the total mass depends on the assumed $\fgv$ relation as:

\begin{equation}
\Mv = \frac{M_{\rm g}(<\Rv)}{\fgv(M)}  \equiv  \frac{M^{2/5}}{\fgv(M)} 
\end{equation}
and, thus, for the gas mass fraction, we have:
\begin{equation}
\Mv^{\rm Mantz} =  \Mv^{\rm REX}\left(\frac{\Mv}{2.5\times10^{14}{\rm M}_{\odot}}\right)^{0.21*5/3=0.33}.
\label{eq:mmc}
\end{equation}
Along with (second-order) aperture corrections, the above fully explains the difference in mass estimates between the two studies.

\subsection{Specific treatment of EMSS clusters}\label{sec:lm500_emss}

As discussed by \citet{2011A&A...534A.109P}, the {\it Einstein} data reported by \citet{1994ApJS...94..583G} does not allow for the estimation of aperture luminosity and we used their estimated total flux in the $[0.3$--$3.5]\ \keV$ band as input. When available, we relied,  as in \citet{2011A&A...534A.109P}, on the more reliable ASCA measurements of \citet{2004ApJ...609..603H}. In that case, we used the total luminosity in the $[0.5$--$2]\ \keV$ band, after correction for the different cosmologies. The conversion to the $\bando$ luminosity was done with the measured ASCA temperature. The derived values are $\sim 1.61$ higher than the MCXC values. This corresponds to the $\KL$-correction, which was apparently missed for this one catalogue in the MCXC-I publication.

\begin{figure*}[!th]
    \centering
	\resizebox{\textwidth}{!} {
    \includegraphics[]{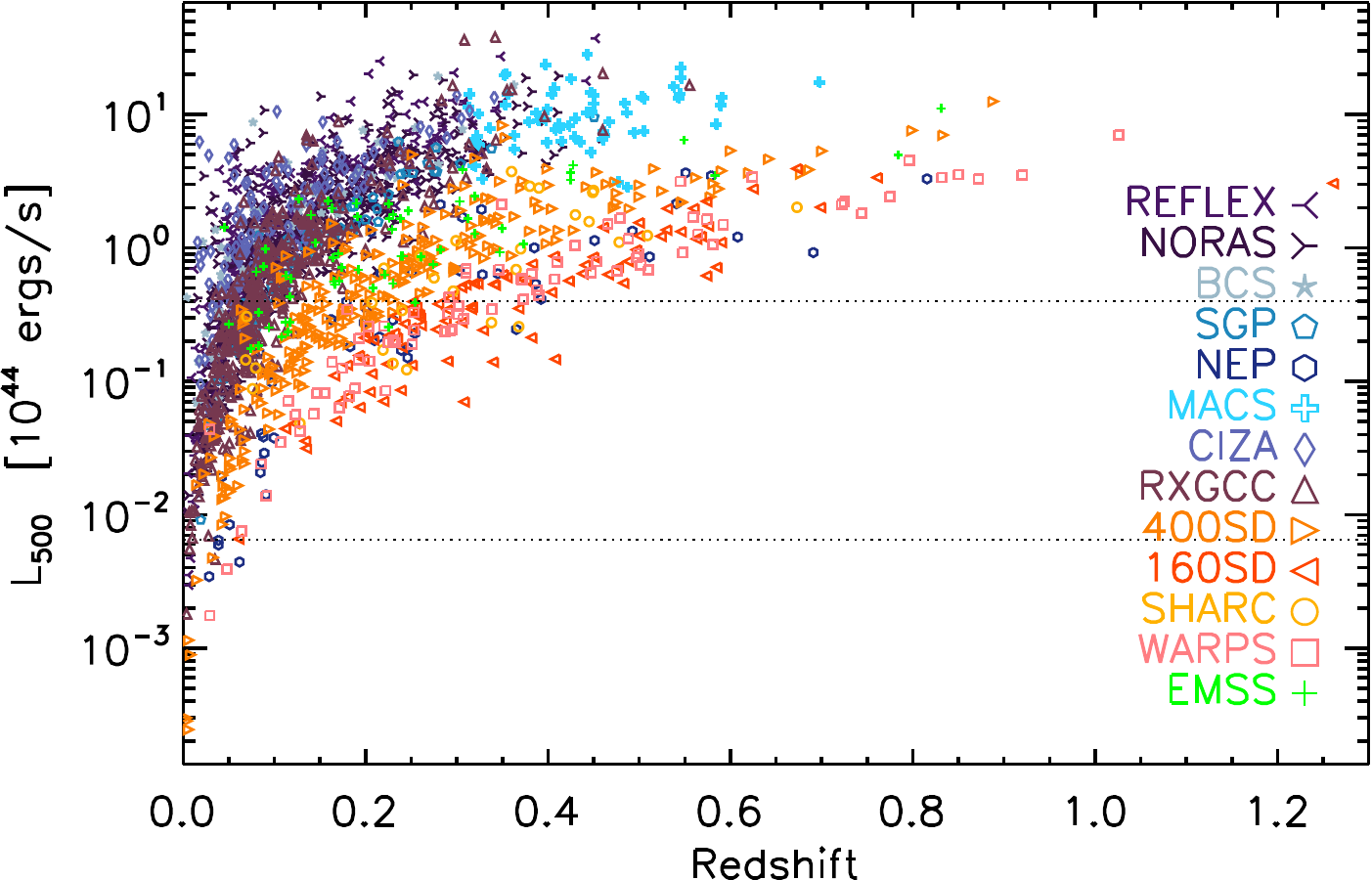}
    \hspace{0.1cm}
    \includegraphics[]{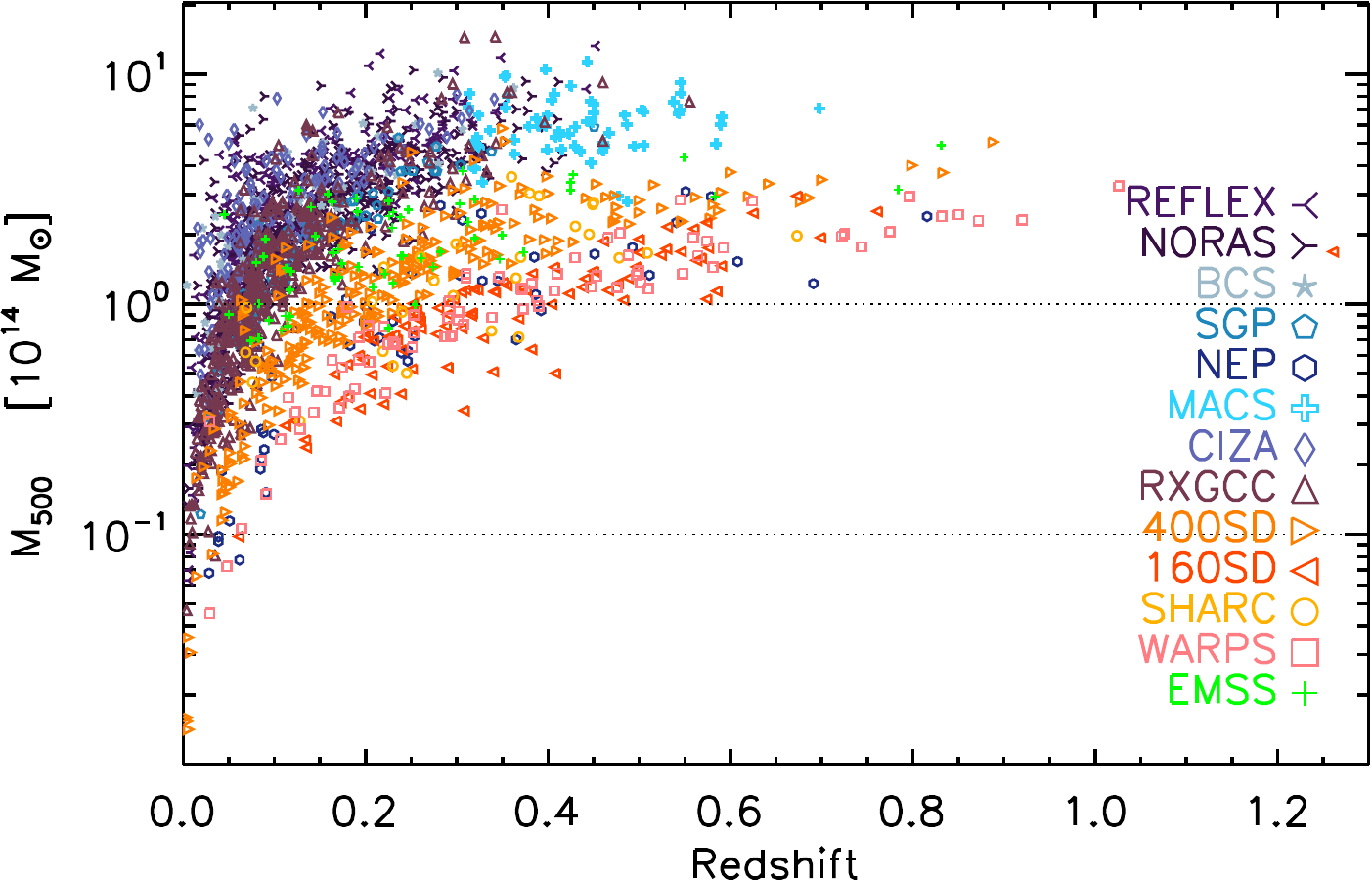}
    }
    \caption{ Luminosity, $\Lv$, in the energy band $[0.1-2.4]$ keV (Left) and mass $ \Mv$ (Right) of the 2221 MCXC-II clusters, plotted as a function of redshift. Source catalogues are identified by different symbols and colors as labelled in the Figure.  The horizontal lines roughly delimit the group and cluster mass regimes. }  
    \label{fig:L500M500_all}
\end{figure*}
\begin{figure}[t]
\begin{centering}
   \includegraphics[width=0.97\columnwidth]{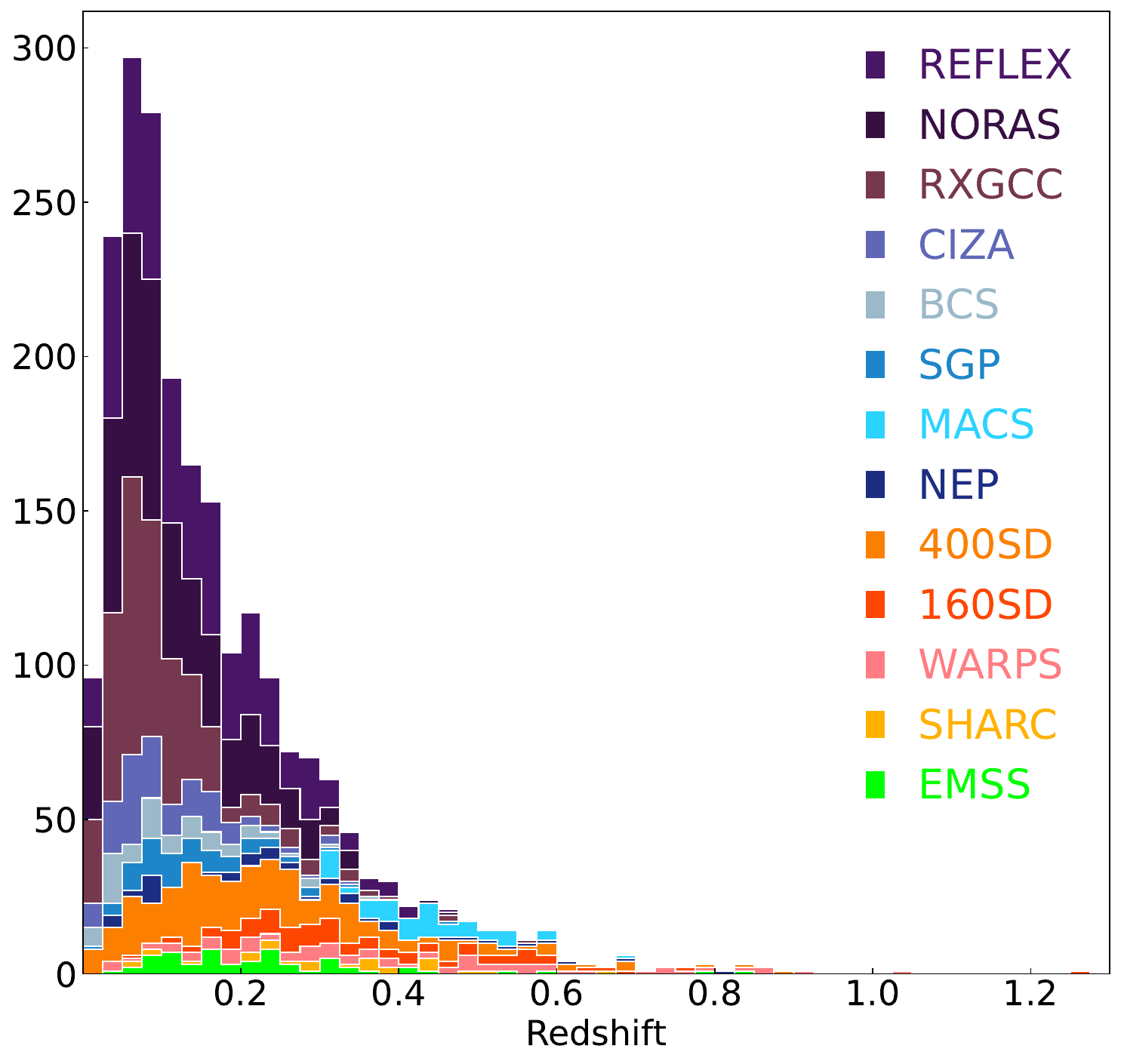}
\caption{MCXC redshift distribution, with histograms coloured by source catalogue as given in the legend. }  
\label{fig:zcats}
\end{centering}
\end{figure}

\begin{figure*}[!th]
\centering
\includegraphics[width=0.95\textwidth]{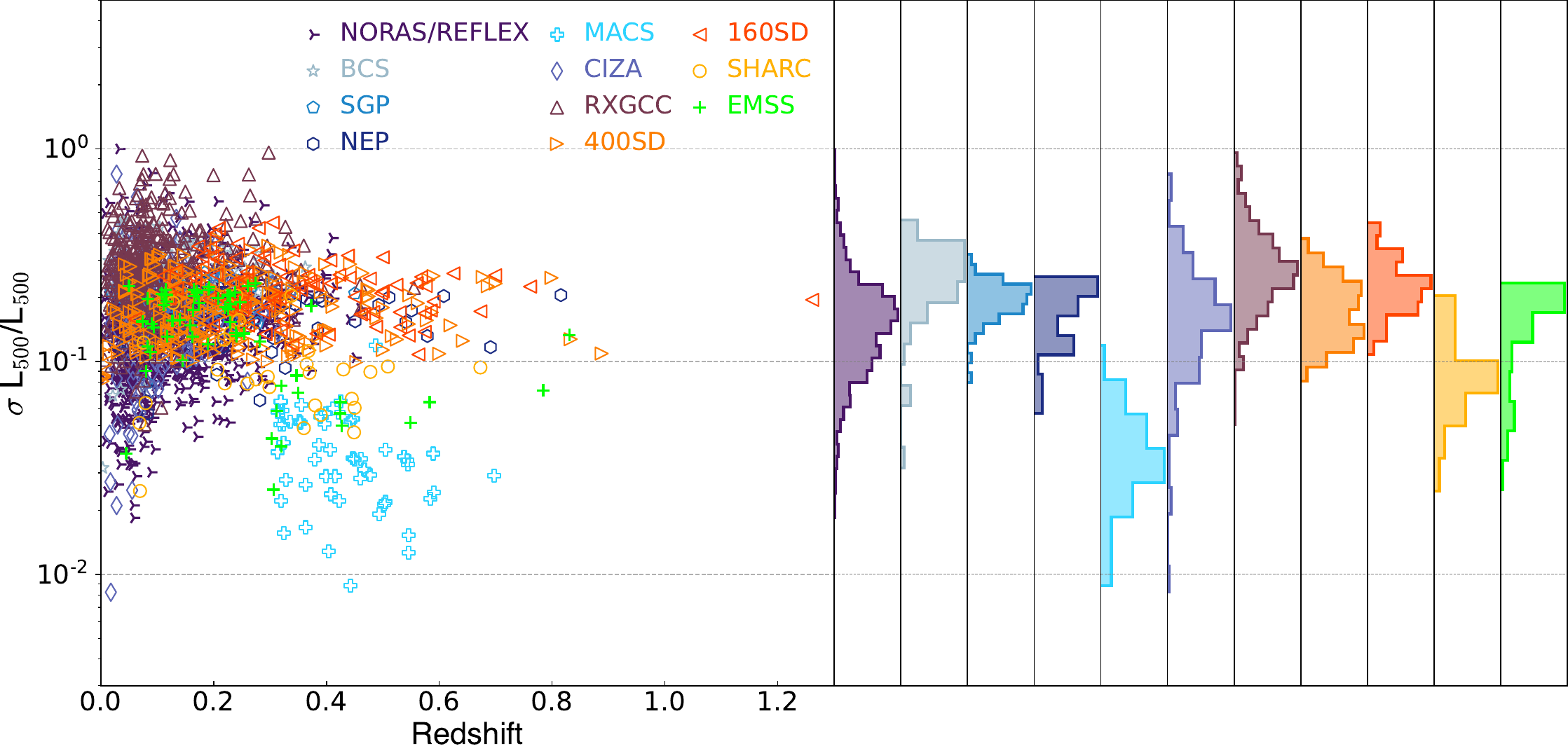}
\caption{Relative statistical $1\ \sigma$ error on the $\Lv$  luminosity. Input catalogues for each cluster are identified by different symbols and colours, as labelled in the Figure. WARPS clusters are not shown since no uncertainties were published in the source catalogue. Left scatter plot of the relative uncertainty as a function of redshift. Right: Corresponding distribution of uncertainties for each input catalogue. The MACS catalogue includes a significant number of systems for which the luminosities come from \chandra\ follow-up observations, hence the smaller statistical uncertainties.
}  
\label{fig:sigmaL_L}
\end{figure*}

\section{The MCXC-II catalogue}  \label{sec:mcxcII}

\subsection{MCXC-II global characteristics}

After completing the catalogue with new entries as discussed in Sect.~\ref{sec:cat}, updating the redshift information as described in Sect.~\ref{sec:zrev}, and re-estimating the X-ray quantities as described in Sect.~\ref{sec:lm},  we constructed MCXC-II, the revised version of the MCXC-I catalogue. The differences with  MCXC-I are discussed in Sect.~\ref{sec:comp_mcxc}. 

MCXC-II comprises a total of 2221 clusters. The catalogue provides  45 parameters for each cluster and the catalogue field names, units, and descriptions are given in Appendix~\ref{ap:mcxcfield}.  The main information includes the cluster position and updated redshift, $\Lv$ and $\Mv$ values, together with statistical errors on these quantities (which were not provided in MCXC-I). 

The fractional contribution of each X-ray survey to MCXC-II is shown graphically in the pie-chart form in Fig.~\ref{fig:piechart}. Of the 2221 objects in the final MCXC-II catalogue, more than $76\%$ were detected in the RASS-based surveys, with REFLEX and NORAS clusters making up $42\%$ of the sample. The remaining clusters are from RXGCC and the serendipitous surveys. 

In Figure \ref{fig:skymap} we show the distribution on the sky of the MCXC-II catalogue in Galactic coordinates.  The distribution of the clusters in the $\Lv$--$z$ plane is shown in Fig.~\ref{fig:L500M500_all}, which is a revised version of Fig.~4 of \citet{2011A&A...534A.109P}.  We also show the corresponding distribution in the $\Mv$--$z$ plane (right panel). Here, and in the following figures, the different input surveys are identified by different colours or symbols. When a cluster appears in several catalogues, the symbol refers to the catalogue that was used to derive the physical parameters for the object in question.

\subsection{Redshift, luminosity and mass distribution}

The distribution of the MCXC-II in the $\Lv$--$z$ plane (Fig.~\ref{fig:L500M500_all}), and consequently its distribution in the $\Mv$--$z$ plane,  illustrate the well-known effect of the X-ray selection. For essentially flux-limited surveys,  the lowest detectable luminosity increases with increasing redshift, following the variation of the luminosity distance. On the other hand, the number of high mass/luminosity clusters decreases due to the evolution of the mass function, an effect amplified by limited sky coverage. 

The MCXC-II cluster distribution reflects the different depth and sky coverage of the RASS and serendipitous survey, respectively. Based on deeper, pointed observations, serendipitous surveys detect lower flux/mass objects at all redshifts and use lower flux-cut selections. On the other hand, the limited sky coverage of the serendipitous surveys does not allow for the detection of the rarest most massive clusters, which are detected in a wide-area survey such as RASS (e.g. REFLEX/NORAS clusters). This effect is also apparent in the MCXC-II sky distribution plot for the different mass bins shown in Fig.~\ref{fig:skymap}.  

The MACS catalogue, based on systematic follow-up of detections in the RASS Bright Source Catalogue (including unresolved), extends an all-sky coverage to a deeper depth, detecting high mass clusters up to higher $z$. This survey is the main contributor to the MCXC-II in the high mass range ($\Mv > 4\times10^{14}\ \msun$) at $z>0.3$ (39 MACS only clusters out of 49 X-ray detected clusters in total). 

 The  MCXC-II redshift histogram is shown in Fig.~\ref{fig:zcats}. Most ($ 92 \%$) of the clusters are $z<0.4$ redshift systems, with a small fraction reaching up to $z=1$. The redshift distribution of the RASS catalogues is dominated by low-$z$ systems, with a median at $z=0.11$, and the most significant fraction in the $z=0.05$ bin.  In contrast, the serendipitous surveys have a flatter redshift distribution, scattered over all $z$, with a slight excess in the middle-$z$ range $[0.1-0.3]$. This redshift difference between RASS and serendipitous surveys is again due to the greater depth of the serendipitous surveys. 

As shown in the right-hand panel of Fig.~\ref{fig:L500M500_all}, the MCXC-II covers a three-decade mass range. The twenty objects with $\Mv \leq 10^{13}\ \msun$ correspond to very nearby massive galaxy-scale systems. A total of 602 objects (27\%) are group-scale systems with $10^{13} < \Mv < 10^{14}\ \msun$, while the remaining 1599 entries are cluster-scale systems with $\Mv \geq 10^{14}\ \msun$. 

\subsection{Uncertainties on $\Lv$}
\begin{table}[]
 \caption{\label{tab:mediandl_L} Median  statistical  error on $\Lv$ of clusters from  the different input catalogue.  }
\centering
\begin{tabular}{lc}
\toprule
\toprule
Catalogue & $\errLv (\%)$  \\
\midrule
          NORAS/REFLEX&   15.2 \\
                CIZA&   16.3 \\
                 BCS&   25.3 \\
                 SGP&   20.1 \\
                 NEP&   17.7 \\
                 MACS&    3.3 \\
                 RXGCC&   26.9 \\
               400SD&   17.5 \\
               160SD&   22.2 \\
               SHARC&    8.9 \\
           EMSS\_1994&   18.5 \\
           EMSS\_2004&    6.5 \\
\bottomrule
\end{tabular}
 \end{table}

Figure~\ref{fig:sigmaL_L}  shows the relative statistical uncertainty on $\Lv$, $\errLv$, plotted as a function of redshift\footnote{WARPS errors are not shown since the flux errors were not given in the source catalogues.}. 
 The histograms of the errors for each are shown on the right panel of Fig.~\ref{fig:sigmaL_L}\footnote{Note that these histograms do not include overlaps, and so are not necessarily representative of the full range of uncertainties for each survey.}. The median uncertainty values for each survey are given in Table~\ref{tab:mediandl_L}.  
Errors are typically about $\pm20\%$.  

The median errors range from  $10\%$ to $25\%$, with no systematic difference between RASS and serendipitous surveys. This is likely due to the fact that survey flux limits of RASS and serendipitous surveys are defined in a similar fashion, usually based on completeness or by setting a minimum signal-to-noise ratio (S/N) for the  detection (the relative flux uncertainty).

The exception is the error for MACS and  EMSS\_2004 clusters ($\errLv \sim 3\%$ and $6.5\%$, respectively), as luminosities are mostly based on more precise \chandra\ or ASCA follow-up data (see Sect.\ref{sec:lm500_macs} and Sect.\ref{sec:lm500_emss}).  

The width of the  $\errLv$  distribution reflects the luminosity range of each survey, as, above the flux limit,   $\errLv$  decreases with increasing $\Lv$. This is the largest for NORAS/REFLEX, which covers the full mass/luminosity range in the local ($z<0.1$) Universe.  

\begin{figure*}[t]
    \centering
	\resizebox{\textwidth}{!}{
    \includegraphics[width=0.925\columnwidth]{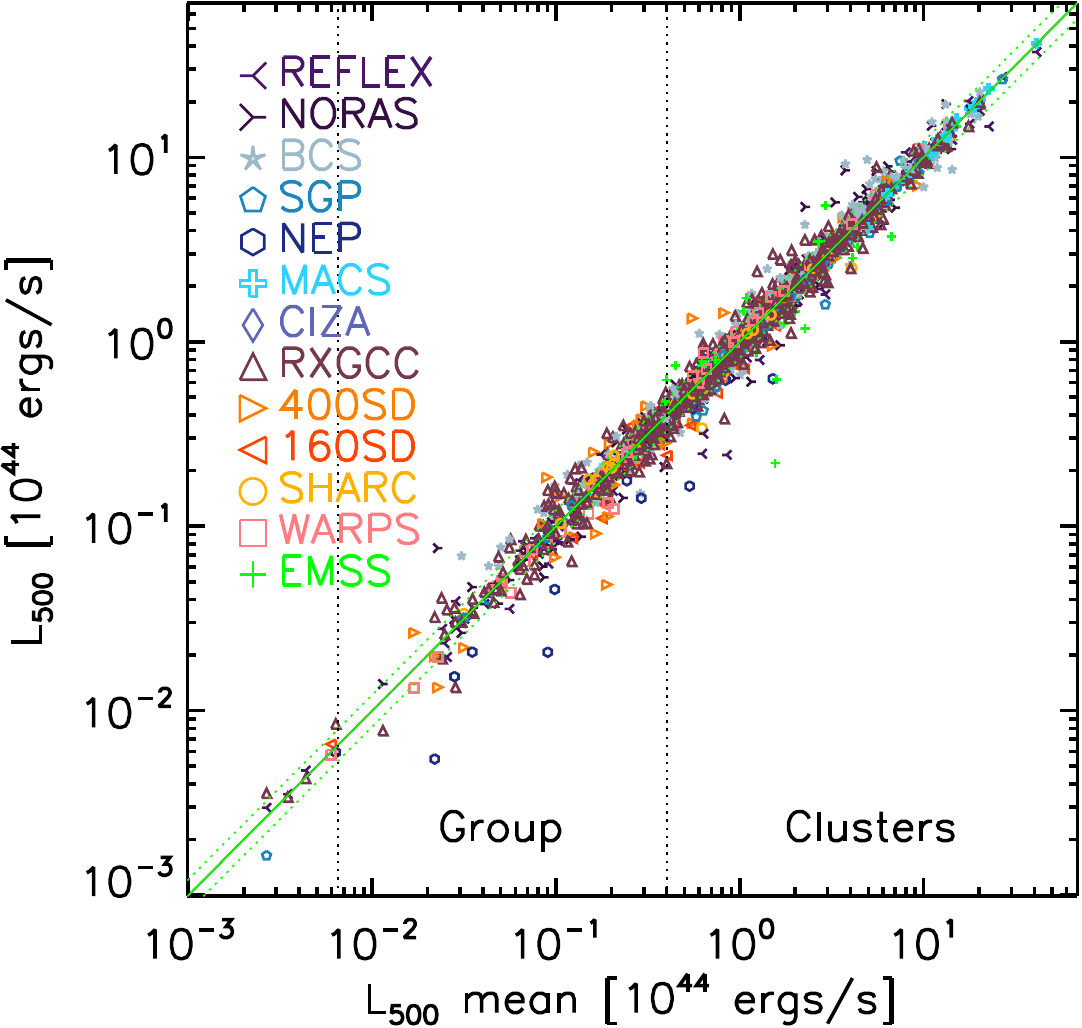}               \hspace{0.5cm}
    \includegraphics[width=0.9\columnwidth]{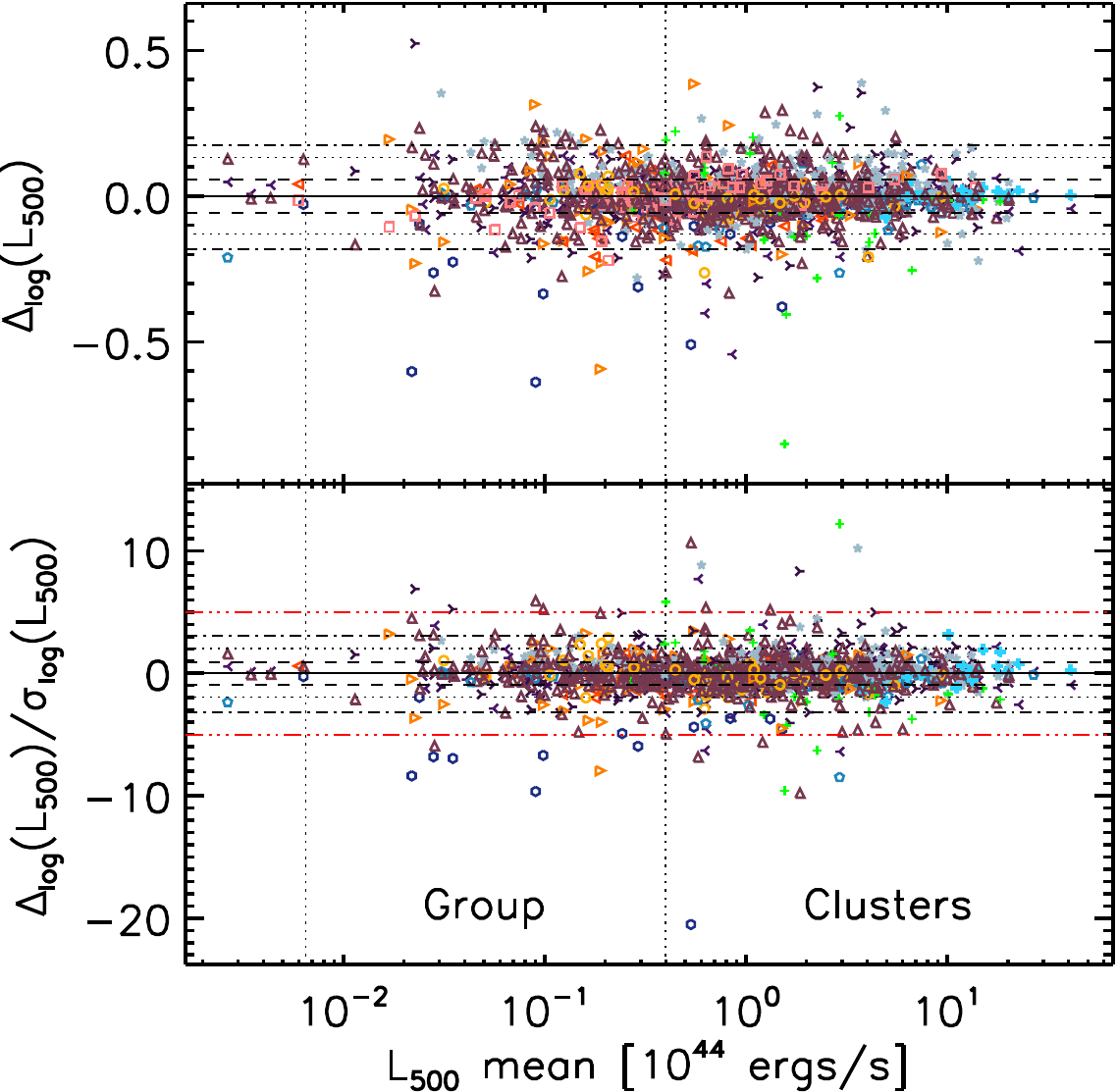}
    }
\caption{Overlap luminosities. Left: Comparison between the luminosity $\Lv$ for each detection of an MCXC-II cluster, computed from the primary or overlapping catalogue parameters,  and their weighted mean value. Overlaps between 400SD and 160SD clusters have been excluded since the data are almost identical. Symbols and colours, as labelled in the Figure,  identify the input catalogue (primary or secondary). Right: Difference in luminosity $\Lv$, in dex (top panel) and relative to the errors (bottom panel).  Full horizontal line: median value; the dashed, dotted and dash-dotted horizontal lines marked the  $68\%$, $90\%$ and $95\%$ dispersion around the median, respectively. } 
    \label{fig:ovlpl500}
\end{figure*}

\subsection{Overlap}\label{sec:overlap} 

A total of 883 objects appear in more than one catalogue. For these clusters, we computed the luminosities  $\Lv$ using the flux from each overlap catalogue (i.e. the overlap luminosity) as input, and using the same redshift and iterative method as described in Sect.~\ref{sec:method}. The resulting luminosities and their ratios are shown in Fig.~\ref{fig:ovlpl500}.

 Not surprisingly, we find the same trends as \citet{2011A&A...534A.109P} who also compared the overlap luminosities to the MCXC-I value (their Figures 7 and 8). There are no systematic differences in $\Lv$, with a median ratio of $1.00\pm0.007$ and a 68\% standard deviation of $[-0.07,+0.09]$ dex ($\sim20\%$), excluding the new  RXGCC data (676 points). The dispersion is slightly smaller than the  $27\%$  derived by \citet{2011A&A...534A.109P}, using input catalogue luminosities simply scaled by the distance-luminosity ratio when catalogue redshifts differed. Using the catalogue flux as input ensures more homogeneous $\Lv$ estimates, through the exact treatment of any redshift dependence and the use of the same $K$-correction models. Although the final effect is small, as discussed Sect.~\ref{sec:lmcompMCXC}, this likely contributes to the decrease in scatter. When including the RXGCC overlap luminosities (1226 ratios), we find a median ratio of $1.00^{+0.005}_{-0.007}$ and a slightly larger standard deviation of $[-0.09,+0.10]$ dex ($\sim24\%$).

We further quantified the differences between catalogues by making use of our estimate of the statistical errors.  For the 883 objects with different luminosity estimates, we computed the weighted mean, $\Lvo$,  and the difference in dex,  of each measurement to the weighted mean $\Delta_{\rm log}(\Lv) =\log(\Lv)-\log(\Lvo)$.  This comparison is therefore fully independent of the exact choice of primary catalogue.  We also considered  $\dLv$  normalised  to the  estimated statistical error, $\dLv/\sigma_{\rm log}(\Lv)$.  The 400SD is an extension of the extraction method developed for 160SD to all ROSAT pointings. As expected,  the corresponding $\Lv$ for clusters in common are in excellent agreement, with a standard deviation of $4\%$. To avoid biasing the difference statistics, we only retain the 400SD value for these clusters, leaving us with 1926 luminosity estimates in total. 

Figure~\ref{fig:ovlpl500} compares all the luminosities to the mean luminosity (left panel), while the differences are plotted on the right panel. The median value of $\dLv$ is zero by construction. The overall  $68\%$ and  $95\%$  dispersion of $\dLv$ are $0.05$ and $0.18$ dex, respectively. For purely statistical errors on independent measurements, $\dLvrel$ should be well-represented by a Gaussian distribution with unit variance. Correlations between luminosities will decrease this scatter, and the presence of systematics will add to it. We do expect error covariance on luminosities derived from the same observations, as in the majority of cases (e.g. RASS data). The covariance depends in particular on exact data selection (e.g. choice of centre, point source excision, etc.) and it is impossible to estimate from the available published information in catalogues. $\dLvrel$ therefore only provides an upper limit on systematics in units of statistical errors. The $68\%$ and  $95\%$  dispersion of $\dLvrel$ are  $[-0.8,+0.9]$  and  $[-3.1,+2.9]$, respectively. While the former is consistent with that of a Gaussian distribution, the latter is larger and may indicate a contribution from systematic errors in the tail of the distribution. 

 We did not find any trend of $\dLv$ or $\dLvrel$ with luminosity or redshift,  or any marked difference between survey types. Only the  EMSS values are more discrepant on average: there is a small and marginally significant underestimate of luminosities (median  $\dLv/\sigma_{\rm log}(\Lv)=-0.40\pm0.3)$, with a $68\%$ dispersion ($[-1.8,+2.2]$) that is twice as large. 
 
 There is some dependence on the aperture when this information is available. The ratio $\dLvrel$ increases when the aperture differs significantly from $\Rv$, for  $\Rap/\Rv\lesssim0.5$ (mostly for NEP clusters) or $\Rap/\Rv\gtrsim2$ (mostly RXGCC clusters).  For small apertures, the aperture correction is large and is sensitive to core properties, resulting in an overestimation of $\Lv$ for highly peaked cool-core clusters. For large aperture values, the aperture correction is small but increased point source contamination may be an issue. 

\begin{table*}[t]
\caption{ Synthetic comparison table listing differences between MCXC-II and MCXC-I. }
\centering
\begin{tabular}{lllll}
\toprule
\toprule
 &  MCXC-I & MCXC-II & Note  & Section\\
\midrule
Cluster content  &  1743 clusters & 2221 clusters & Added: &\\
		&	  & 	&~~~22 REFLEX-II & Sect.~\ref{sec:reflexII}\\    
		&        &	&~~~24 MACS & Sect.~\ref{sec:newmacs}\\
		&        &	&~~~393 RXGCC & Sect.~\ref{sec:rxgcc}\\
		& 	 & 	&~~~47 new redshifts or  $\LX$ & Sect.~\ref{sec:newmcxc} \\
		&	 &	& Removed: 7 duplicates and one false & \\
Revised redshift  &    -    &   39  clusters  & & Sect.~\ref{sec:zrev}\\
\hdashline[0.1pt/2pt]
\\
$\Lv$ $\Mv$ derivation & $\LX$ input  &   $ \FX$ input & Flexibilty & Sect.~\ref{sec:method}\\
				&		      & Fully iterative method & & Sect.~\ref{sec:method}\\
				& $ E(z)^{-7/3}$\LM\    &  $E(z)^{-2}$\LM\    &  Small effect & Sect.~\ref{sec:lmr} \\
\hdashline[0.1pt/2pt]
\\
New information &   & $z$ type (spec, phot ...) & & Sect.~\ref{sec:ztyperef}\\
			 &    & $z$ reference &  & Sect.~\ref{sec:ztyperef}\\
			 &     &Standardised  $\FX$ and  errors & \\
			 &     & Errors on $\Lv$ and $\Mv$  & \\
\bottomrule
\end{tabular}
\label{tab:mcxcIIvsmcxc}
\end{table*}
 In this connection, we also compared the luminosity measured from RASS and from deeper serendipitous (SER) catalogues, when both were available. Owing to higher resolution and generally deeper pointed observations, the SER luminosity estimates are expected to be less sensitive to point source contamination. For the 51 available pairs of values, we estimated the ratio of SER to RASS luminosities in log space, $\Delta_{\rm log}(\Lv)=\log(L_{\rm 500,SER}) - \log(L_{\rm 500,RASS})$. There is a slight offset  of the luminosity ratio of $ -0.10^{+0.03}_{-0.02}$ dex, implying that SER values are 20\% smaller on average, with a large 68\% dispersion of  $[-0.24,+0.15]$ dex. The error on  $\Delta_{\rm log}(\Lv)$, $\sigma(\Delta_{\rm log}(\Lv))$ was obtained from the quadratic sum of the RASS and SER luminosities, these measures being independent. Normalised by the statistical error,  the median ratio offset is not very significant:  $\Delta_{\rm log}(\Lv)/\sigma(\Delta_{\rm log}(\Lv)) = -0.87^{+0.5}_{-0.2}$.
  
As already noted by  \citet{2011A&A...534A.109P} for MCXC-I,  there are several notable outliers. There are 30 cases of $>5\sigma$ deviation (see Fig.~\ref{fig:ovlpl500}, bottom right).  This includes four EMSS measurements, seven pairs of RASS estimates between RXGCC and REFLEX/NORAS or NEP values.

\begin{figure*}[t]
    \centering
	\resizebox{0.95\textwidth}{!} {
    \includegraphics[width=\columnwidth]{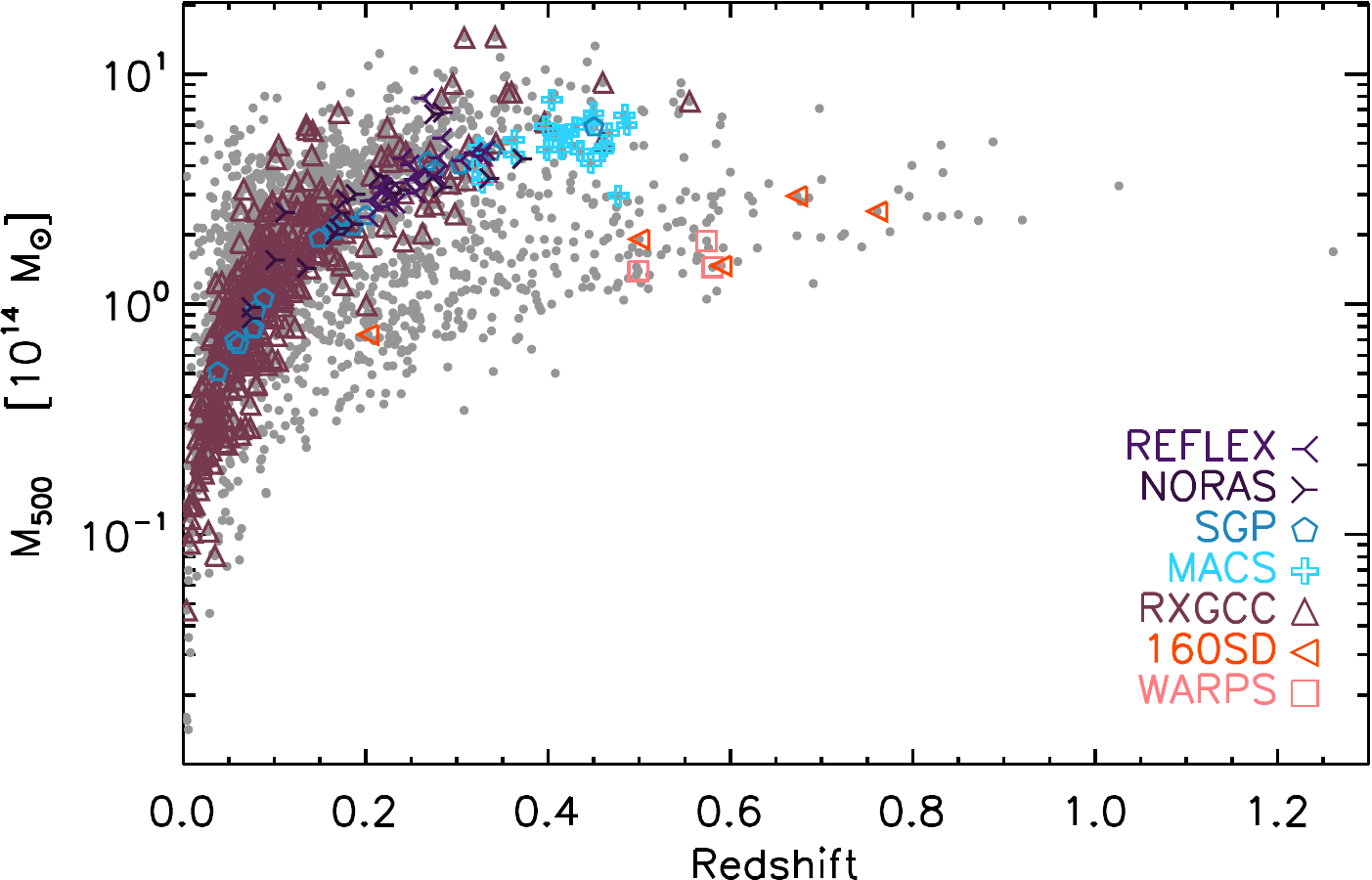}
    \hspace{0.5cm}
     \includegraphics[width=0.98\columnwidth]{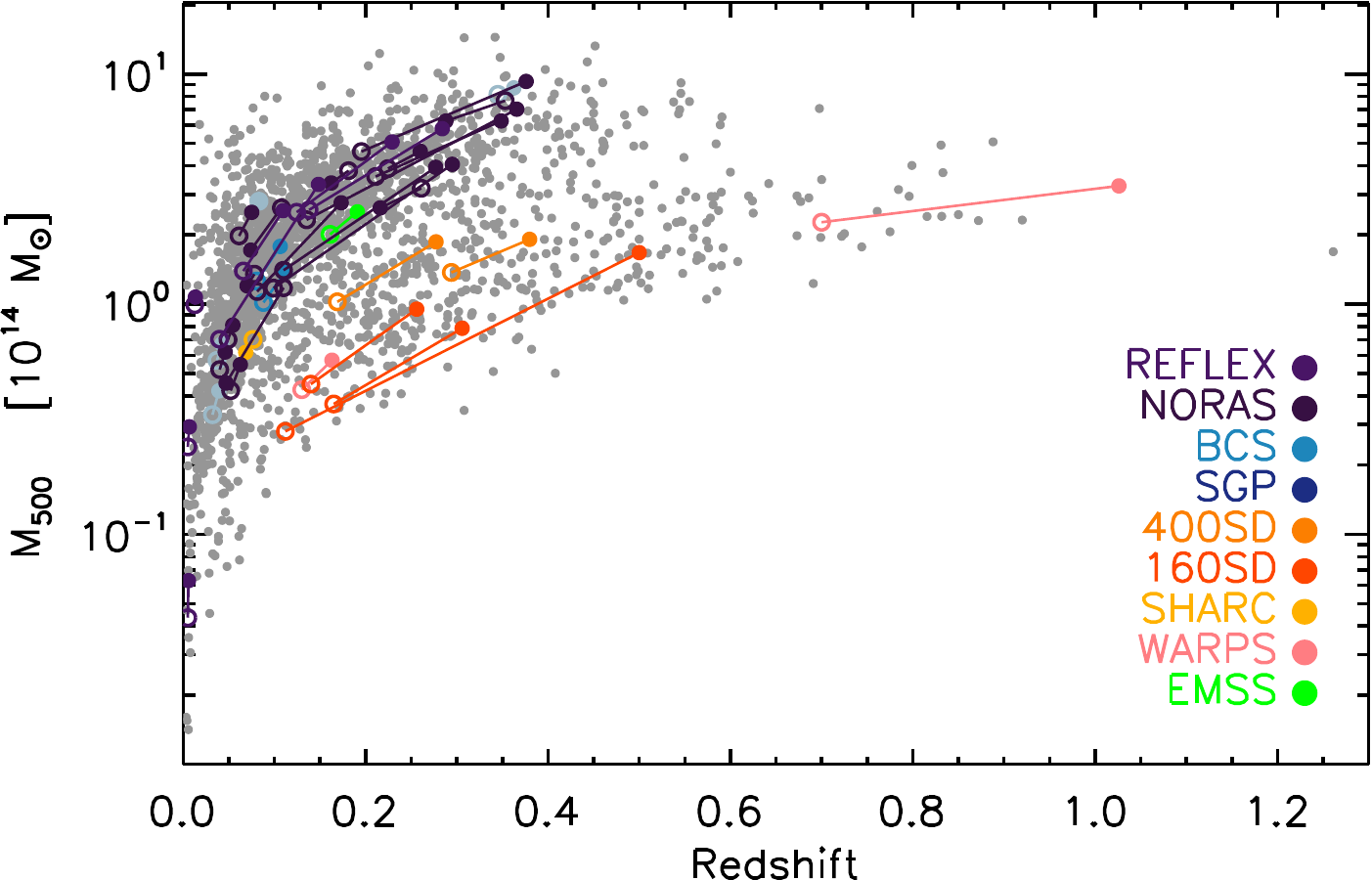}
 }
      \caption{Clusters in MCXC-II shown in the  $z$--$\Mv$ plane. Left panel: New objects marked in colour, which include new REFLEX-II clusters, new MACS clusters above $z=0.3$, and new RXGCC clusters, as well as NORAS, 400SD, and SGP clusters that were previously without redshift. Right panel:  Objects with revised redshifts (see Table.~\ref{tab:zrev}) are marked in colour, as labelled in the Figure. The revised position (filled circle) is connected to the MCXC-I position (open circle).}    
    \label{fig:zM_newcatzrev}
\end{figure*}

\section{Discussion}\label{sec:discussion}

\subsection{Comparison with MCXC-I \citep{2011A&A...534A.109P}} \label{sec:comp_mcxc}

A summary overview of the differences between MCXC-I and MCXC-II content and methods is given in  Table~\ref{tab:mcxcIIvsmcxc}. 

\subsubsection{Cluster content and redshift distribution}
%
The MCXC-II cluster content and new redshift information is detailed in Sect.~\ref{sec:cat} and summarised in Fig.~\ref{fig:sum_diagram}. MCXC-II includes 486 new clusters compared to MCXC-I: 22 from REFLEX-II, 24 from new MACS releases, 393 from RXGCC, and 47 from MCXC-I input catalogues (45 previously without redshift and two previously without luminosity). 
Figure~\ref{fig:zM_newcatzrev} shows the distribution of the new clusters in the $z$--$\Mv$ plane. Clusters with newly-available redshifts lie primarily at or near the flux limits of the corresponding catalogues.  Inclusion of the RXGCC catalogue increases the number of clusters at low mass and low redshift: the new RXGCC clusters are at a median redshift of $z=0.080$, with a median mass of $1.1\times10^{14}\ \msun$ (see also Fig.~\ref{fig:zcats}). On the other hand, the new MACS and REFLEX clusters significantly improve the coverage of the sparsely-represented high $z$, high mass region. 

One also notes the eight very massive new  RXGCC clusters with $\Mv>7.5\times10^{14}\msun$ at $z>0.25$.
One of these (\object{MCXC\,J0600.1-2007}, \object{RXGCC\,229}) coincides with an already known cluster detected by \planck\ (\object{PSZ2\,G225.93-19.99}), SPT (\object{SPT-CLJ0600-2007}), and ComPRASS (\object{PSZRX\,G225.95-20.01}), and with a source from the ROSAT Bright Source Catalogue which is not included in the MACS catalogue. The remaining seven do not coincide with any sources from the Bright Source Catalogue, so they cannot be MACS clusters. However, they all have a very large aperture radius in RXGCC ($R_{sig}/\Rv > 3$). It is therefore possible that the RXGCC flux (and mass) may have been overestimated due to the inclusion of additional sources in the aperture under consideration. For example, \object{MCXC\,J1320.0-3556} (\object{RXGCC\,496}, $R_{sig}/\Rv = 4.8$) may have some contamination from the nearby cluster \object{Abell\,S\,729} ($z= 0.0499$), as mentioned on the RXGCC catalogue website. Also, three of these seven objects coincide with known clusters from other catalogues, in which their estimated mass is lower than that derived by RXGCC. This is the case for \object{MCXC\,J0957.4+6048} (\object{RXGCC\,333}, $R_{sig}/\Rv$ = 7.5), \object{MCXC\,J1249.6+4952} (\object{RXGCC\,466}, $R_{sig}/\Rv$ = 5.9), and \object{MCXC\,J1508.0+4033} (\object{RXGCC\,590}, $R_{sig}/\Rv$=3.1), for which the mass derived from the RXGCC detection is higher than that derived from the ComPRASS or \planck\ detections: 14.45 vs 3.51 (ComPRASS); 7.83 vs 4.38 (\planck) and 3.44 (ComPRASS); and 7.59 vs 5.34 (ComPRASS), respectively, in $\ 10^{14}\msun$ units.

We have systematically revised the cluster redshifts, which in MCXC-I were simply taken from the input catalogues. Using the NED and Simbad databases, in addition to new optical catalogues from large galaxy surveys, MCXC-II includes 39 revised redshifts, located across the full $z$--$\Mv$ plane, as shown in Fig.~\ref{fig:zM_newcatzrev}. An interesting case is \object{MCXC\,J1415.1+3612}, previously estimated at $z\sim 0.7$ in the WARPS catalogue, which now has a spectroscopic redshift of $z=1.026$. 

The MCXC-II redshift revisions can also be compared to those adopted in the recent study by \citet{2020A&A...636A..15M}, which focussed on the subsample of 313 clusters in common between MCXC-I and their sample. 
Their critical review of the MCXC-I values was undertaken using spectroscopic redshifts from various galaxy surveys available in NED, or redshifts from X-ray spectroscopic analysis (denoted $z_{\rm X}$ in the following).
They found generally good agreement between safe optical and independent X--ray redshifts (see Fig.~2 in \citealt{2020A&A...636A..15M}). They decided to adopt $z_{\rm X}$ for 41 clusters (with $0.03\,<z\,<0.3$) where there was insufficient optical spectroscopic data. The MCXC-II values for these clusters are in excellent agreement with $z_{\rm X}$. Taking into account statistical errors on $z_{\rm X}$ kindly provided by K. Migkas (private communication) and potential intrinsic scatter, a BCES orthogonal power-law fit gives a slope of $1.01\pm0.04$ and a normalization of  $1.005\pm 0.014$ at a pivot of $z=0.1$. The intrinsic scatter is $~5\%$. Two of the clusters with revised MCXC-II values, \object{MCXC\,J1359.2+2758} (\object{Abell\,1831}) and \object{MCXC\,J1520.9+4840} (\object{Abell\,2064}), with new redshifts  $z=0.0751$ and $z=0.0740$, respectively, are now in good agreement with the X-ray values ($z_{\rm X}$= $0.078\pm0.010$ and $0.065\pm0.006$, respectively); the MCXC-I values, $z_{\rm MCXC}=0.061$  and $z_{\rm MCXC}=0.108$ were significantly different. There are eight other clusters of the \citet{2020A&A...636A..15M}  sample with updated MCXC-II values (two revised values and six consolidated), but the changes are small and remain in good agreement with the values chosen by \citet{2020A&A...636A..15M}.

\begin{figure*}[t]
    \centering
	\resizebox{\textwidth}{!} {
    \includegraphics[width=\columnwidth]{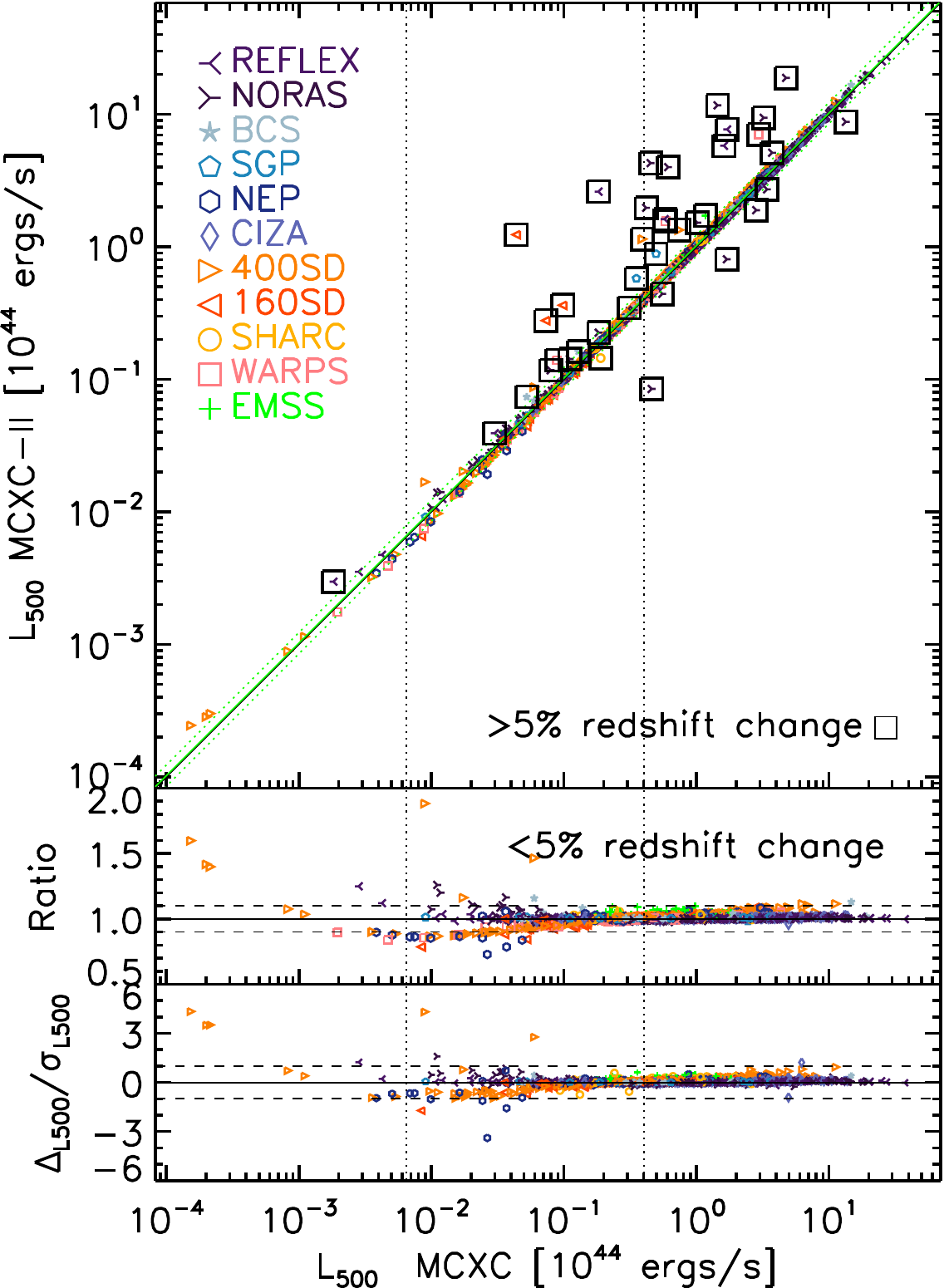}
    \hspace{0.5cm}
    \includegraphics[width=\columnwidth]{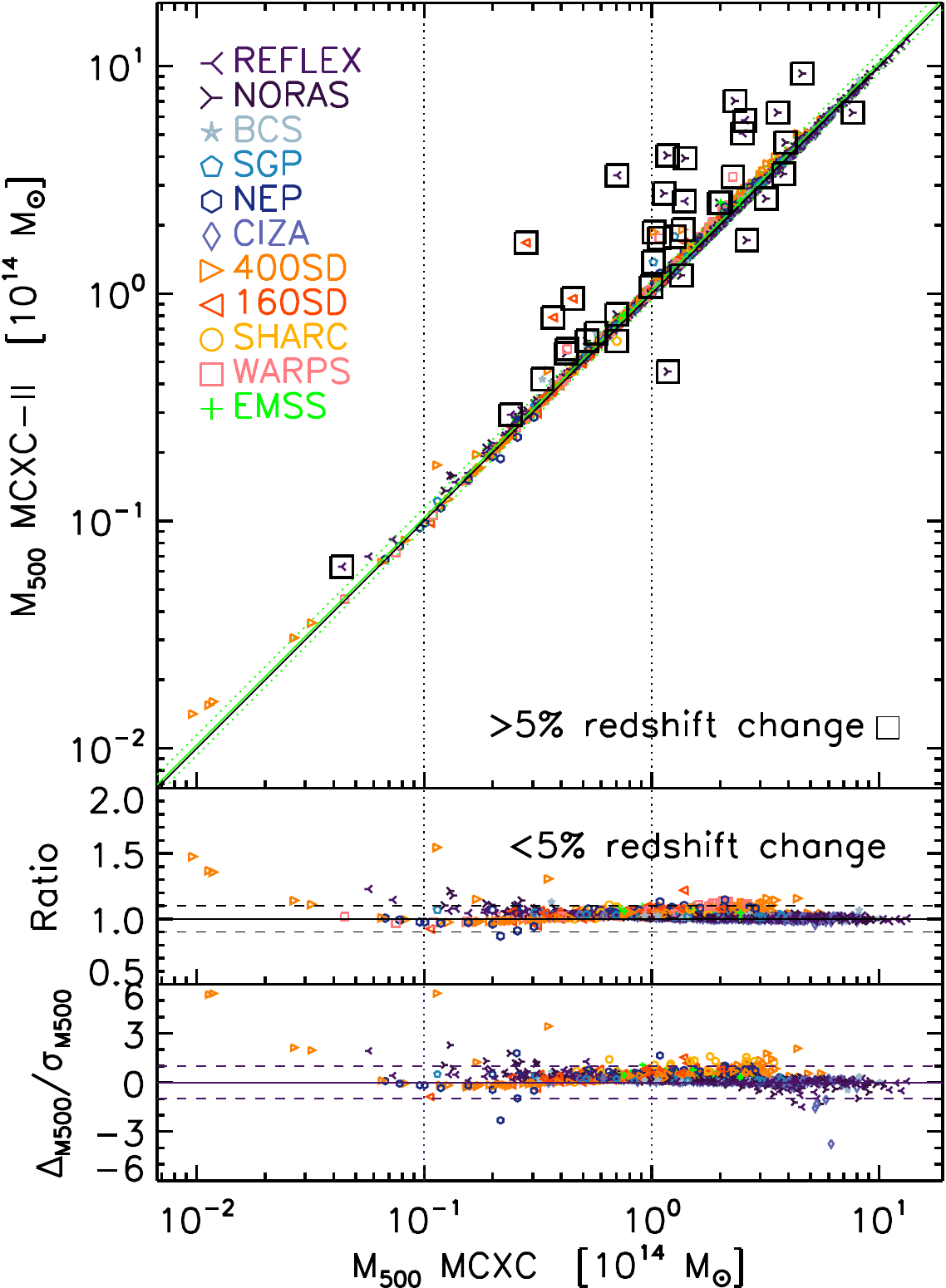}}
    \caption{Left: Comparison between the MCXC $\Lv$ luminosities and the updated luminosities. Source catalogues are identified by different symbols and colours as labelled in the Figure. MACS and EMSS\_1994 clusters are not shown (see Sects.~\ref{sec:lm500_macs} and~\ref{sec:lm500_emss} for a description of specific treatment and comparison with MCXC-I). Clusters with redshift changed by more than $5\%$ are marked by a box. Middle panel: Ratio of luminosities for clusters with redshift changed by less than $5\%$.
    Bottom panel:  Same for the relative difference in terms of statistical error. The WARPS clusters without errors are not shown. Left: Same for the mass $\Mv$. }
    \label{fig:l500m500comp}
\end{figure*}

Most of the redshift revisions have a large impact on the $\Lv$ and $\Mv$ estimates (see Fig.~\ref{fig:l500m500comp}). However, the redshift revision affects only a small fraction, $\sim 2\%$, of the  MCXC-I clusters, illustrating the robustness of the original catalogue values. There is a caveat though:  the revision is largely supported by mostly Northern SDSS data, and the revision might be incomplete in the South (in particular for REFLEX clusters). 

MCXC-II further provides information on the nature of the redshift (photometric, spectroscopic, estimated...) through {\tt Z\_TYPE}  and the bibliographic reference for the redshift through {\tt Z\_REF}.

\subsubsection{$\Lv$ and $\Mv$ derivation }

A major difference between MCXC-II and MCXC-I is the procedure to derive homogenized $\Lv$ and $\Mv$ values. Rather than starting from published X-ray luminosities, we used the catalogue flux information and a fully consistent iterative procedure, with the use of the same \LT\ and \LM\ relations at all steps\footnote{The published X-ray luminosities in each catalogue rely on $K$-corrections based on different \LT\ relations, sometimes established in different energy bands than that of the extracted flux, and often with unclear $\KL$ conversion.}. This approach allows us to include new redshifts and to propagate the statistical errors on directly measured quantities, therefore providing errors on $\Lv$ and $\Mv$. More importantly, the major advantage of this new approach is its flexibility. Redshift updates or change of underlying models and scaling laws can be easily undertaken, in a fully consistent way. The method, including the corresponding equations, is detailed in  Sect.~\ref{sec:method}. In the MCXC-II catalogue, we provide the standardised input flux and its uncertainty, together with related information necessary for the $\Lv$ estimates (e.g. energy band, aperture). 

For instance, the user interested in updating the \LM\ relation or the cluster density profile (or to study their impact) can start from the flux information given in MCXC-II and derive new $\Lv$ and $\Mv$, using Eq.~\ref{eq:lapl500} (combined with Eq.~\ref{eq:KK}, and Eq.~\ref{eq:KL} if the flux energy band is different from $\bando$), Eq.~\ref{eq:lt}, and Eq.~\ref{eq:lm}. This requires a double iteration, as described at the end of the Section. The $K$ and $\KL$-correction can be computed with standard software such as {\tt XSPEC} and the aperture correction from the equations given in \ref{ap:apcor}. For convenience, tabulated values of $K$ and $\KL$, for the energy bands used in the MCXC-II, and the aperture correction for the \rexcess\ density profile, are given in Appendix \ref{ap:kcor}. 
 
\subsubsection{$\Lv$ and $\Mv$  values}
\label{sec:lmcompMCXC}

In Figure~\ref{fig:l500m500comp}, we compare the MCXC-II $\Lv$ and $\Mv$ estimates to the MCXC-I values (excluding the MACS and EMSS\_1994 clusters for the reasons discussed above).  

The main differences arise from redshift changes, as clearly visible at the top of both panels. The residual differences are shown in the middle and bottom panels, where we have excluded clusters with redshift changed by more than  $5\%$. These differences in $\Lv$ and $\Mv$ for these systems reflect the cumulative effects of the model changes (Sect.~\ref{sec:method}).  

The differences in $\Lv$ arise  essentially from the estimate of the aperture luminosity:
\begin{itemize}[noitemsep,topsep=0pt]
\item The increasing scatter at low luminosity ($\Lv\!\lesssim\!10^{43}\ \ergs$) reflects the scatter in the ratio between the aperture luminosity computed from the flux (used in MCXC-II) to the luminosity value given in the input catalogues (used in MCXC-I). This is essentially due to truncation errors in the published values at low luminosities (Sect.~\ref{sec:comp_laplapcat} and Fig.~\ref{fig:comp_laplapcat}). In addition, the effect of the more accurate $\KL$-correction at low luminosities is also apparent for input catalogues with flux measured in the $\bandi$ energy band (compare for example the ratio shape for 400SD clusters to that of Fig.~\ref{fig:KL}). 
\item At higher luminosities and in the cluster regime, the scatter in the MCXC-II to MCXC-I $\Lv$ ratio again reflects the ratio of computed to published $\Lap$ luminosities, but with the systematic differences now being due to the details of the $K$-correction. It is most prominent for the 400SD sub-catalogue (see Sect.\ref{sec:comp_laplapcat}).
\item The other model change, the \LM\ relation, has a negligible effect on $\Lv$ as shown in Sect.~\ref{sec:lmr}. 
\end{itemize}
 Overall the effects of these changes are small, making up less than $10\%$ in the cluster regime and across the majority of the group regime. They are nearly always smaller than the statistical errors (bottom panels), with only seven clusters showing a difference larger than $2\sigma$, including the two strong 400SD outliers mentioned in Sect.~\ref{sec:comp_laplapcat}.

The mass $\Mv$ differences first reflect the differences in $\Lv$. The new \LM\  only slightly affects the differences, as its general impact is small (Fig.~\ref{fig:comp_lmr}). At low mass (low redshift) the small increase in mass due to the new \LM\ relation transfers the effect of the updated $\KL$-correction onto $\Lv$. At high mass,  high redshift, the mass increase amplifies the luminosity increase, when present. 

\subsection{Systematic uncertainties}
\label{sec:systematic}
\
The generic approach to homogenisation described above in Sect.~\ref{sec:method} has notably allowed us to include uncertainties on the $\Lv$ estimates. These were derived from the fractional count rate or flux errors given in the original catalogues. In most cases, these represent purely statistical (Poisson) uncertainties. Exceptions are REFLEX/NORAS, where count rate errors include some small additional systematic error linked to the uncertainty in the determination of the plateau in the growth curve analysis, and BCS, which includes an additional 5\% systematic uncertainty on the count rate error. 

However, additional systematic uncertainties may still be present, due to certain modeling assumptions. These include:
\begin{itemize}
    \item Estimation of the plasma temperature, which is needed for the $K$ and $K_L$ -corrections (Sect.~\ref{sec:method}). This is defined by the choice of $\Lv - T$ relation, for which different studies have found widely varying slopes. This is particularly important at low $\Lv$ or $T$, where the slope is most uncertain \citep[see e.g. the review by][and references therein]{lov21}. 
    \item The plasma metallicity, which is an issue below about 2~keV, where line emission starts to become significant with respect to the bremsstrahlung continuum. We have assumed a standard metallicity of $0.3\  Z_\odot$ for our analysis, which is typical in the region outside the core for both clusters and groups \citep[see e.g. the recent review by][and references therein]{gas21}. The low-temperature group regime is also where the plasma emission codes are known to evolve significantly, owing to the increasing refinement of our knowledge of the atomic lines over time \citep[e.g.][]{lov21}.
    \item The aperture correction, which depends on the choice of $\Lv-M$ relation used to relate the luminosity to the mass, and thus define the $\Rv$ aperture, and the density profile model assumed for the emission. As argued in \citet{2011A&A...534A.109P}, the choice of $\Lv-M$ relation, and the effect of intrinsic scatter about that relation, has a negligible impact on the final luminosity due to the steep drop of the density profile model with radius. However, the choice of luminosity profile model does have an impact, which is discussed at length in Appendix~A of \citet{2011A&A...534A.109P}. Defined on a representative X-ray selected sample, the \rexcess\ profile is clearly more representative of the X-ray cluster population than the commonly adopted $\beta$ model. Additional systematic uncertainties associated with the adopted profile model will become particularly important when the aperture correction is large, or when the object is at low luminosity.
    \item A final issue is related to possible sources of contamination. This will depend on the ratio of the aperture radius to $\Rv$, and will be particularly problematic when the aperture is too large compared to $\Rv$ or when cluster angular size is large. The latter could be an issue for very nearby systems.   
\end{itemize}

The comparison between luminosities derived in different sub-catalogues in Sect.~\ref{sec:overlap} and the discussion of trends with input parameters (e.g. fluxes, energy bands, apertures) has allowed us to estimate the systematics on luminosities to some extent. The $\Lv$ measurement is found to be very robust on average, with a median of $1.00^{+0.005}_{-0.007}$, with a standard deviation of [$-0.09, +0.10$]. However, while our computation of $\Lv$ does not depend on the adopted $\Lv-M$ relation (see discussion above), the $\Mv$ obviously does. We have followed \citet{2011A&A...534A.109P}, using the \rexcess\ \LM\ relation, and assuming that the Malmquist bias of the individual sub-samples used to construct the MCXC is the same as that of \rexcess\ on average. If needed, the computation of different total masses using an alternative $\Lv-M$ relation is straightforward.

%
\section{Conclusions}\label{sec:conclusion}

We have presented the construction and properties of the MCXC-II, the second release of the Meta-Catalogue of X-ray detected Clusters of galaxies. MCXC-II was constructed from publicly available RASS-based (NORAS, REFLEX, BCS, SGP, NEP, MACS, CIZA, and RXGCC) and serendipitous (160SD, 400SD, SHARC, WARPS, and EMSS) cluster catalogues (see Sect.~\ref{sec:cat}). Clusters with newly available redshifts from the NORAS, SGP, WARPS, and 160SD surveys were included (47 systems), in addition to new objects from the REFLEX II, MACS, and RXGCC catalogues (439 systems). MCXC-II additionally contains a complete sub-catalogue of all published MACS objects (Sect.~\ref{sec:newmacs}). Duplicates and overlaps were exhaustively identified, leading to a final total entry count of 2221 objects. As in the first release, the information from these input catalogues was standardised and homogenised using a self-consistent approach that takes advantage of our knowledge of the structural and scaling properties of the cluster population (Sect.~\ref{sec:lm}). 

In this second release, we have significantly changed the homogenisation procedure. While in MCXC-I the published luminosities were homogenised, for MCXC-II, we have used when possible the original flux measurements to calculate a new luminosity for each object in the catalogue. This approach allows for more flexibility when adding new sub-catalogues or when redshift changes are introduced, and allows for the same input models to be used for the luminosity calculation. A comparison of our newly derived luminosities to those in the original catalogues shows an excellent agreement (Fig.~\ref{fig:comp_laplapcat}), with any variations essentially attributable to differences in the choice of input models. We would encourage future publications to systematically include flux measurements, errors, and associated radii in their catalogue tables, as this is essential for the recalculation of luminosities. 

Two significant extensions compared to MCXC-I are the inclusion of additional redshift information (and redshift updates when deemed necessary), and uncertainties on the luminosity values. The additional redshift information includes: {\tt Z\_TYPE} (spectroscopic, photometric, etc), {\tt Z\_REF} (the reference for the redshift value), and {\tt Z\_FLAG} (indicating whether the redshift has changed from the value published in the original sub-catalogue). The luminosity uncertainties (Fig.~\ref{fig:sigmaL_L}) were computed from the flux errors in the input catalogues. A summary of the differences between MCXC-I and MCXC-II can be found in Table~\ref{tab:mcxcIIvsmcxc}. For each object, MCXC-II now contains 46 catalogue fields, which are listed in Appendix~\ref{ap:mcxcfield}. MCXC-II will be accessible electronically at the M2C database\footnote{\url{https://www.galaxyclusterdb.eu}} and at the CDS.

MCXC-I has already been widely used by the community. MCXC-II now comprises essentially all ROSAT-based cluster catalogues where the cluster detection is based on measurements of X-ray source flux and extent. As an all-sky metacatalogue, it should therefore be useful for a better understanding of systematics and survey selection when compared to other X-ray (e.g. eROSITA, for which the first catalogue was recently released in \citealt{2024arXiv240208452B}), SZ, and optical catalogues. In this connection, we are currently working on a companion meta-catalogue of SZ-detected clusters, the MCSZ, which will shortly be made available.

\begin{acknowledgements}
We acknowledge funding from the European Research Council under the European Union’s Seventh Framework Programme (FP72007-2013) ERC grant agreement no. 340519, and from the French space agency, CNES.
 We thank  E. Silva and D. Chapon for their contributions to the M2C database, and H. Ebeling for useful discussions. We thank K. Migkas for providing unpublished errors on the X-ray redshift errors, and for additional useful discussions.
This research has made use of the SIMBAD and VizieR databases, operated at the Centre de Données astronomiques de Strasbourg (CDS\footnote{\url{https://cds.unistra.fr}}), Strasbourg,  France, and the NASA/IPAC Extragalactic Database (NED\footnote{\url{https://ned.ipac.caltech.edu/}}), operated by the Jet Propulsion Laboratory, California Institute of Technology. We made extensive use of data and software provided by the High Energy Astrophysics Science Archive Research Center (HEASARC\footnote{\url{https://heasarc.gsfc.nasa.gov/}}) which is a service of the Astrophysics Science Division at  NASA/GSFC, data from the \xmm\ Science Archive\footnote{\url{https://www.cosmos.esa.int/web/xmm-newton/xsa}}, the \chandra\ Data Archive\footnote{\url{https://cxc.cfa.harvard.edu/cda/}} provided by the \chandra\ X-ray Center (CXC), the SWIFT Data Archive\footnote{\url{https://swift.gsfc.nasa.gov/sdc/}}, and the ROSAT all-sky  Data Archive\footnote{\url{https://heasarc.gsfc.nasa.gov/docs/rosat/rass.html}} at HEASARC. We acknowledge the use of the 'Aladin sky atlas' developed at CDS \citep{2000A&AS..143...33B}, Strasbourg Observatory, France and of the NASA's SkyView facility\footnote{\url{http://skyview.gsfc.nasa.gov}}, and the TOPCAT\footnote{\url{http://www.starlink.ac.uk/topcat/}} software package \citep{taylor2005astronomical,2022arXiv221116913T}.
We acknowledge the use of the RXGCC catalogue website\footnote{\url{https://github.com/wwxu/rxgcc.github.io}} and the X-CLASS website \footnote{\url{https://xmm-xclass.in2p3.fr/}}.
This research has made use of data from the Sloan Digital Sky Survey (SDSS\footnote{\url{http://www.skyserver.sdss.org/dr18/}}), which is funded by the Alfred P. Sloan Foundation, the Heising-Simons Foundation, the National Science Foundation, and the Participating Institutions. SDSS is managed by the Astrophysical Research Consortium for the Participating Institutions of the SDSS Collaboration.
It has also made use of data from 2dF\footnote{\url{http://www.2dfgrs.net/}}, and 6dF\footnote{\url{http://www.6dfgrs.net/}} surveys, 
The results reported in this article are partly based on data obtained from the \xmm\ observatory, an ESA science mission with instruments and contributions directly funded by ESA Member States and NASA.
\end{acknowledgements}

\bibliographystyle{aa}
\bibliography{bibl.bib}

\begin{appendix} 
\section{Note on individual clusters} 
Objects with names starting with WHL, WHY and  RM are from the optical cluster catalogues of \citet{2015ApJ...807..178W,2018MNRAS.475..343W,2016ApJS..224....1R}  respectively and their quoted redshifts are the spectroscopic values from these sources. The richness of WHL and WHY objects, $\RL$, are taken from the respective catalogues. For WHL objects, the richness is defined from the total $r$-band luminosity within $\Rv$ \citep[][eq.~16]{2015ApJ...807..178W}. The richness of WHY objects are estimated consistently within $\Rv$, using the total $r$-band luminosity within 1~Mpc, with a calibration from WH data \citep[][Eq.~6]{2018MNRAS.475..343W}. 

Clusters with names starting with X-CLASS and RXGCC are from the X-ray cluster catalogues of \citet{kou21} and \citet{2022A&A...658A..59X}. Their latest redshift values were retrieved from the corresponding databases \footnote{ Used databases: \url{https://xmm-xclass.in2p3.fr} \citep{kou21} and \url{https://github.com/wwxu/rxgcc.github.io/tree/master/} \citep{2022A&A...658A..59X}}. For better readability,  we do not repeat the reference(s) in each section. 

\subsection{New clusters from MCXC source catalogues}\label{app:new_znorasssgp}

Below are individual notes on NORAS or SGP clusters included in MCXC-II, for which the distance to the counterpart used for redshift estimates is large. In all cases, the association is supporting by converging evidence and/or a physical explanation of the offset.

\subsubsection{NORAS clusters}\label{app:noras}
\begin{figure}[t]
    \centering
   \includegraphics[width=\columnwidth]   {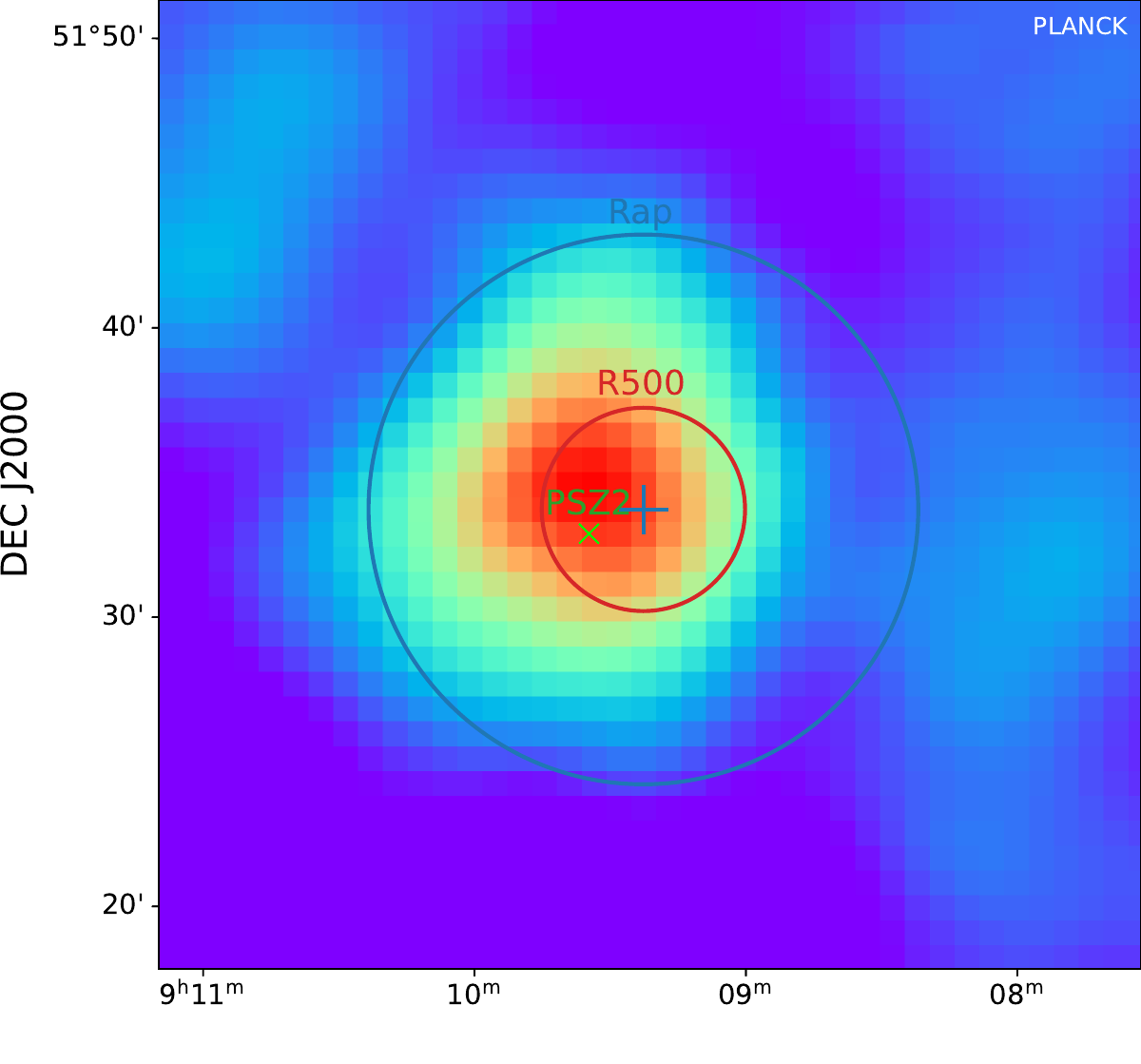}
   \vspace{0.1cm}
\includegraphics[width=\columnwidth]{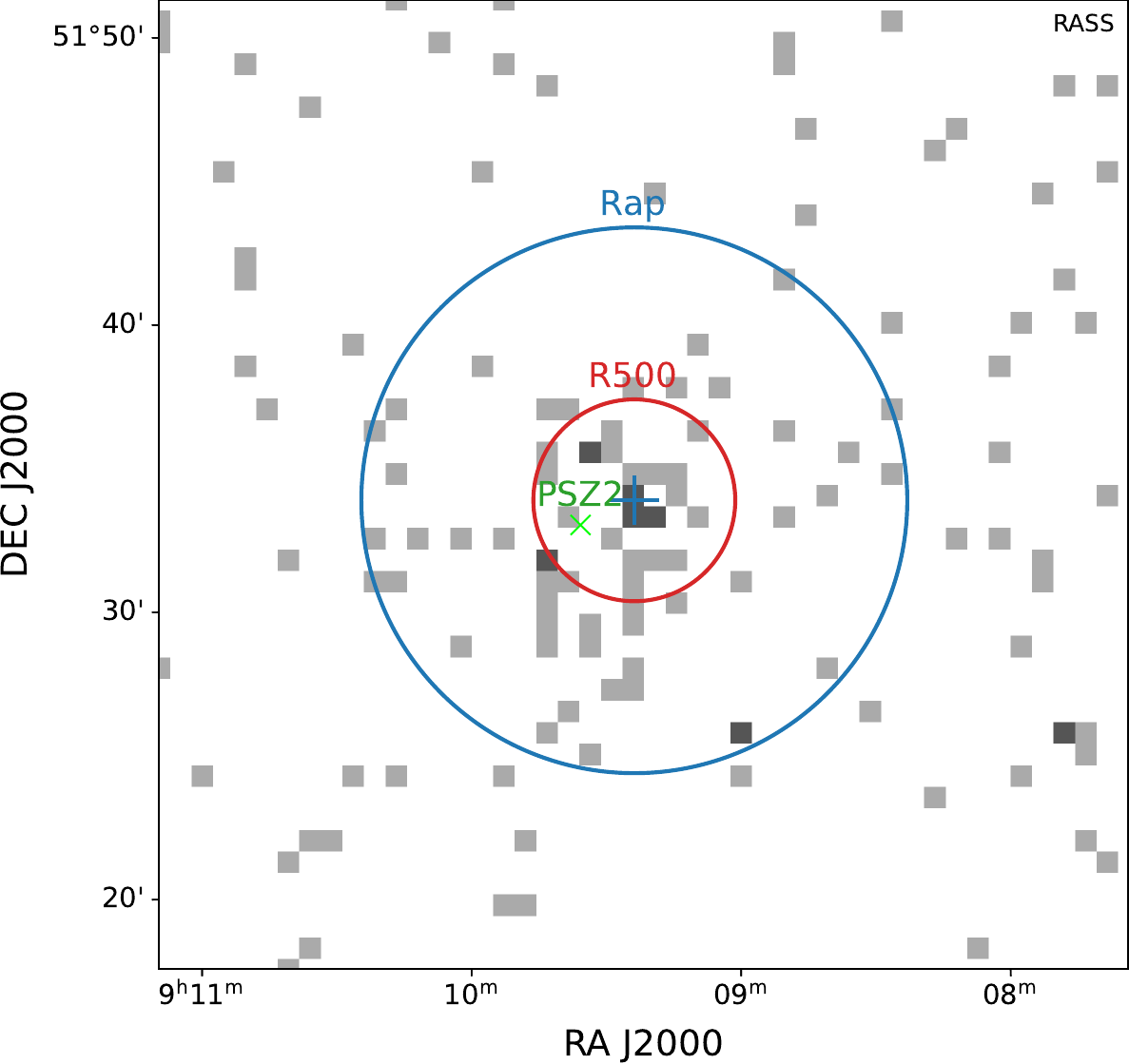}

    \caption{Planck MMF3 (top) and RASS (bottom) images centred on the \object{MCXC\,J0909.3+5133} position. A green cross indicates the position of \object{PSZ2\,G166.62+42.13}. The cluster apertures $\Rap =9.5\arcmin$, for the flux extraction,  and $\Rv=3.5\arcmin$,  are indicated in blue and red, respectively. The offset between the X-ray position and the SZ counterpart is due to the limited resolution of \planck.}
 \label{fig:ovrl_J0909}
\end{figure}

\noindent{\bf \object{MCXC\,J0909.3+5133}:} This source is associated with \object{PSZ2\,G166.62+42.13}. The separation distance, $2\arcmin$,  is typical of the \planck\ resolution. Fig.~\ref{fig:ovrl_J0909}. \\

\begin{figure*}[]
    \centering
\resizebox{\textwidth}{!} {
   \includegraphics[width=0.72\columnwidth]{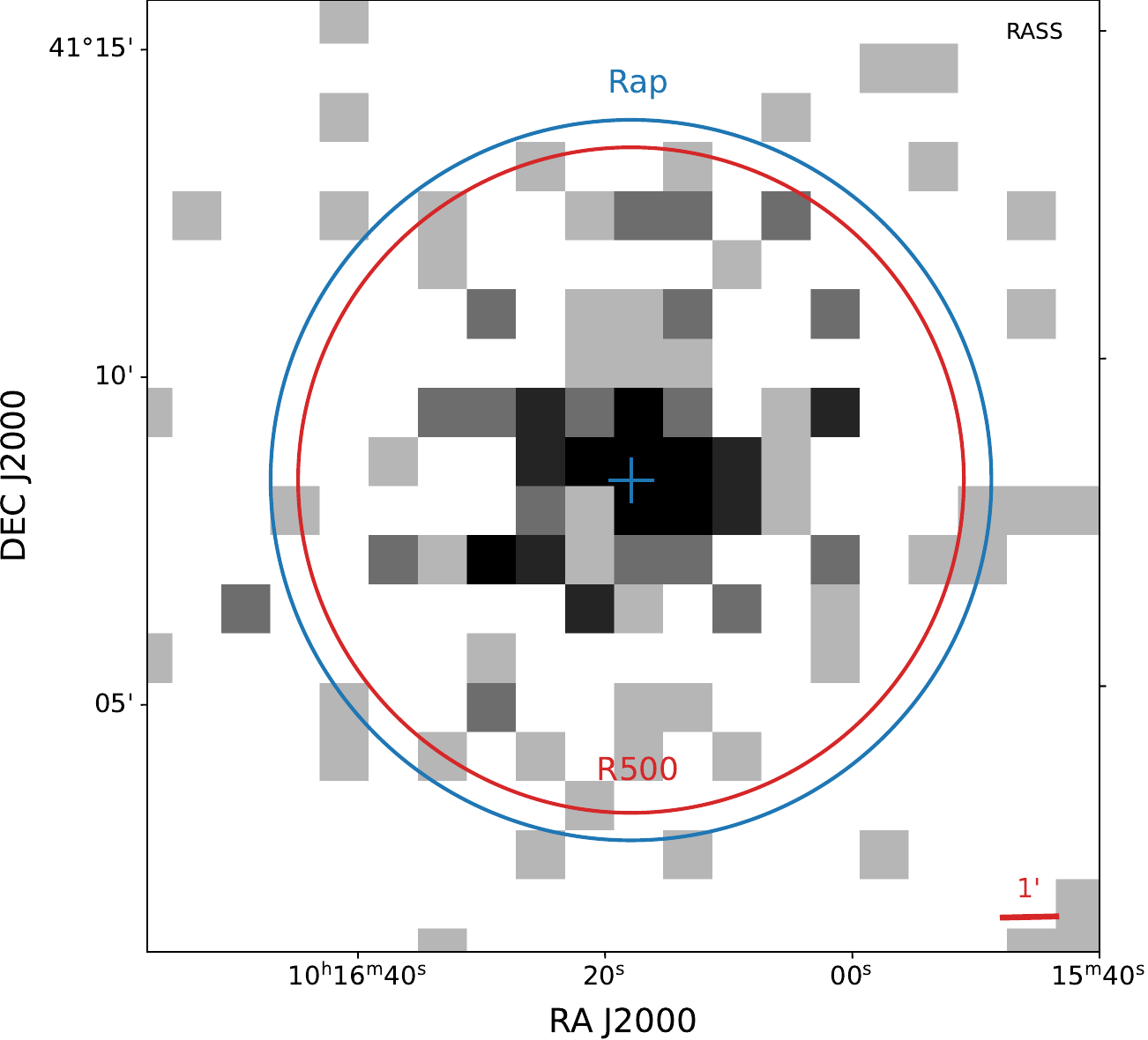}
   \includegraphics[width=0.7\columnwidth]{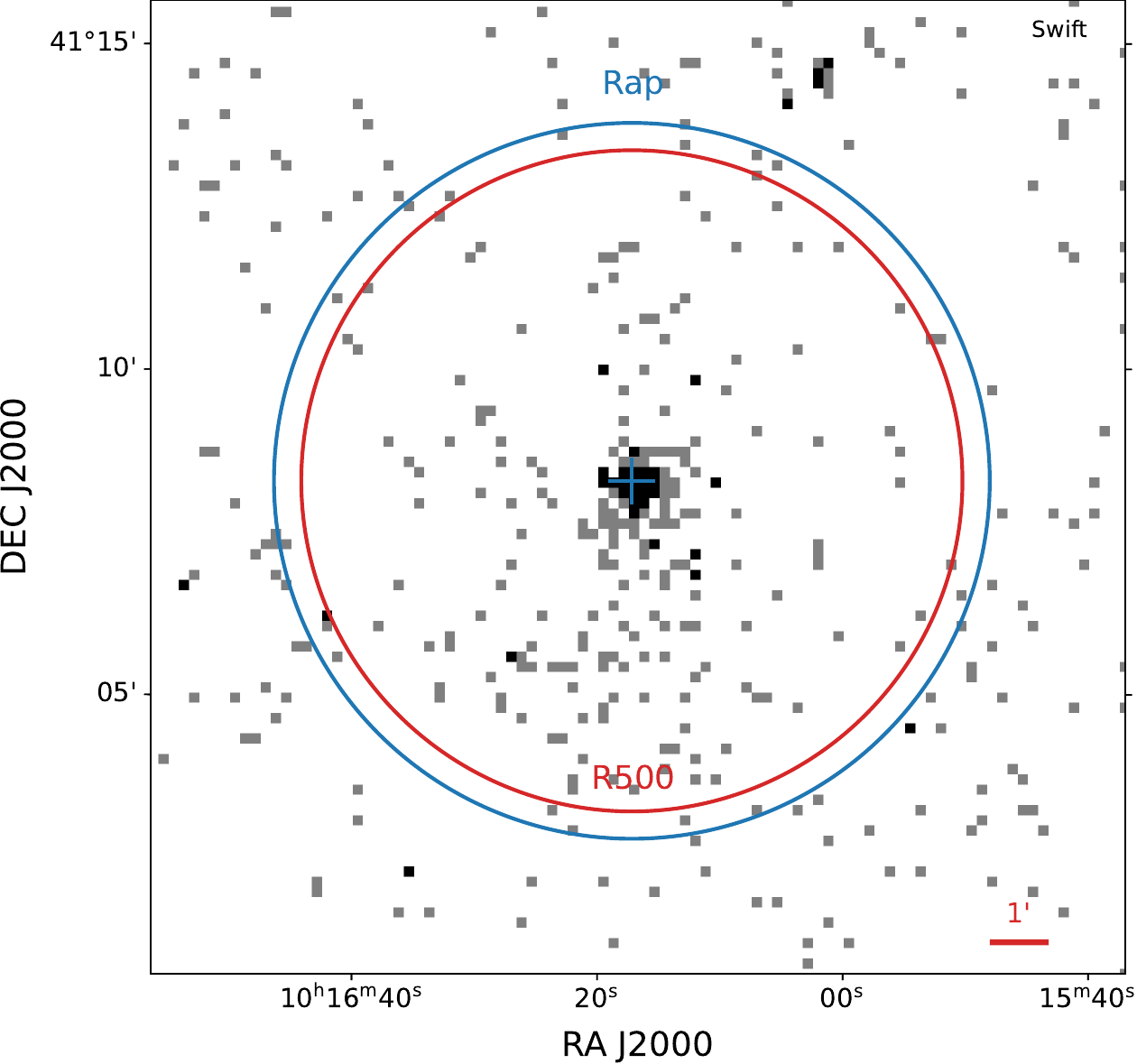}
   \includegraphics[width=0.7\columnwidth]{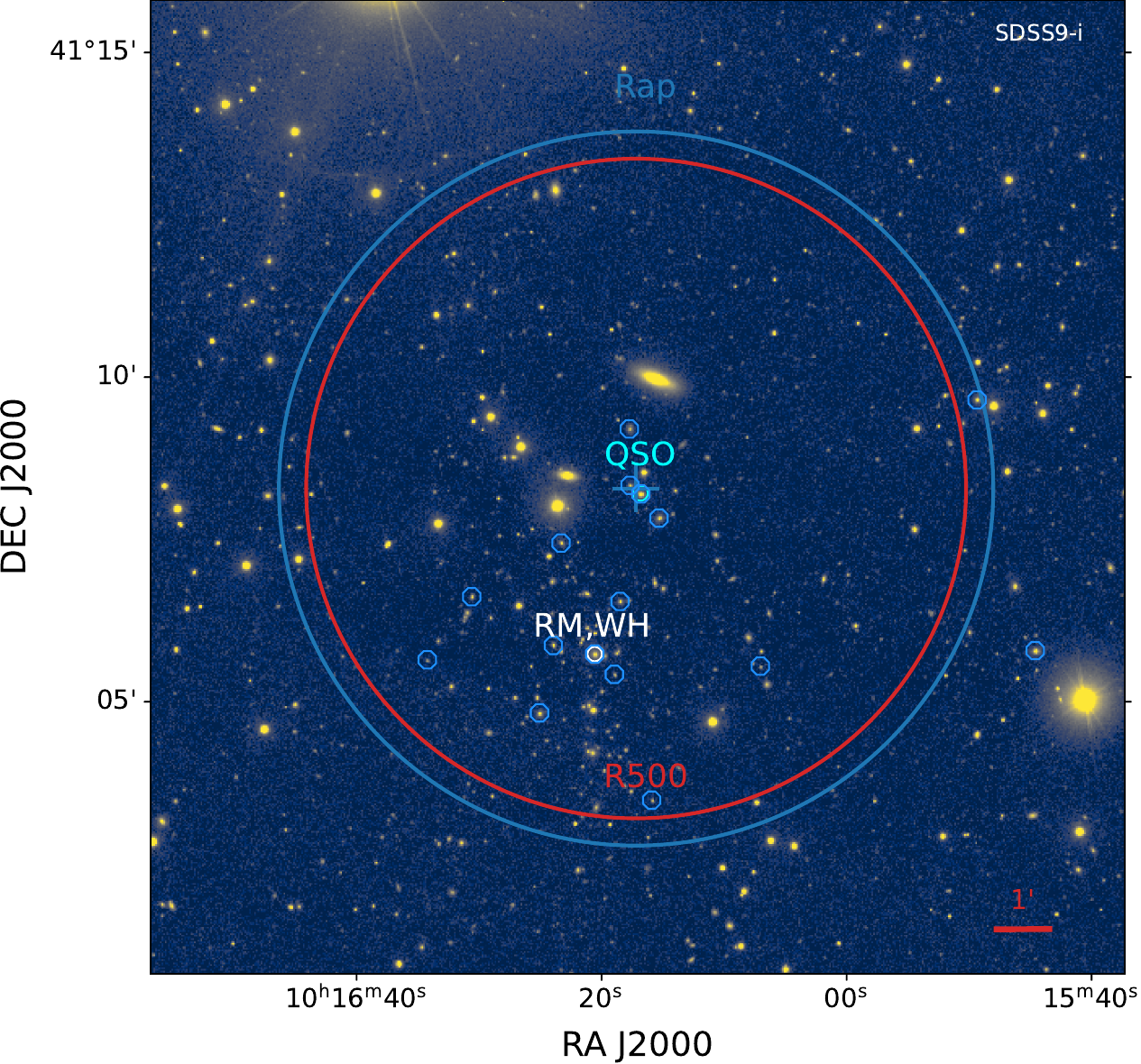}
 }
    \caption{Image cutouts centred on the position of \object{MCXC\,J1016.2+4108} from RASS (left), {\it Swift} (centre), and SDSS9-i (right). The cluster apertures $\Rap = 5.5\arcmin$ and $\Rv=5\arcmin$  are indicated in blue and red, respectively. We identified this NORAS cluster, previously without redshift, with the SDSS cluster  \object{WHL\,J101620.6+410545} (WH) or \object{RM\,J101620.6+410544.7} (RM) at $\zs = 0.2782$. The offset between the X--ray peak (blue cross) and the optical centre (white circle, in the right image), $D = 2.63\arcmin$, is due to contamination by AGN emission from a cluster galaxy member, the QSO \object{FBQS\,J101616.8+410812}. 
    The X-ray emission is peaked on this QSO, while diffuse emission extends mainly to the South (resolved in the {\it Swift} image), coincident with the concentration of the WHL spectroscopic redshifts of cluster members (blue octagons).}
    \label{fig:ovrl_J1016}
\end{figure*}

\noindent{\bf \object{MCXC\,J1016.2+4108}:} We associated this source with the SDSS cluster \object{WHL\,J101620.6+410545}  or  \object{RM\,J101620.6+410544.7} ($\zs=0.2782$ from BCG) at a distance of $D=2.63\arcmin=0.55\,\Tv$. The cross-identification with \object{Abell\,0958} in the NORAS catalogue might be incorrect in view of its distance ($8.4\arcmin$).  \object{WHL\,J101620.6+410545}  is a rich cluster with  $\RL = 69$. The SWIFT image and SDSS overlay confirm the association (see Fig.\ref{fig:ovrl_J1016}).\\

\begin{figure*}[!h]
    \centering
    \vspace{0.5cm}
\resizebox{\textwidth}{!} {
    \includegraphics[width=0.7\columnwidth]{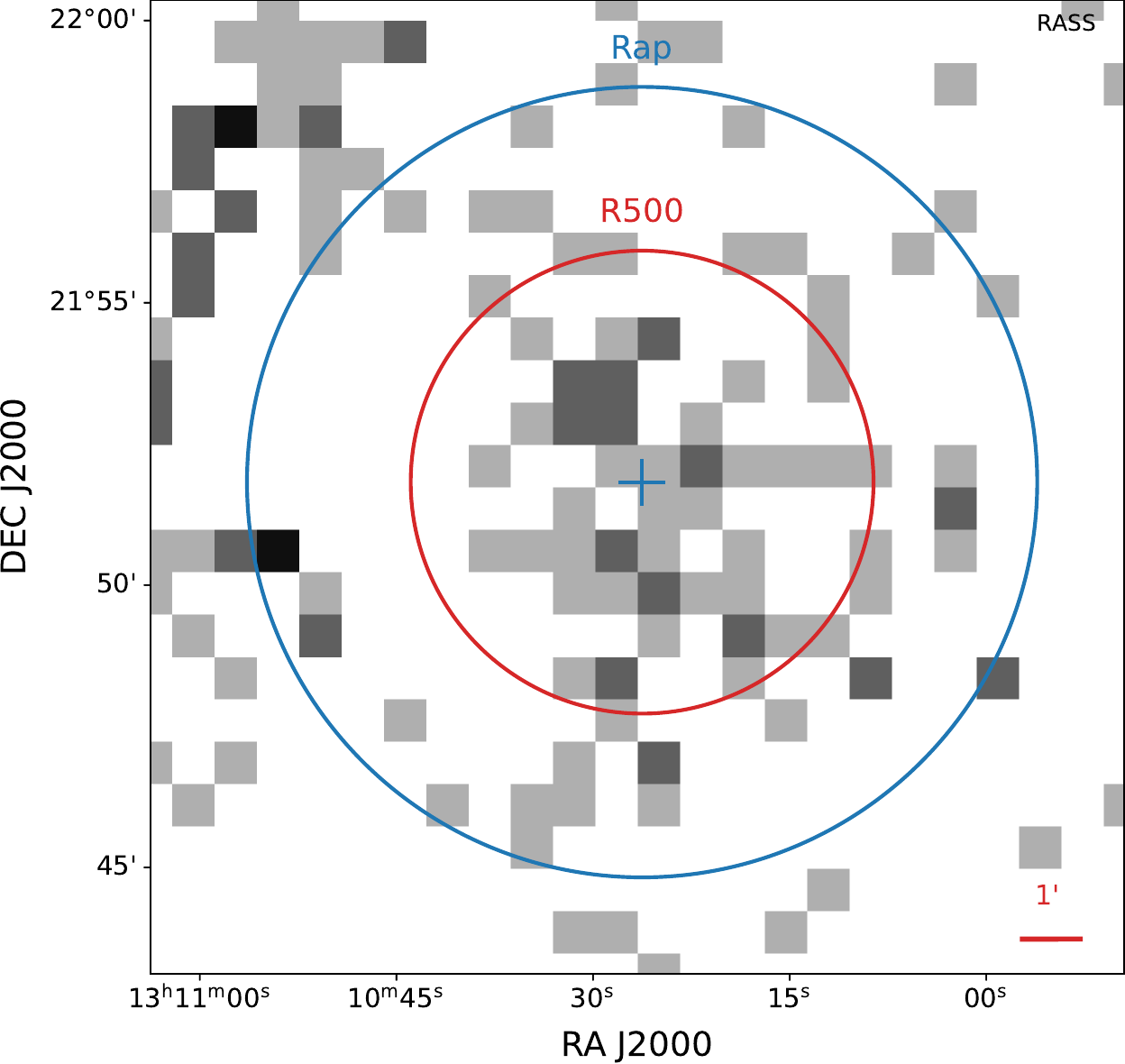}
    \includegraphics[width=0.7\columnwidth]{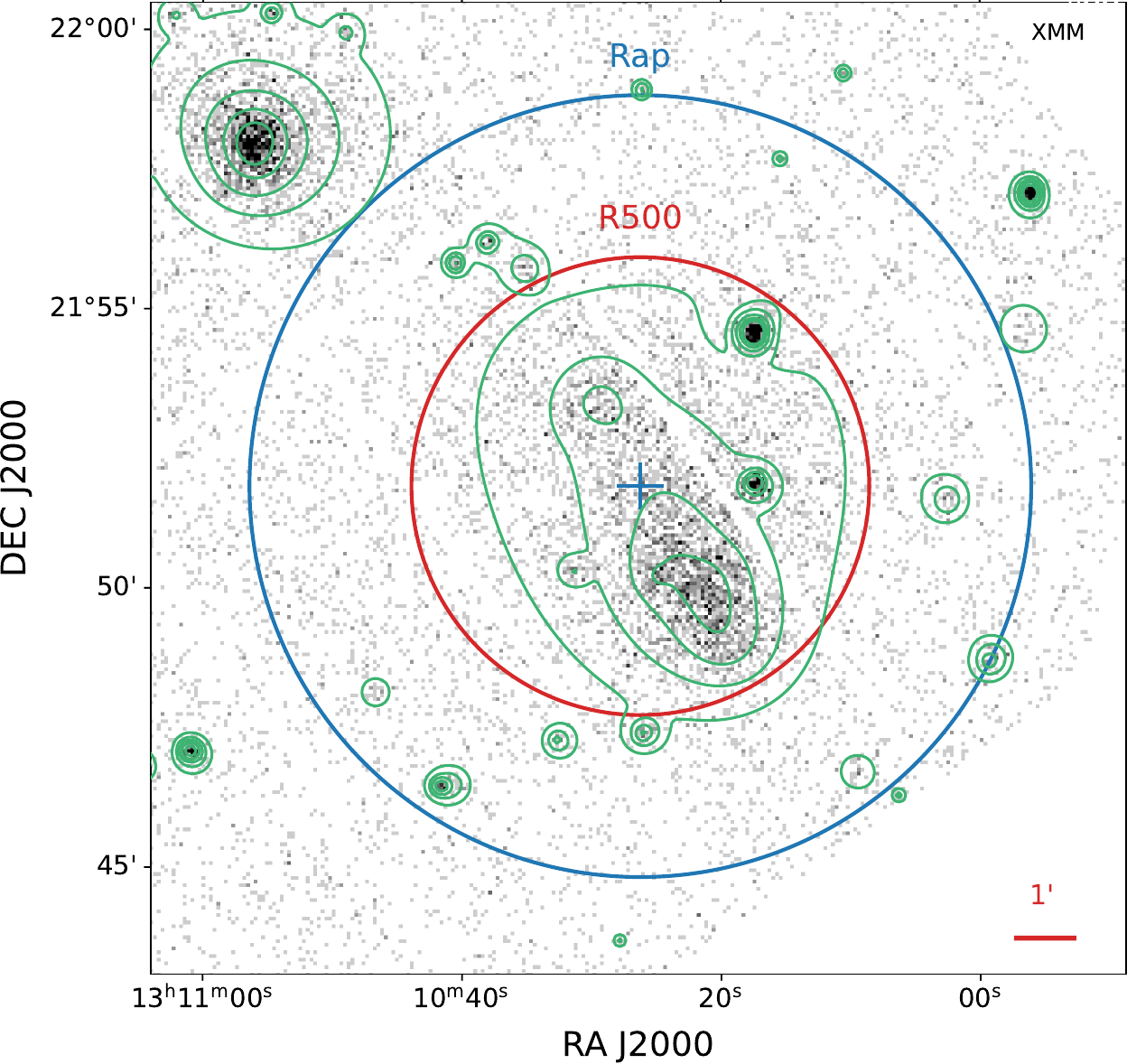}
    \includegraphics[width=0.7\columnwidth]{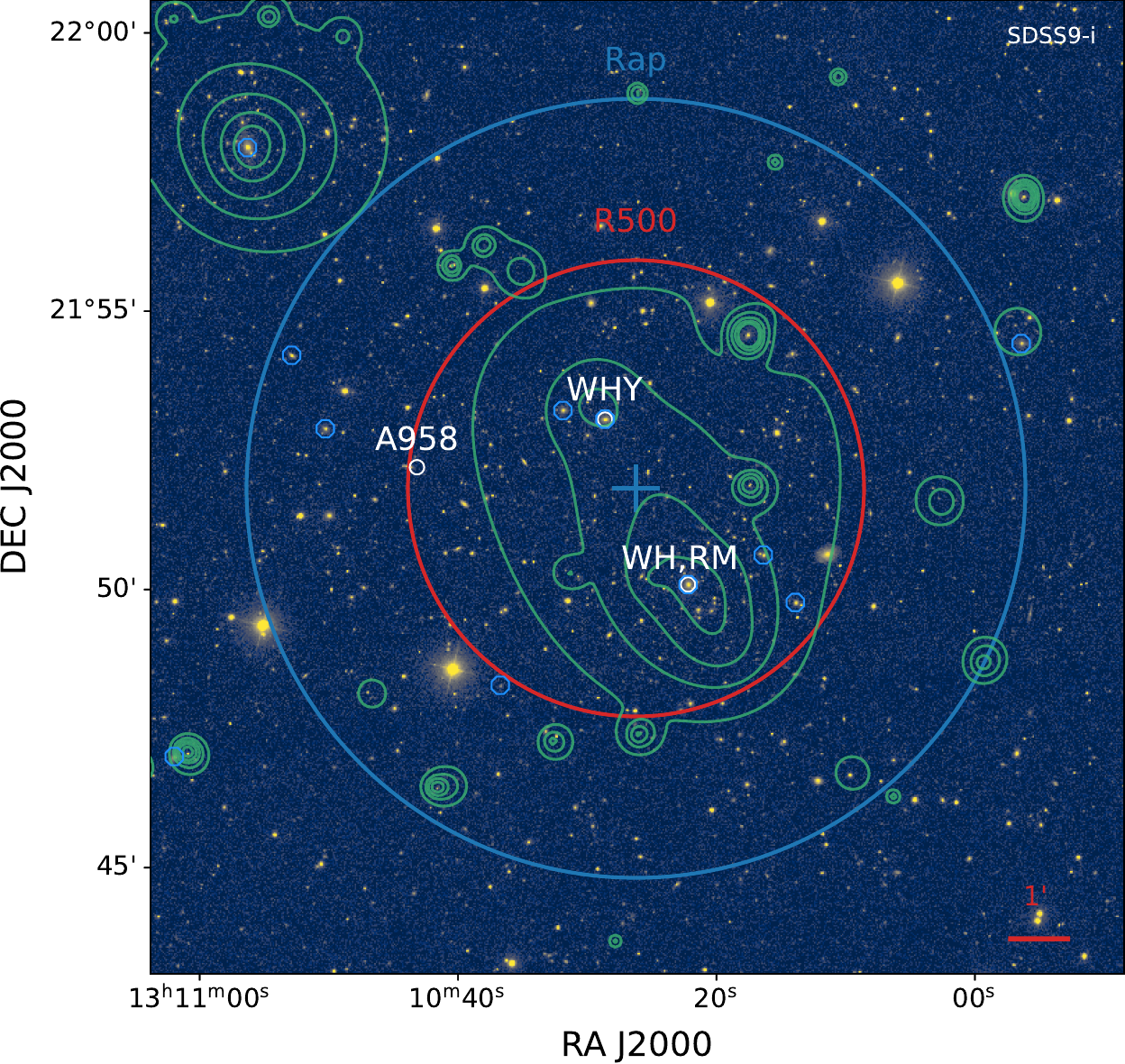}
    }
 \caption{Image cutouts centred at the position of \object{MCXC\,J1310.4+2151} from RASS (top), \xmm\  (centre), and SDSS9-i (bottom). The cluster apertures $\Rap = 7\arcmin$ and $\Rv=4\arcmin$ are indicated in blue and red, respectively. The contours from the wavelet filtered \xmm\ image are shown in green.  We identified this NORAS cluster with the SDSS cluster \object{RM\,J131022.2+215005.6} (RM) or \object{WHL\,J131022.2+215006} (WH) at $\zs=0.2730$. The distance between the optical position (white circle labelled WH,\,RM, in the right image) and the NORAS position (blue cross), $D=1.97\arcmin=0.5\Rv$, is due to the cluster morphology, which is unresolved in RASS. The cluster has an offset-centre, possibly bimodal, morphology which is well resolved with \xmm. The main peak is at the position of the optically-identified cluster. There is a secondary peak at the position of the cluster \object{WHY\,J131028.6+215304} (white circle labelled WHY), located at a very similar redshift of $\zs=0.2691$. The cluster may be in a pre- or post-merger state.  The \object{Abell\,0958} cluster lies at $3.9\arcmin$ to the East from the MCXC position.
 }
     \label{fig:ovrl_J1310}
\end{figure*}

\noindent{\bf \object{MCXC\,J1310.4+2151}:}  We cross-identified this source with \object{RM\,J131022.2+215005.6} at a distance $D=1.97\arcmin=0.5\Tv$. It also matches the \planck\ cluster \object{PSZ2\,G343.33+83.19}. The \xmm\ image shows that the cluster has an offset centre morphology, with a peak at the optical position. This, in combination with the RASS resolution,  explains the positional offset of $\sim 2 \arcmin$. See Fig. \ref{fig:ovrl_J1310}.\\

\noindent{\bf \object{MCXC\,J1320.6+3746}:} This system is identified with  \object{Abell\,1715} in the NORAS catalogue at a distance of $2.3\arcmin$. It also coincides with  \object{RM\,J132028.2+374623.3} (also known as \object{WHL\,J132028.2+374623}, with $\zs=0.2381$ from the BCG) at a distance of $2.4\arcmin = 0.56\,\Tv$. The SWIFT image shows a very diffuse morphology with no clear peak, which may explain the offset.  \\

\noindent{\bf \object{MCXC\,J1331.5+0451}:}  This source is identified with  \object{Abell\,1715}, at $1.82\arcmin$ separation in the NORAS catalogue.  We also cross-identified \object{MCXC\,J1331.5+0451} with the rich SDSS cluster \object{RM\,J133141.3+045315.5}, or  \object{WHL\,J133141.3+045316}  ($\zs=0.1710$ from five galaxies) and the SZ cluster \object{ACT-CL\,J1331.6+0452} at the same redshift ($z=0.1723$), at a distance of $2.4\arcmin=0.46\,\Tv$.  There is no \xmm, \chandra, or SWIFT archival image.  

\begin{figure*}[t]
    \centering
    \includegraphics[width=1\textwidth]{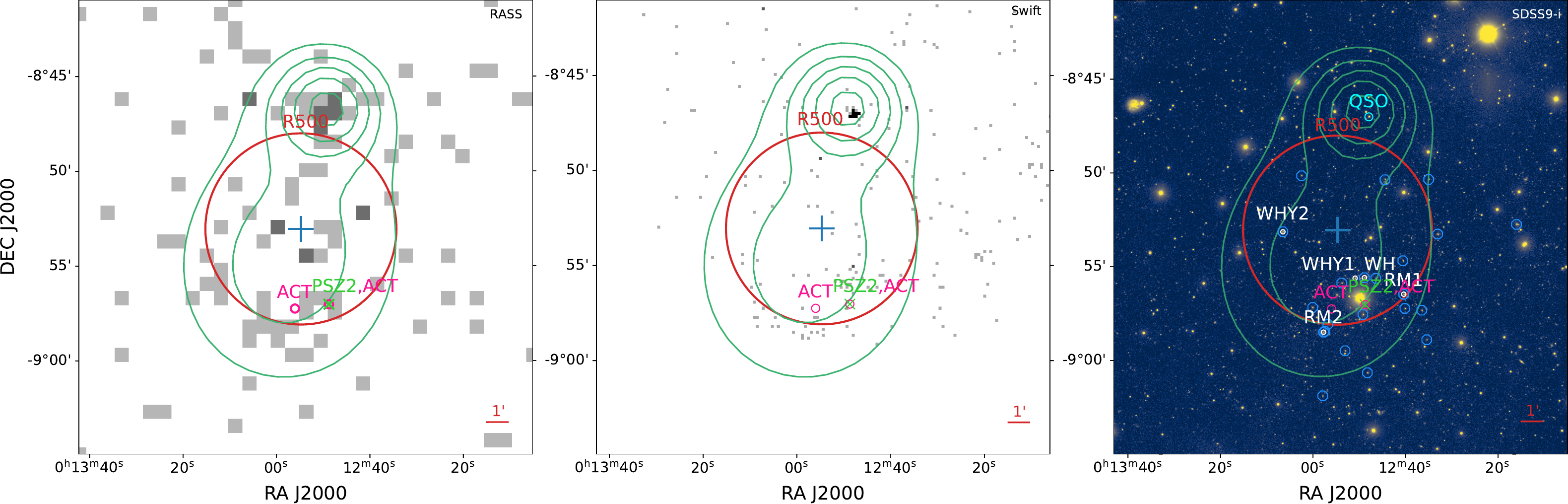}    
     \caption{Image cutouts centred at the position of \object{MCXC\,J0012.9-0853},  from RASS (left), Swift (centre), and SDSS9-i (right). We identified this SGP cluster, previously without redshift, with the SZ cluster \object{ACT-CL\,J0012.8-0855} at $\zs=0.3377$ (ACT, magenta circle). This is a case of a complex structure with  SZ clusters (PSZ2/ACT, green cross and magenta circles), and optical clusters (white circles) at very close redshifts, just South of the SGP position. In addition, there is bright QSO (\object{WISEA\,J001247.92-084700.6} at $z=0.22$ from NED, cyan circle in the right panel) well resolved in RASS (as 1RXS J001248.7-084705), to the North, slightly beyond the $\Rv$ aperture (red circle). The contour of the wavelet-filtered image shows extended emission toward the cluster counterpart(s), with the SGP position likely biased North by the emission from the QSO. This would explain the offset. The SDSS image further illustrates the complexity of the cluster structure(s). In the right-hand panel, the   
     galaxies with spectroscopic redshift from SDSS around $\zs=0.338$  are marked with blue circles, as well as the ACT cluster (magenta circles), the PSZ2 detection (green cross) and the optical clusters (white circles).  The richest optical cluster is \object{WHL\,J001248.9-085535} (WH, $\zs=0.3356$, $\RL=118.2$, likely corresponding to \object{WHY\,J001250.9-085537} (WHY1) at $\zp=0.3791$, located slightly North of the position of  \object{ACT-CL\,J0012.8-0855}, with a consistent redshift. Another ACT cluster, \object{ACT-CL\,J0012.9-0857} ($\zs=0.3520$), slightly less massive, coincides with the redMaPPer cluster, \object{RM\,J001240.4-085628.3} (RM2), which lies at $\zs=0.3370$ from the BCG. Both components likely contribute to the X-ray emission, and also to the \planck\ source \object{PSZ2 G094.46-69.65} ($\zp=0.3521$). The PSZ2 redshift comes from cross-identification with the second redMaPPer cluster, \object{RM\,J001257.7-085829.5} (RM1) in the West. In addition, there is a fourth optical cluster, \object{WHY\,J001306.5-085308} (WHY2) at $\zp = 0.3655$ lying to the East. All components likely contribute to the SGP cluster detection. For MCXC-II, the redshift is taken from that of  the most massive component.}
     \label{fig:ovrl_J0012}
\end{figure*}

\begin{figure}[!h]
    \centering
    \includegraphics[width=0.8\columnwidth]{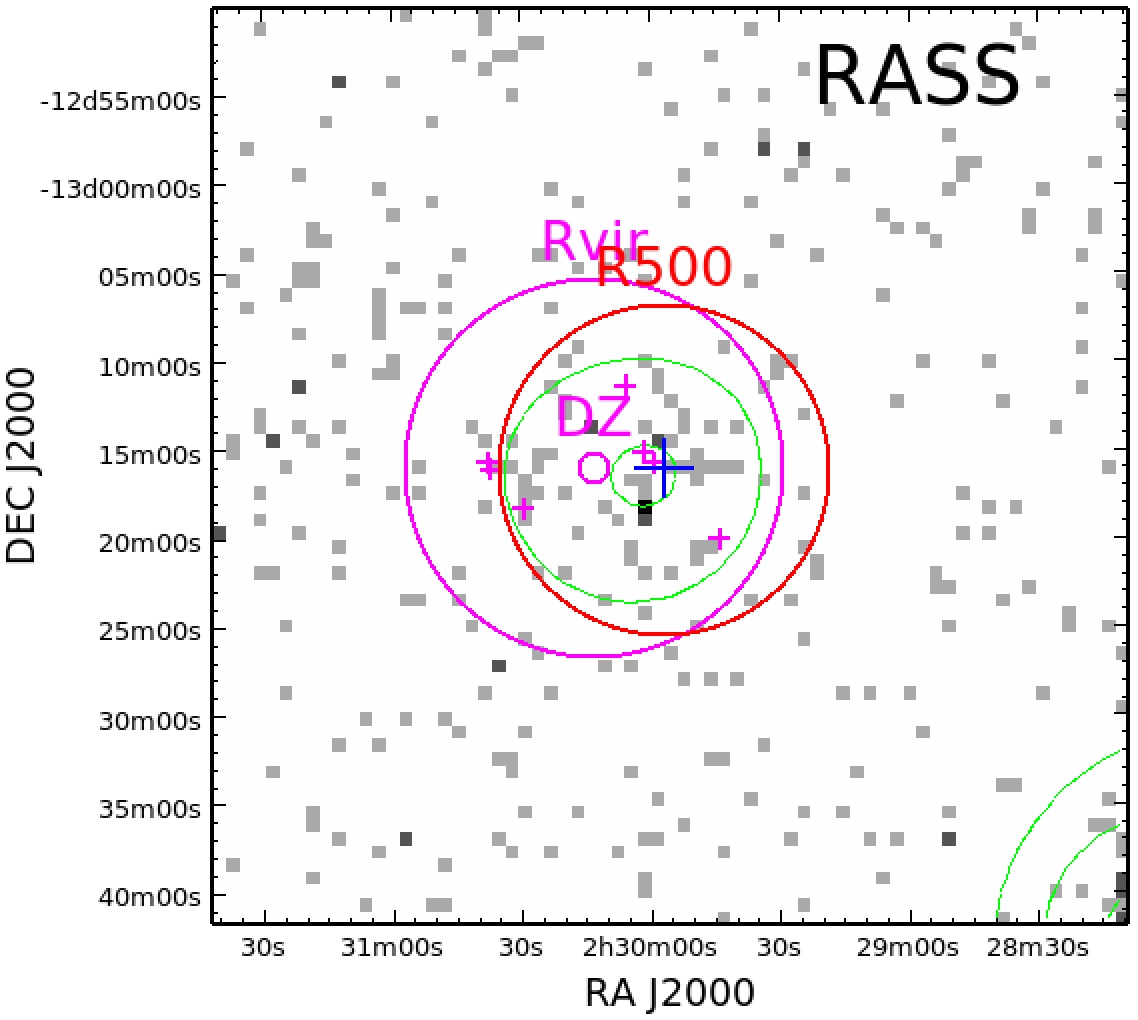}
    \caption{RASS image centred on the position of \object{MCXC J0229.9-1316}, an SGP cluster, previously without redshift. The green lines are the iso-contours from the wavelet-filtered RASS image, showing that the emission is extended. We identified this cluster with the \object{[DZ2015]\,408} group at $z=0.0565$ \citep{dia15}. Group spectroscopic members are marked by magenta plusses and the group virial radius ($R_{rm vir}=13\arcmin$) by a magenta circle. 
    The group members are concentrated within $\Rv$ (red circle). The relatively large offset between the optical and X--ray centre of $D = 3.97\arcmin = 0.42 \Rv$ is due to the large angular size of the group, which lies at a very low redshift.}
    \label{fig:J0229}
\end{figure}
\newpage

\subsubsection {SGP clusters}\label{app:sgp}
\noindent{\bf \object{MCXC\,J0012.9-0853}:} This source is matched to \object{ACT-CL\,J0012.8-0855}. The RASS  position is likely biased (and the flux probably contaminated) by a point source to the North, which is well resolved in the SWIFT image, see Fig. \ref{fig:ovrl_J0012}.\\

\begin{figure}[!h]
    \centering
    \includegraphics[width=0.8\columnwidth]{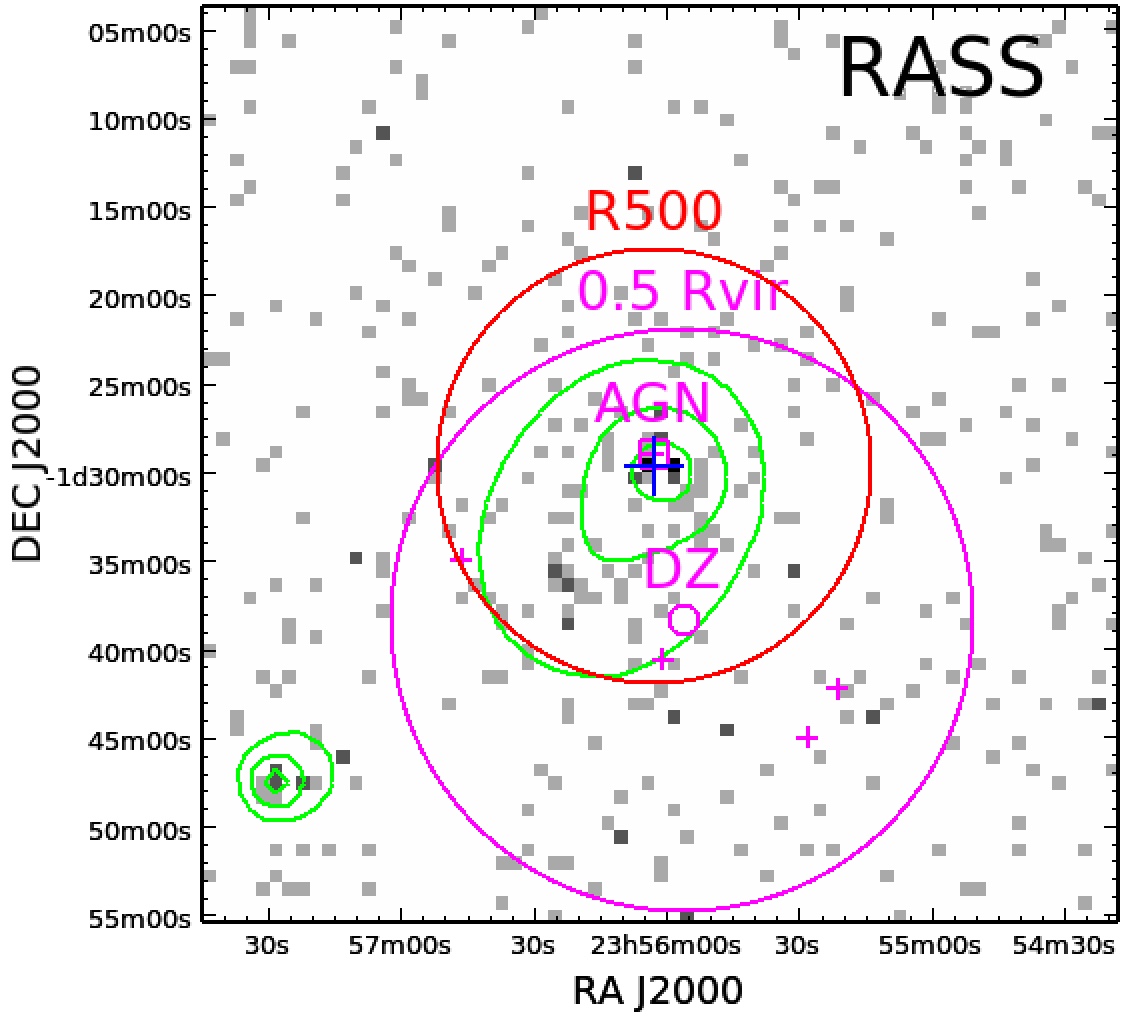}
    \caption{RASS image  centred on  \object{MCXC J2356.0-0129} position, an SGP cluster previously without redshift.  The X-ray emission is peaked on the AGN \object{MCG\,+00-01-003} at $\zs = 0.0370$, a member of the \object{[DZ2015]\,540} group at $z=0.038$ \citep{dia15}. Other group members are marked by magenta plusses and the half-virial radius by a magenta circle.
    The contours from the wavelet filtered RASS image (green lines) show that the emission is extended, in the direction of the group centre, $D = 8.8\arcmin$ from the MCG galaxy member. We identified  \object{MCXC J2356.0-0129}  with the \object{[DZ2015]\,540} group, noting that the emission is likely highly contaminated by a non-central AGN group member, which explains the offset between the X--ray and optical positions.
 }
    \label{fig:J2356}
\end{figure}

\noindent{\bf  \object{MCXC\,J0229.9-1316}:} We cross-identified this source with the [DZ2015]\,408 group, from the catalogue of compact groups extracted by \citet{dia15} from the 2MASS catalogue of galaxies with spectra.  The distance between the X-ray and optical position is $D=3.97\arcmin=0.42\,\Tv$. As shown in Fig.\,\ref{fig:J0229}, the RASS image displays extended diffuse emission with all group galaxies inside $\Tv$ or the virial group radius  $R_{\rm vir} = 13\arcmin$, estimated by \citet{dia15}. This confirms the association.\\

\noindent{\bf  \object{MCXC\,J2355.1-2834}:} This source was identified with  \object{Abell\,4054} in the NORAS catalogue. The redshift is  $z=0.1866$ from \citet{2002MNRAS.329...87D}, and the separation distance  $D=2.2\arcmin = 0.48\,\Tv$. The cluster also matches \object{WHY\,J235510.1-283212}  ($z=  0.1884$), a very rich cluster with  $\RL= 95.3$. The offset, although large in terms of $\Tv$, is likely explained by the RASS resolution.  \\

\noindent{\bf \object{MCXC\,J2356.0-0129}:} The source was identified with the \object{[DZ2015]\,540} group  at $z=0.038$ \citep{dia15}, which has a virial radius estimated to $R_{\rm vir} = 33\arcmin$. One of the galaxy members is the AGN \object{MCG\,+00-01-003} at $z=0.0370$, located at $D=8.8\arcmin$ from the group centre.  The RASS image shows diffuse emission, centred on the AGN, which dominates the emission. This explains the offset of $D=8.82\arcmin= 0.72 \Tv$. See Fig.~\ref{fig:J2356}.

\subsection{Duplicates in MCXC}  \label{app:dupl}

We discuss here duplicate cases for which the distance and/or the redshift or luminosity difference between the two clusters is large.\\ 

\begin{figure*}[t]
    \centering
	\resizebox{\textwidth}{!} {
    \includegraphics[width=\columnwidth]{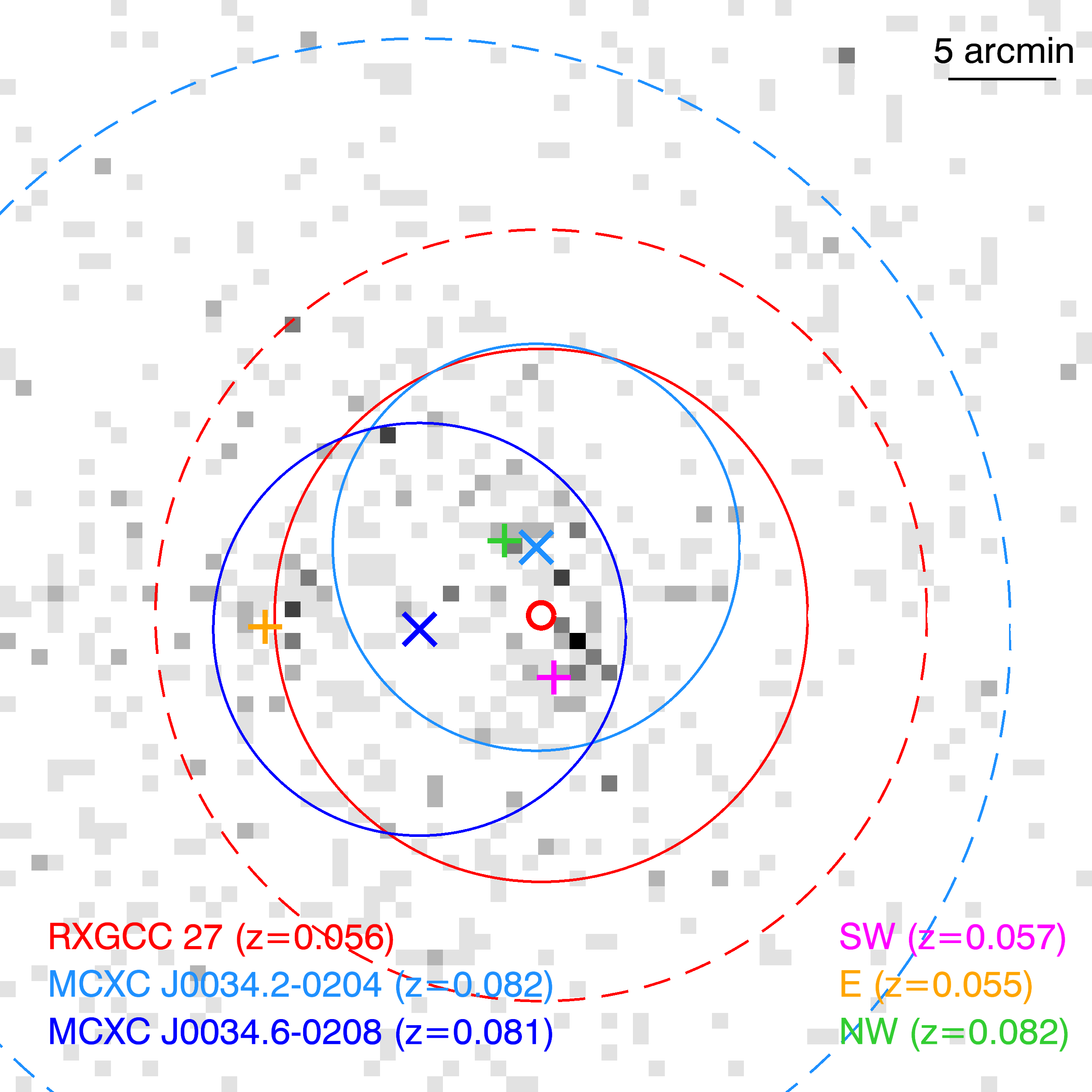}
    \hspace{0.5cm}
    \includegraphics[width=\columnwidth]{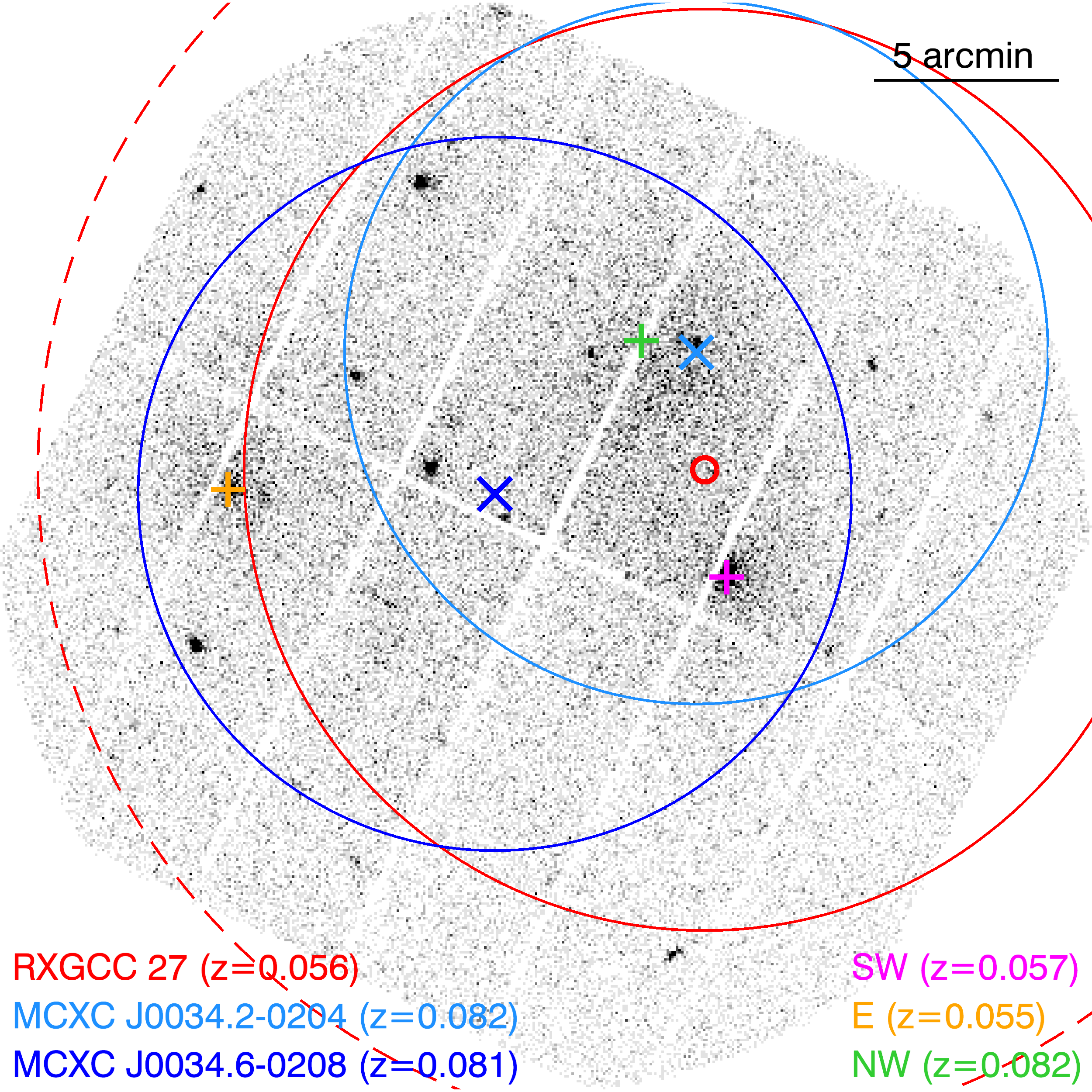}}
    \caption{\object{MCXC\,J0034.2-0204}, a complex structure with three components at two different redshifts. We have associated the two MCXC-I clusters and the RXGCC cluster because all of them cover the whole system. The left and right panels show the RASS and \xmm\ images, respectively. The positions of the MCXC-I clusters \object{MCXC\,J0034.2-0204} ($z=0.0822$, SGP) and \object{MCXC\,J0034.6-0208} ($z=0.0812$, REFLEX) are marked with a light and dark blue cross, respectively. The position of \object{RXGCC\,27} ($z=0.056$) is marked with a small red circle. The corresponding solid and dashed circles represent the size $\theta_{500}$ of the clusters and the aperture radius of the detection (RAP in MCXC, RSIG in RXGCC). The green, orange, and magenta pluses represent the position of the NW, E, and SW components studied in \citealt{RamosCeja2019}.}
    \label{fig:RXGCCimages_MCXC J0034.2-0204}
\end{figure*}
\noindent{\bf \object{MCXC\,J0034.2-0204}:} Complex system with three components: E and SW at $z\sim0.056$, and NW (the most massive) at $z\sim0.082$ (see Fig. \ref{fig:RXGCCimages_MCXC J0034.2-0204} and discussion on the \xmm\ observation of this cluster in \citealt{RamosCeja2019}). MCXC-I included two different clusters around this position: \object{MCXC\,J0034.2-0204} ($z=0.0822$, SGP) and \object{MCXC\,J0034.6-0208} ($z=0.0812$, REFLEX). The two of them are merged in Simbad as cluster \object{ZwCl\,0031-0220}. Both have a similar flux ($\sim8\times10^{12}$ erg s$^{-1}$), which corresponds approximately to the sum of the NW and SW components measured by \xmm. The two clusters seem to be the same RASS detection covering the three components, with the only difference being the cluster position: in REFLEX the detection is centred on the centroid while in SGP it is centred on the NW peak. Therefore, we have decided to merge them in MCXC-II. We have chosen \object{MCXC\,J0034.2-0204} (SGP) as the primary detection because it is well centred on the most massive component (NW). RXGCC also includes a detection that covers the three components of this system (\object{RXGCC\,27}), which is centred in between the NW and SW components, and has an assigned redshift of $z=0.056$, corresponding to the SW component. Despite the redshift difference, the detection covers the whole system, so we have decided to associate them. \\

\noindent{\bf \object{MCXC\,J1010.2+5430}:} \object{MCXC\,J1010.2+5429} ($z=0.047$, NORAS) and \object{MCXC\,J1010.2+5430} ($z=0.045$, 400SD), are both associated with \object{WARP\,J1010.1+5430} ($z=0.047$) in NED. The ROSAT pointing observation clearly shows that there is only one cluster, with the NORAS position only slightly offset, by 0.9\arcmin. The much higher  NORAS  luminosity (nearly a factor of 6), is likely due to point source contamination.  There are several point sources, including one as bright at the cluster, with the 10\arcmin\ NORAS aperture. They are not resolved in the RASS image. In this case, the 400SD catalogue is chosen as the source catalogue. \\

\noindent{\bf \object{MCXC\,J1311.7+2201}:} We associated the  cluster \object{MCXC J1311.7+2201}   ($z=0.1716$, NORAS) and \object{MCXC J1311.5+2200} ($z=0.266$, eBCS). 
The \xmm\ image shows that there is a single cluster, centred at the NORAS position. The eBCS position is offset by $\sim 0.6\Rv$ but the aperture well encompasses the bulk of the cluster emission as seen on the \xmm\ image. We cross-matched \object{MCXC\,J1311.7+2201} with the SDSS cluster,   \object{WHL\,J131146.2+220137} at $0.7\arcmin$ ($\zs=0.1704$ from  7 galaxies) or \object{RM\,J131146.2+220137.2} ($\zs=0.1715$ from BCG). This confirms the NORAS redshift value. \\

\noindent{\bf \object{MCXC\,J1652.9+4009}:} Both \object{MCXC\,J1652.9+4009} ($z=0.1492$, NORAS) and \object{MCXC\,J1652.6+4011} ($z=0.1481$, eBCS) are associated with the same optical cluster \object{NSC\,J165252+400906} ($z=0.1492$) in NED. \object{MCXC J1652.9+4009} can also be matched to the SDSS cluster \object{WHL\,J165253.2+400913} or  \object{RM\,J165253.2+400912.9} at  $0.46\arcmin$ ($\zs= 0.1493$ from BCG) and \object{RXGCC\,679}. 
There is no \xmm\ or SWIFT image and only incomplete coverage at the border of a \chandra\ pointing. A single extended source, with a bimodal morphology, is apparent in the RASS image. The  NORAS source position is well centred at the maximum, whereas the eBCS (and RXGCC) coordinates better match the centroid position. This likely explains the large position offset of $\sim 0.6\Tv$. \\

\begin{figure*}[t]
    \centering
	\resizebox{\textwidth}{!} {
    \includegraphics[width=\columnwidth]{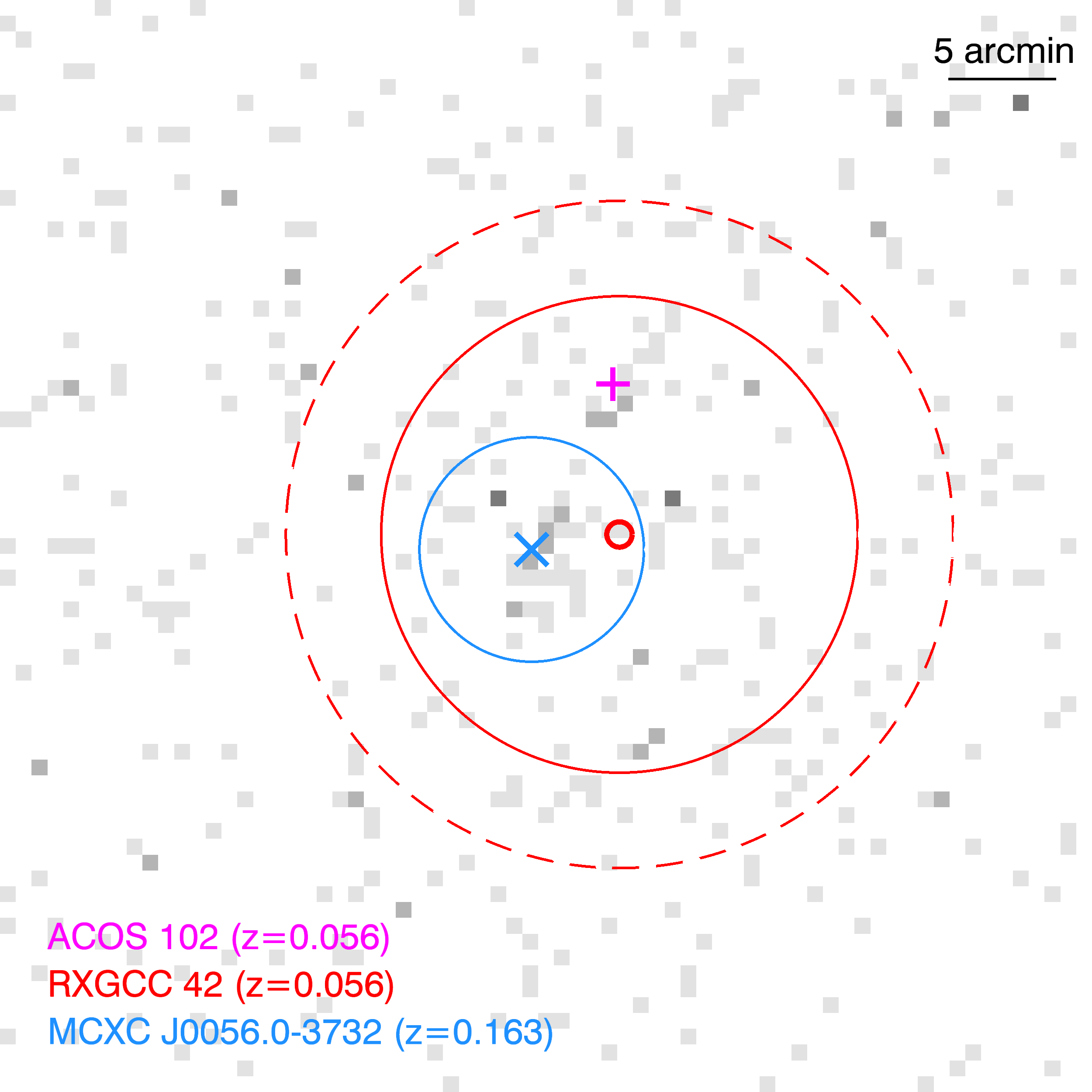}
    \hspace{0.5cm}
    \includegraphics[width=\columnwidth]{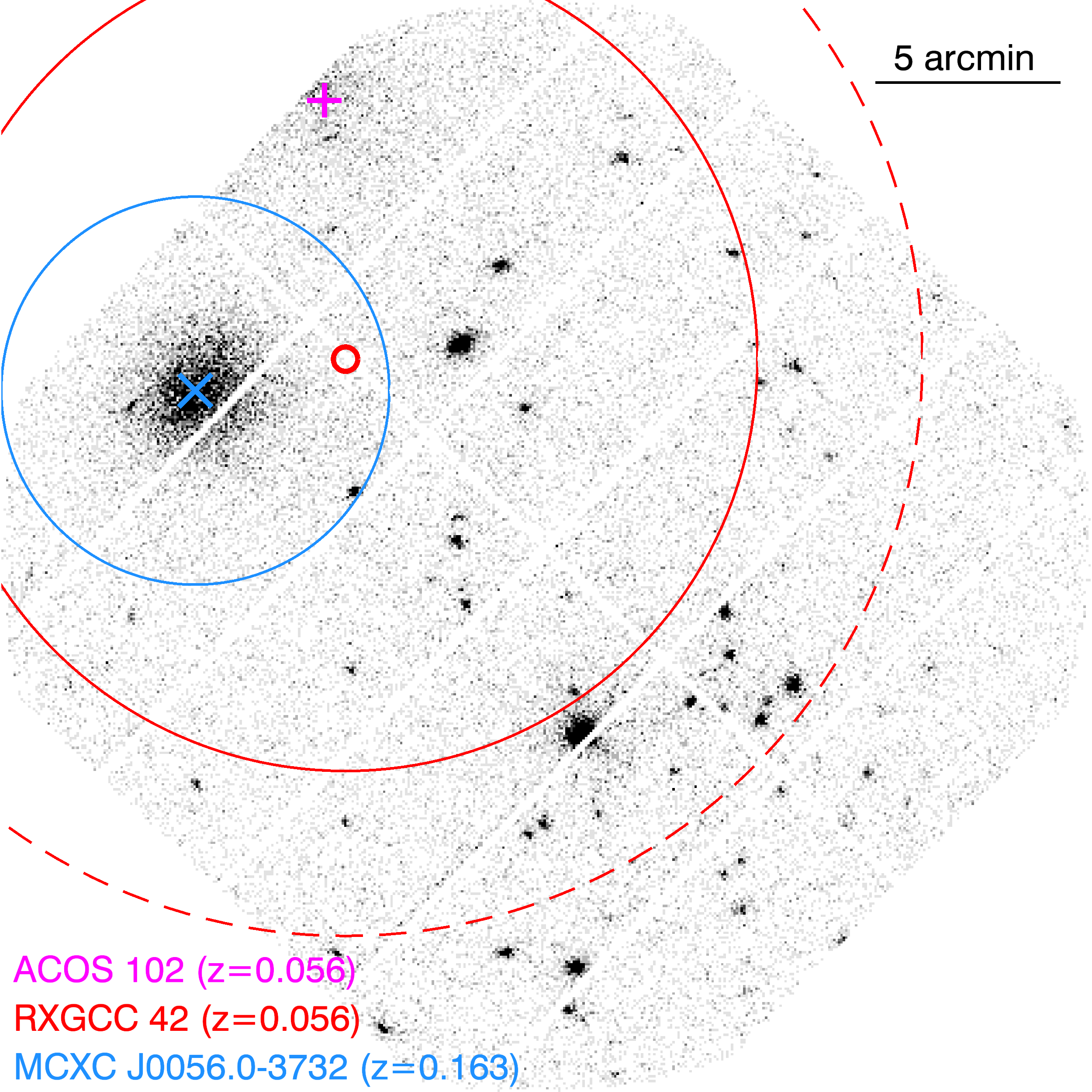}}
    \caption{\object{MCXC\,J0056.0-3732}, a case in which the RXGCC position has an offset with respect to the X-ray peak due to a nearby point source.  We have associated the MCXC and RXGCC clusters because both detections come from the same X-ray emission, although the RXGCC catalogue has chosen an incorrect  redshift (from the nearby cluster \object{Abell\,S\,102}) and its flux may be contaminated. The left and right panels show the RASS and \xmm\ images, respectively. The positions of \object{MCXC\,J0056.0-3732} ($z=0.163$) and \object{RXGCC\,42} ($z=0.056$) are marked with a light blue cross and a small red circle, respectively. The corresponding solid and dashed circles represent the size $\theta_{500}$ of the clusters and their aperture radius. The magenta plus represents the position of Abell cluster \object{Abell\,S\,102} ($z=0.056$).}
    \label{fig:RXGCCimages_MCXC J0056.0-3732}
\end{figure*}

\noindent{\bf \object{MCXC\,J2350.5+2929}:} \object{MCXC\,J2350.5+2929} ($z=0.1498$, NORAS) and \object{MCXC\,J2350.5+2931} ($z=0.095$, BCS), are both associated with \object{ZwCl\,2348.4+2908} ($z=0.15$) in NED. The RASS image shows a regular extended emission, with a peak at the NORAS position, while the BCS position is slightly offset.  However, there is a significant difference between the published MCXC redshifts. The NORAS value is confirmed by the cross-match with \object{WHL\,J235035.1+292944} at $0.35\arcmin$ with $\zs=0.1526$ from 15 galaxies or the recent study of \citet[][$z=0.1542$]{2016ApJ...819...63R}. The cluster is also identified with \object{MS\,2348+2913} and the BCS redshift is based on the EMSS reference. As discussed by \citet{2001AJ....121.1294B} this is the redshift of a foreground galaxy.

\begin{figure*}[t]
    \centering
	\resizebox{\textwidth}{!} {
    \includegraphics[width=\columnwidth]{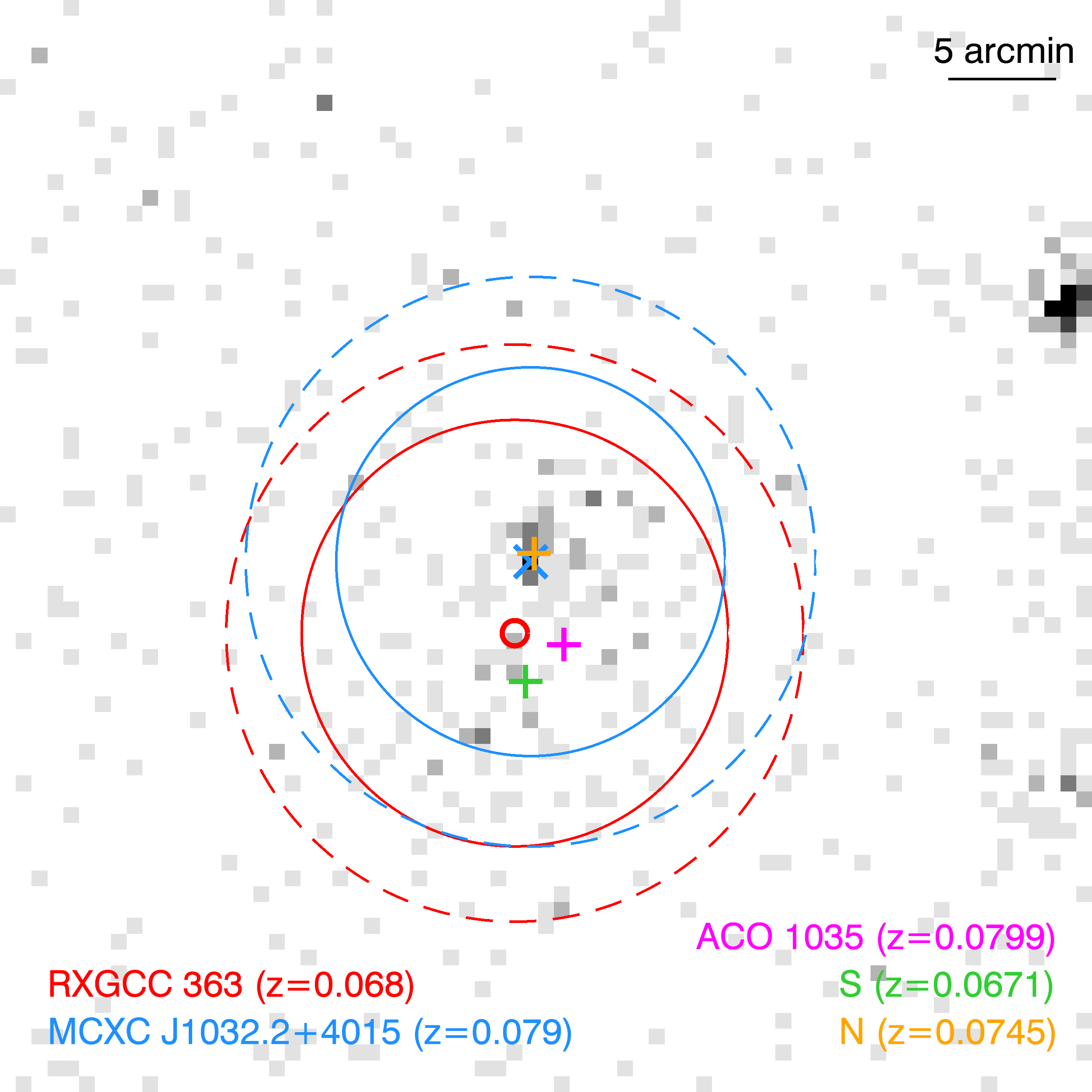}
    \hspace{0.5cm}
    \includegraphics[width=\columnwidth]{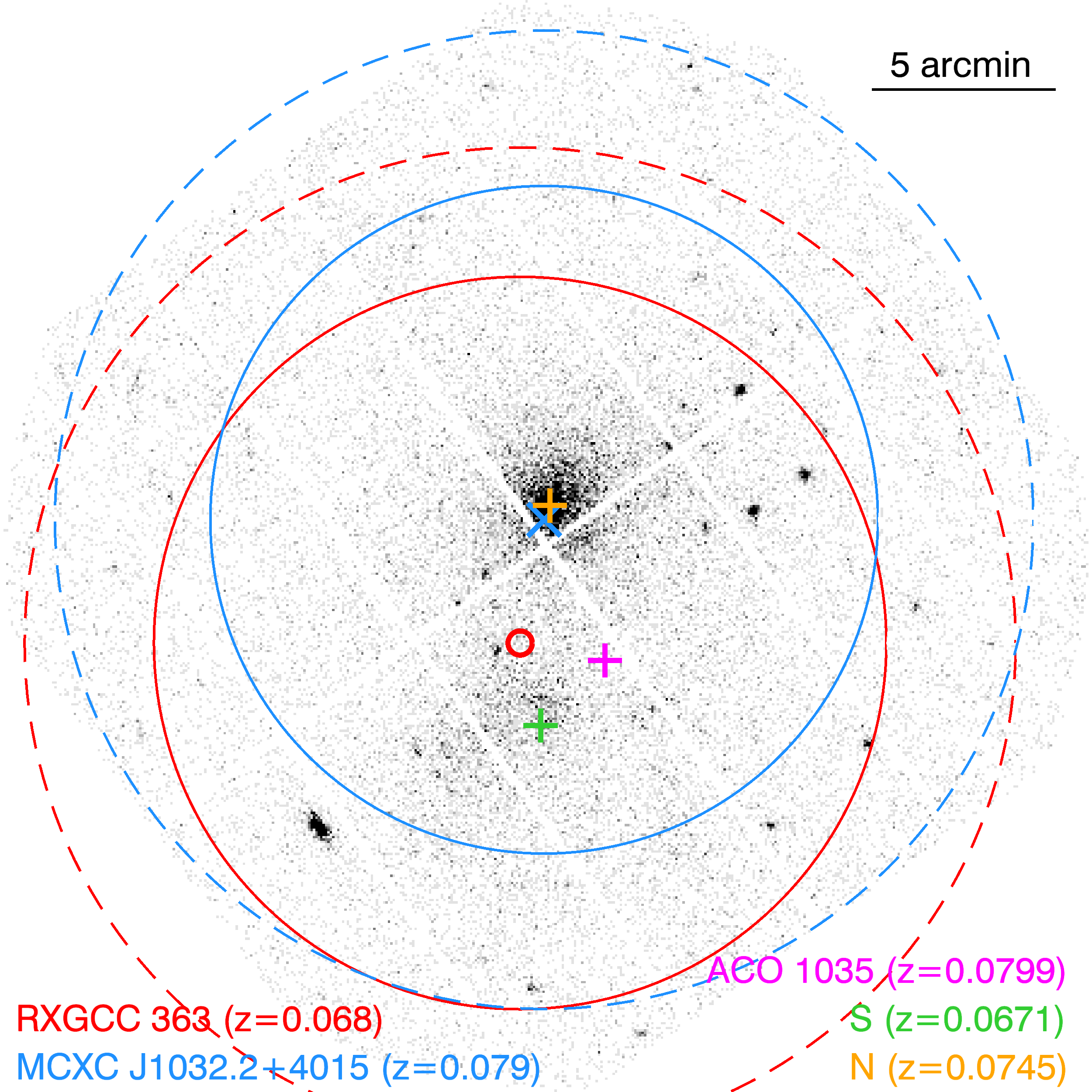}}
    \caption{\object{MCXC\,J1032.2+4015}, a system with two components at different redshifts. We have associated the MCXC and RXGCC detections since both cover the whole system. The left and right panels show the RASS and \xmm\ images, respectively. The positions of \object{MCXC\,J1032.2+4015} ($z=0.079$) and \object{RXGCC\,363} ($z=0.068$) are marked with a light blue cross and a small red circle, respectively. The corresponding solid and dashed circles represent the size $\theta_{500}$ of the clusters and their aperture radius. The orange and green pluses represent the position of two optical clusters: \object{WHL\,J103214.0+401616}, in the North with a spectroscopic redshift of $z=0.0745$ \citep{2015ApJ...807..178W}, and \object{WHL\,J103215.3+401012}, in the South with a spectroscopic redshift of $z=0.0671$ \citep{2009ApJS..183..197W,2010ApJS..187..272W}. The magenta plus shows the position of Abell cluster \object{Abell\,1035}.}
    \label{fig:RXGCCimages_MCXC J1032.2+4015}
\end{figure*}
\begin{figure*}[t]
    \centering
	\resizebox{\textwidth}{!} {
    \includegraphics[width=\columnwidth]{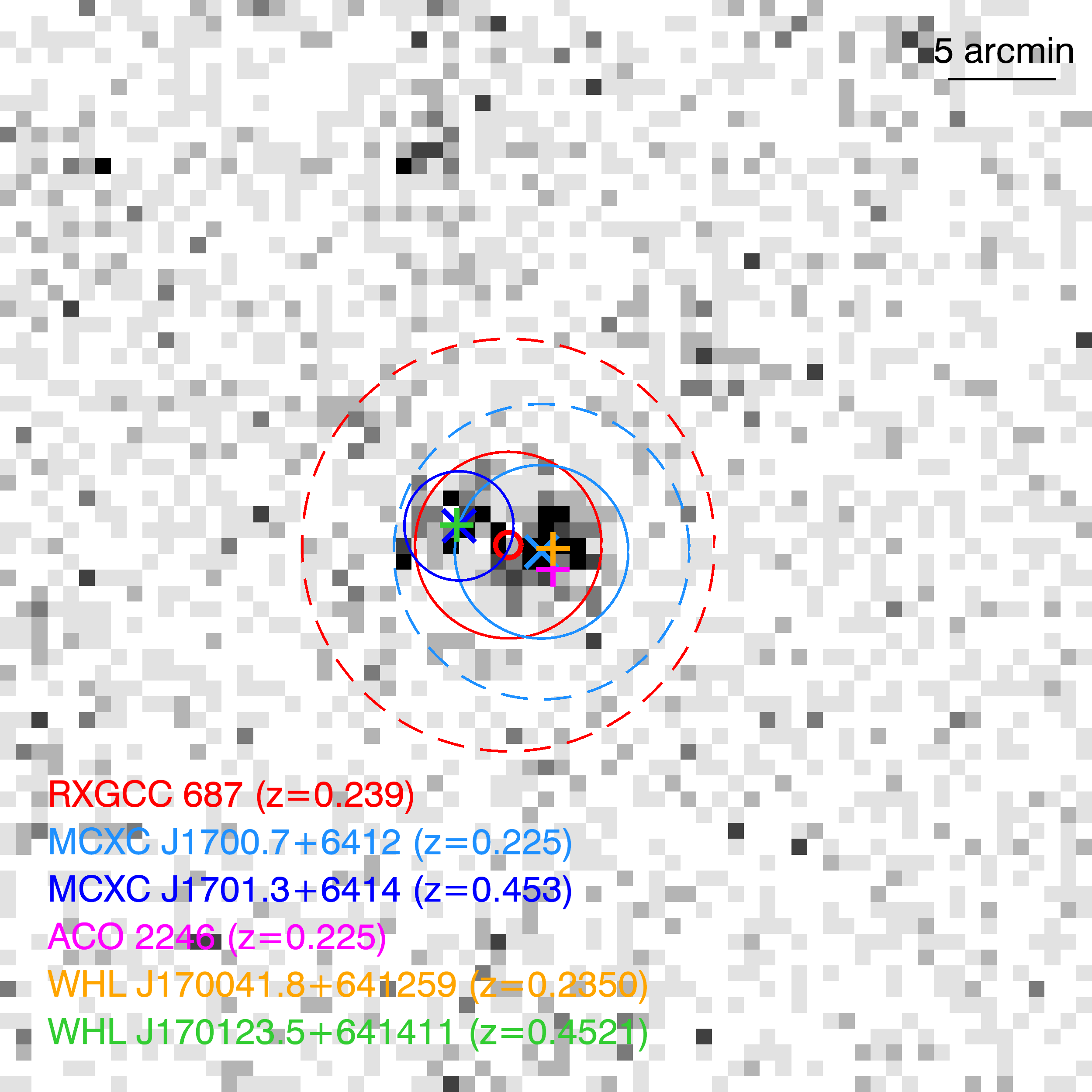}
    \hspace{0.5cm}
    \includegraphics[width=\columnwidth]{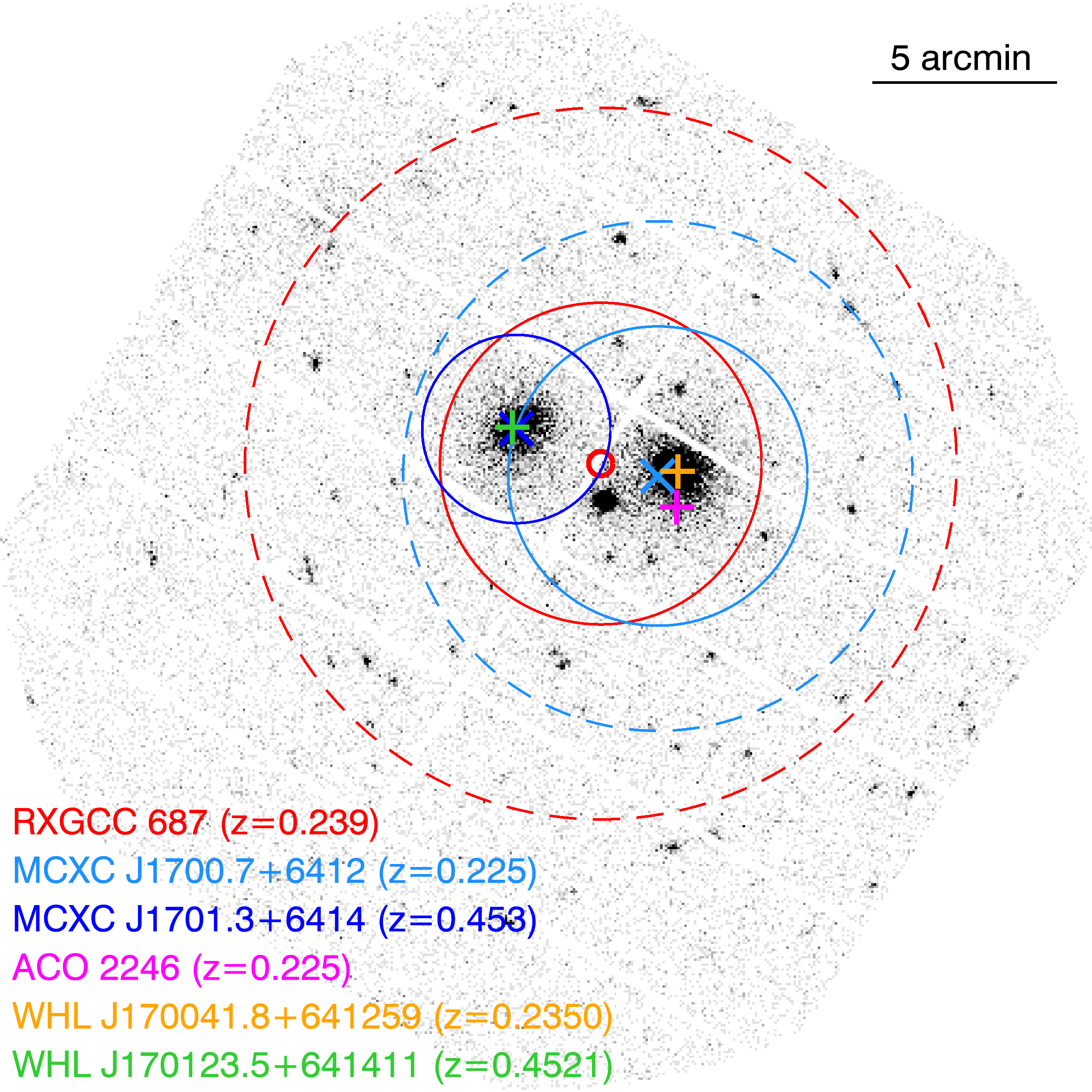}}
    \caption{\object{MCXC\,J1700.7+6412}, a system with two components at different redshifts. MCXC detects the two components separately, while RXGCC detects the system as a whole. We have associated RXGCC to the MCXC with a stronger contribution to the X-ray emission. The left and right panels show the RASS and \xmm\ images, respectively. The positions of \object{MCXC\,J1700.7+6412} ($z=0.225$) and \object{MCXC\,J1701.3+6414} ($z=0.453$) are marked with a light and dark blue cross, respectively. The position of \object{RXGCC\,687} ($z=0.239$) is marked with a small red circle. The corresponding solid and dashed circles represent the size $\theta_{500}$ of the clusters and their aperture radius. The green and orange pluses represent the position of two optical clusters: \object{WHL\,J170041.8+641259}, with an spectroscopic redshift of $z=0.2350$ \citep{2015ApJ...807..178W}, and \object{WHL\,J170123.5+641411}, with an spectroscopic redshift of $z=0.4521$ \citep{2009ApJS..183..197W,2010ApJS..187..272W}. The magenta plus shows the position of Abell cluster \object{Abell\,1035}.}
    \label{fig:RXGCCimages_MCXC J1700.7+6412}
\end{figure*}

\subsection{Cross-match between RXGCC  and MCXC-II} \label{app:rxgcc}

\noindent{\bf \object{MCXC\,J0056.0-3732}:} This cluster at $z=0.163$ is close to another object at $z=0.056$, \object{Abell\,S\,102} (see Fig. \ref{fig:RXGCCimages_MCXC J0056.0-3732}). There is an RXGCC detection very close to \object{MCXC\,J0056.0-373}, slightly off-centred due to the presence of a nearby point source, and with an aperture radius that is large enough to also cover \object{Abell\,S\,102}. Of the two redshift peaks found by RXGCC, the one assigned to the cluster ($z=0.056$) corresponds to the galaxies situated around \object{Abell\,S\,102}, not to those around \object{MCXC\,J0056.0-3732}. Due to the difference in redshifts, the RXGCC catalogue does not include the match with \object{MCXC\,J0056.0-3732}. However, taking into account that the X-ray emission producing the RXGCC detection is from \object{MCXC\,J0056.0-3732}, we have associated them. We note that the RXGCC redshift is not correct and that its mass estimation, even with the correct redshift, may still be inaccurate due to the large aperture radius that includes \object{Abell\,S\,102}.\\

\noindent{\bf \object{MCXC\,J0909.1+1059}:} There are two likely pre-merging clusters visible in the \xmm\ image at very close redshifts: $z\sim0.175$ (main contribution) and $z\sim0.165$ (fainter). Two ACT clusters are approximately centred at the two X-ray peaks: \object{ACT-CL\,J0909.2+1058} ($z_{\rm spec}=0.1764$) and \object{ACT-CL\,J0908.8+1102} ($z_{\rm spec}=0.1636$). NORAS, BSC, EMSS, and RXGCC all detect the system as a single object covering both clusters, so we have decided to associate all of these detections. However, RXGCC provides a redshift of $z=0.165$, while the other catalogues give $z=0.175$. We have taken NORAS as the primary detection, as its redshift corresponds to the main X-ray contribution. The RXGCC redshift is likely different due to its being photometric, and it may have contamination from the secondary cluster. \\

\noindent{\bf \object{MCXC\,J0956.4-1004}:} this is the supercluster \object{Abell\,0901}-\object{Abell\,0902}, which has three components at $z = 0.163$: \object{Abell\, 0901a} (N-E), \object{Abell\,0901b} (N-W) and \object{Abell\,0902} (S). It is detected by REFLEX and RXGCC. The REFLEX position is in between the three components, with an aperture radius that covers all the emission. The RXGCC position is centred on component \object{Abell\,0901a} with a smaller aperture radius that does not cover all the emission. Since the morphology is different from an AB model, the derived $\Lv$ is not accurate.  According to \cite{Gilmour2007}, the main diffuse emission is around \object{Abell\,0901b}, and there is also a fainter diffuse emission around \object{Abell\,0902}. However, the emission in the region of \object{Abell\,0901a} is a very bright point source rather than an extended emission. Therefore, the RXGCC detection seems to be driven by a point source and, for this reason, we have taken the REFLEX measurements as prime. \\

\noindent{\bf \object{MCXC\,J1032.2+4015}:} There are two nearby clusters visible in the XMM image (see Fig. \ref{fig:RXGCCimages_MCXC J1032.2+4015}): a brighter cluster to the North, where SDSS shows many galaxies with $z_{spec}\sim0.079$, and a fainter cluster to the South, where SDSS shows many galaxies with $z_{spec}\sim0.067$. Both are detected as a single object by NORAS, eBCS, and RXGCC, so we have decided to associate all these detections. However, RXGCC provides a redshift of $z=0.068$, while the other catalogues give $z=0.079$. Since the main X-ray contribution is the cluster in the North, we have taken NORAS as the primary detection in this case.  \\

\noindent{\bf \object{MCXC\,J1601.3+5354}:} This is a line-of-sight superposition of two clusters at $z=0.065$ and $z=0.107$. They were detected as a single system by NORAS/eBCS and RXGCC, with different assigned redshifts ($z=0.1068$ for NORAS, $z=0.065$ for RXGCC). The GalWeight cluster catalog \citep{2020ApJS..246....2A} finds both clusters and provides an estimation of their mass. Taking into account these masses and the luminosity distances, it seems that the most distant cluster would contribute more to the X-ray flux. Therefore, we have kept $z=0.1068$ as the preferred redshift in MCXC-II.  \\

\noindent{\bf \object{MCXC\,J1659.6+6826}:} This detection is a collection of point sources dominated by two AGNs, which were classified as a cluster by NORAS and RXGCC with different assigned redshifts ($z=0.05$ for NORAS and $z=0.037$ for RXGCC). The two brightest point sources are a blazar at $z=0.05$ at the NORAS/RXGCC position and \object{NGC\,6289} at $z=0.036$ at the position where RXGCC measured the redshifts of the galaxies. There may be a very poor group (three members) around \object{NGC\,6289} which is a potentially extended faint source (with a very small extension of $80\arcsec$). The sum of the three brightest faint sources (from the ROSAT FSC) in the aperture radius agrees with the count rate of the NORAS cluster. \\

\noindent{\bf \object{MCXC\,J1700.7+6412}:} This system at $z=0.225$ is very close to another MCXC cluster, \object{MCXC\,J1701.3+6414}, at $z=0.45$ (see Fig. \ref{fig:RXGCCimages_MCXC J1700.7+6412}). The RXGCC detection is centred in between the two clusters and in extent it covers both of them plus a nearby point source. We have associated the RXGCC cluster (which has an assigned redshift of $z=0.239$) to \object{MCXC\,J1700.7+6412}, because it is the stronger contributor to the X-ray emission. Since the RXGCC detection includes multiple sources, its flux and mass estimation may be inaccurate. \\

\subsection{Revised redshift} \label{app:zrev}

We discuss below the rationale for the redshift revision of certain clusters, focussing on cases where the revision is particularly large (Sect.~\ref{sec:zrev} and Table~\ref{tab:zrev}). \\

\noindent{\bf \object{MCXC\,J0016.3-3121}:} 
This is a SGP cluster identified with \object{Abell\,2571}, located at $2.2\arcmin$ from the X-ray centre ($0.3\Rv$). The catalogue redshift is based on the long-slit spectroscopic measurement of one galaxy close to the X-ray centre ($1.9\arcmin$) by \citet{1994MNRAS.269..151D}. The study by \citet{2002MNRAS.329...87D} of known clusters within the 2dF Galaxy Redshift Survey gives a higher value,  $z\,=\,0.106$, based on 36 redshifts. We also cross-identified \object{MCXC\,J0016.3-3121} with \object{WHY\,J001617.2-311938} (at 2.3\arcmin\ offset) and \object{RXGCC\,16}, both at consistent redshift, $\zs=0.1060$ and  $z\,=\,0.1065$, respectively.  In view of the redshift difference the original measurement is likely that of a foreground galaxy. \\

\noindent{\bf \object{MCXC\,J0019.6+2517}:} The catalogue redshift $z\,=\,0.1353$ is based on the dedicated NORAS follow-up \citep{2000ApJS..129..435B}.  There are two WHL clusters very close to the X-ray centre:  \object{WHL\,J001941.3+251807} ($z\,=\,0.3657$ from five galaxies) at $0.85\arcmin\ (0.22\Tv)$ and \object{WHL\,J001939.4+251647}  ($z\,=\,0.1336$ from two galaxies) at $0.66\arcmin\ (0.22\Tv)$. Their respective richnesses are $\RL=130$ and $\RL=23$. 
The \xmm\ archive image shows a regular single cluster, with more than 21 DR17 galaxies at $z\sim0.36$ within the X-ray contours, and two galaxies at $z\sim0.14$. \object{MCXC\,J0019.6+2517} is clearly a superposition along the line of sight of a rich cluster at $z\,=\,0.365$,  and a more nearby group at $z\,=\,0.1353$. \object{MCXC\,J0019.6+2517} is also detected at high ${\rm S/N}$ by \planck\ and is coincident at $0.85\arcmin$ distance with \object{RM\,J001941.3+251807.3} at $\zp=0.3709$. We thus revised the redshift to $z\,=\,0.365$, that of the likely main component.    \\

\noindent{\bf \object{MCXC\,J0210.4-3929}:}
\object{MCXC\,J0210.4-3929} is a 160SD/WARPS cluster. It is a case of redshift mismatch listed by \citet{2011A&A...534A.109P}, who cross-identified  \object{WARP\,J0210.4-3929}  at $z\,=\,0.273$ in the WARPS catalogue \citep{2002ApJS..140..265P} with \object{RXJ\,J0210.4-3929} (\object{VMF\,25}) at  $z\,=\,0.165$ in the 160SD catalogue \citep{2003ApJ...594..154M}. The close-by system \object{MCXC\,J0210.2-3932}, located 4.1\arcmin\ distance to the South-West, is also a case of redshift mismatch between \object{WARP\,J0210.2-3932} (uncertain $z\,=\,0.19$ ) and \object{[VMF98]\,024} ($z\,=\,0.168$).  \citet{2011A&A...534A.109P} adopted the 160SD values: $z\,=\,0.168$ (\object{MCXC\,J0210.4-3929}) and $z\,=\,0.165$ (\object{MCXC\,J0210.2-3932}), while NED assigns \object{MCXC\,J0210.4-3929} a redshift of $z\,=\,0.306$ from cross-identification with the \citet[][X-CLASS catalogue]{2017MNRAS.468..662R}. 

 The field was observed with \xmm\ and both objects are in the X-CLASS catalogue:  \object{X-CLASS\,2078} (\object{MCXC\,J0210.4-3929}) at  $z\,=\,0.306$ from eight galaxies and \object{X-CLASS\,2079} (\object{MCXC\,J0210.2-3932}) at  $z\,=\,0.166$ from three galaxies. Most of the redshifts are from the [MLF2006] catalogue \citep{2006ApJ...646..133M}. An overlay of the position of the spectroscopic galaxies from the X-CLASS database onto the \xmm\ image clearly confirms the respective redshifts of the two MCXC clusters. However, one of the [MLF2006] galaxies at $z\,=\,0.1655$ (i.e. a \object{MCXC\,J0210.2-3932} member) is very close to the centre of  \object{MCXC\,J0210.4-3929}. Therefore it is likely that the 160SD redshift was based on measurement of this foreground galaxy.  \\

\noindent{\bf \object{MCXC\,J0507.7-0915}:} 
This is a REFLEX cluster cross-identified with \object{Abell\,0536}, at $1\arcmin$ distance. The catalogue redshift, $z\,=\,0.0398$, taken from \citet[]{1999ApJS..125...35S}, is  based on one galaxy measurement \citep{1988A&A...206...27V}. We cross-identified the cluster with \object{WHY\,J050744.5-091526} ($z\,=\,0.14830$)  at  $1\arcmin$ distance and \object{RXGCC\,196} at a similarly consistent redshift ($z\,=\,0.1455$). The redshift histogram in the RXGCC database shows a prominent peak at that redshift, with galaxies concentrated inside the X-ray contours.  There is no galaxy peak around the REFLEX redshift value, which is likely that of a foreground galaxy.\\

\noindent{\bf \object{MCXC\,J1011.4+5450}:}
This  is a 400SD cluster, matching the 160SD cluster, \object{[VMF98]\,086},  at $ z\ =\ 0.294$  from  the dedicated 160SD follow-up  \citep{2003ApJ...594..154M}.  It is also a WARPS cluster,  \object{WARP\,J1011.5+5450},  given at the same redshift \citep{2008ApJS..176..374H}.
 
 This is one of the cases of mismatch between the MCXC redshift and redMaPPer photometric redshift ($\zp=0.3841$ from \citealt{2015MNRAS.453...38R}) identified by \citet{2014ApJ...783...80R}. The RM cluster,  \object{RM\,J101135.7+545005.3}, corresponds to \object{WHL\,J101135.7+545005}  with $\zs=0.3798$ from one galaxy, confirming the photometric redshift. The ROSAT/PSPC image shows a bimodal structure, with a secondary, much fainter, peak at $\sim2.4\arcmin$ to the West. The WARPS iso-contours \citep[][their Figure~2]{2008ApJS..176..374H} and position correspond to that of the main peak, while the published 400SD position lies between the two peaks. The spectroscopic galaxy \object{SDSS\,J101135.73+545005.0} ($z=0.3797\pm0.0001$ from the SDSS-DR17 release), the central galaxy of both the WHL and RM clusters in view of its position,  is very close ($0.21\arcmin$) to the X-ray maximum,  further supporting the MCXC-SDSS cross-identification.  The second closest  SDSS spectroscopic galaxy is a brighter object at $z\,=\,0.2956$, located at the western border of the secondary peak. This is likely the galaxy upon which the 400SD redshift is based, explaining the mismatch. \\

\noindent{\bf \object{MCXC\,J1016.6+2448}:}
\object{MCXC\,J1016.6+2448} is a NORAS  cluster cross-identified with \object{Abell\,0964}, located at $0.5\arcmin$ distance. The catalogue redshift, $z\,=\,0.0811$  is from dedicated follow-up but the number of galaxies is not specified. The cross-match with  SDSS cluster catalogues gives \object{WHL\,J101636.5+244803}, a rich cluster ($\RL=83.1$) at $0.45\arcmin$ distance with $z\,=\,0.1732$ (from 21 galaxies), and \object{RM\,J101633.5+244843.9}  at $0.63\arcmin$ distance with  $\zs=0.1788$ from BCG, obviously the SDSS detection of \object{Abell\,0964}. The NORAS value is likely that of foreground galaxies. \\

\noindent{\bf \object{MCXC\,J1017.5+5934}:} 
This is a NORAS  cluster cross-identified with \object{Abell\,0959}, located at $1.3\arcmin$ from the X-ray detection. The catalogue redshift, $z\,=\,0.353$  is based on two galaxy redshifts from \citet{1990ApJ...365...66H}.  This is one of the cases of mismatch between the MCXC redshift and redMaPPer photometric redshift, identified by \citet{2014ApJ...783...80R} ($\zp=0.2891$ from \citet{2015MNRAS.453...38R}). This lower value 
is confirmed by the  dynamical analysis of \citet{2009A&A...495...15B} and the  cross-identification with \object{WHL\,J101734.3+593340} at $0.37\arcmin$,  which lies at $z\,=\,0.2880$ based on 14 galaxies \citep[][]{2015ApJ...807..178W}. \\

\noindent{\bf \object{MCXC\,J1159.2+4947}:} 
\object{MCXC\,J1159.2+4947} is a NORAS cluster identified with \object{Abell\,1430} (located at a distance a $1.85\arcmin$ from the X-ray centre). Its redshift is $\zs=0.211$, taken from \cite{1999ApJS..125...35S}, which refers to \cite{1984ATsir1344....1K}. It is based on two galaxies. 
\object{MCXC\,J1159.2+4947} is located $0.45\arcmin$ away from \object{RM\,J115914.9+494748.4} ($\zs=0.3501$ from BCG) and $0.45\arcmin$ away from \object{WHL\,J115914.9+494748} ($\zs=0.3486$ based on 13 galaxies, richness $\RL=138$). An SDSS search within the X-ray contours yielded 29 spectroscopic redshifts including the BCG. \object{RXGCC\,437} is located at $0.94\arcmin$ distance and shows a clear redshift peak on the line-of-sight at $z=0.3435$, with a second peak at $z\,=\,0.08$. Three galaxies at $z\sim0.2$, the NORAS catalogue value,  are also visible. 
The original NORAS redshift is likely that of the two foreground galaxies. We adopted the redshift $\zs=0.3486$  from \citet[][WHL15]{2015ApJ...807..178W}.\\

\noindent{\bf \object{MCXC\,J1340.9+3958}:}
This 160SD/400SD cluster was cross-identified by \citet{2003ApJ...594..154M} with \object{Abell\,1774}, with a redshift $z=0.1690$ from \citet{1984ATsir1344....1K}.  This publication is not available on ADS but \citet{1991ApJS...77..363S},  quoting this reference for \object{Abell\,1774}, specifies that the redshift is based on two galaxies. However, \object{Abell\,1774} is located 3.5\arcmin\ away from the X-ray peak, while \object{WHL\,J134053.6+395755} is closer, at $0.56\arcmin$, with $z= 0.2772$ from two galaxies. \object{WHL\,J134053.6+395755} is also in the redMaPPer catalogue as \object{RM\,134053.6+395754.8} at the same redshift. The ROSAT/PSPC and SWIFT images show a small cluster extent, with  X-ray contours overlaid on the SDSS image supporting the cross-match with the RM and WHL cluster. There are more than ten galaxies with SDSS redshift $z\sim0.169$ around the \object{Abell\,1774} position, confirming its redshift. \object{Abell\,1774} is a foreground object, but it remains unclear whether it can contaminate the 160SD cluster emission. \\

\noindent{\bf \object{MCXC\,J1343.4+4053}:} 
This is a 160SD cluster with redshift $\zs \ =\ 0.14$ from \citet{2003ApJ...594..154M} based on an unknown number of galaxies. 
\object{MCXC\,J1343.4+4053} is located $0.13 \arcmin$ from \object{WHL\,J134325.6+405318} ($\zs =  0.2546$ based on two galaxies, richness $\RL=18.6$) and $0.13 \arcmin$ from \object{RM\,J134325.6+405317.7} ($\zs=0.2560$ based on the central galaxy), on which the NED and Simbad preferred redshifts are based. We adopted the redshift $\zs=0.2560$ from \citet[][redMaPPer]{2016ApJS..224....1R}. \\

\noindent{\bf \object{MCXC\,J1359.2+2758}:} This MCXC cluster corresponds to \object{Abell\,1831}, in which \cite{2004AJ....128.1558S} identified two components (\object{Abell\,1831A} and B) superposed along the line of sight. \object{Abell\,1831B} seems to be richer with a higher velocity dispersion, and thus a priori \object{Abell\,1831B} is more massive and likely to be the main contributor to the X-ray signal. We have updated its redshift to  $z=0.07507$ from \citet{2020ApJS..246....2A}, the most recent reference, based on 75 galaxies.  \\

\noindent{\bf \object{MCXC\,J1415.1+3612}:} 
The original redshift $z=0.7$ is from \citet{2002ApJS..140..265P} but mentioned by the authors as from noisy data, and the X-ray morphology indicates possible point-source contamination. The Simbad redshift is $\zs =1.027$ from \citet[][X-CLASS]{kou21} based on nine galaxies. We  changed the redshift to $\zs=1.026$ from \citet{2009ApJ...707L..12H} based on its determination from a larger number of galaxies, $Nz=25$.\\

\noindent{\bf \object{MCXC\,J1447.4+0827}:} 
This is a NORAS cluster with original redshift $z =0.1954$ from dedicated follow-up \citep{2000ApJS..129..435B}. This object is now associated with a MACS DR3 cluster located $0.3\arcmin$ away, lying at $z=0.38$ based on optical data from SDSS \citep{2012MNRAS.420.2120M} . 
\object{MCXC\,J1447.4+0827} can be matched to \planck, ACT, ComPRASS, Wen, and redMaPPer clusters. From \citet{2015ApJ...807..178W}, the closest match is \object{WHL\,144726.0+082825} at $0.6 \arcmin$ ($0.92 \arcmin$ from the MACS DR3 cluster) with $\zs=0.376$ based on four spectra. The cluster is rich with $\RL = 83.5700$ and a number of galaxies  $N=45$. 
The ACT cluster has a redshift $\zs=0.375$ based on redMaPPer, which is consistent with the redshift from the Wen et al. cluster. We changed the \object{MCXC\,J1447.4+0827} redshift to $\zs=0.376$ from \citet[][WH15]{2015ApJ...807..178W}.\\

\noindent{\bf \object{MCXC\,J1501.1+0141}:} 
The original redshift of this REFLEX cluster is $z\,=\,0.0050$ \citep{2004A&A...425..367B}, which is an average of two sources: \citet{2000MNRAS.313..469S} at $\zs\sim0.00659$ from spectroscopic data and NED information prior to 1992, without reference. 
The current NED redshift is $z\,=\,0.0065$ from RXGCC 585 \citep[][]{2022A&A...658A..59X}.  Simbad associated the cluster with \object{X-CLASS\,2301}, with  $z\,=\,0.005$ \citep[][]{kou21}. This redshift is based on $\sim 30$ galaxies over a large region, $\sim \Tv$ in radius with a biweight re-estimate of $\zs=0.00553\pm 0.00016$. However, the \xmm\ image shows that the emission is dominated by the \object{NGC\,5813} galaxy (located at 25\arcsec\ from the X-ray peak) and its associated halo. All redshift references in NED and Simbad of \object{NGC\,5813} confirm $z\sim 0.0065$, with the latest being  $z\,=\,0.00677$ from  \citet{2016MNRAS.460.1758H}. Since the X-ray emission is dominated by the \object{NGC\,5813} halo, we adopted the latter value.\\

\noindent{\bf \object{MCXC\,J1520.9+4840}:} 
This is a NORAS cluster with an original redshift $z\,=\,0.1076$. The catalogue redshift references are \citet{1987ApJS...63..543S} (based on two galaxies) and J.P.~Huchra (2000, private communication). The NED redshift is $z\,=\,0.0736$ from \object{RXGCC\,602} \citet[][]{2022A&A...658A..59X} consistent with the  Simbad spectroscopic redshift $z\,=\,0.0738$ from \citet{2006AJ....132.1275R}. We adopted the more recent and well-matched spectroscopic measurement $\zs=0.074$ from \citet[][based on $Nz=38$ galaxies]{2007MNRAS.379..867V}

The cluster also matches the GalWCat cluster \#103 at $z=0.0735$ from 24 galaxies.\\

\noindent{\bf \object{MCXC\,J1544.0+5346}:} 
This is a 160SD cluster with an original redshift $\zs=0.112$ from \citet{2003ApJ...594..154M}. \citet{2011A&A...534A.120T} associated this object with an extended source (35 arcsec extent) from the 2XMM catalogue and with \object{WHL\,J154407.3+534657} at $\zs=0.4970$. In Simbad, \object{MCXC\,J1544.0+5346} is also associated with \object{X-CLASS\,638} at  $\zs \ =\ 0.50$ from SPIDERS spectroscopy \citep[][four galaxies]{kou21}. There is an AGN (\object{SDSS\,J154407.18+534559.6}, \object{2XMM\,J154407.2+534600}) at $z\,=\,0.10911$ \citep{2016MNRAS.455.2551N} very close to the 160SD cluster centre, and a few galaxies around that redshift in the image. Presumably, the 160SD follow-up identified this galaxy as the cluster BCG, hence the incorrect redshift. One can also note that the original $\zp$ from \citet{1998ApJ...502..558V} was $\zp=0.33$. We adopted the redshift $\zs=0.50$ from \citet{kou21}.\\

\noindent{\bf \object{MCXC\,J1730.4+7422}:}
The redshift of $\zs=0.110$ \citep{2000ApJS..129..435B} is from a dedicated follow-up and is thus {\it a priori} better than the previous redshift $z\,=\,0.041$ from \citet{1998ApJS..117..319A}. However the cluster can be associated with \object{RXGCC\,724}  \citep{2022A&A...658A..59X} at $1.12 \arcmin$ distance, with $z\,=\,0.0470$. The histogram of redshifts from RXGCC shows only one galaxy at $z\,=\,0.1$ and $\sim 5$ galaxies around $z\,=\,0.0470$ located near the centre of the X-ray contours. The \object{NGC\,6414} galaxy at $z\,=\,0.054$ is located $0.5 \arcmin$ from the MCXC object and is an optical AGN \citep{2019ApJ...872..134Z}. The cluster was also imaged by \xmm. The emission is dominated by this `BCG' galaxy and it may be possible that the NORAS follow-up includes background objects.
We changed to redshift to $z\,=\,0.047$  based on \citet{2022A&A...658A..59X}.\\

\noindent{\bf \object{MCXC\,J2032.1-5627}:} 
This is a REFLEX cluster with original redshift $\zs=0.138$ in \citet{2011A&A...534A.109P}. From the dedicated spectroscopic follow-up programme for the  REFLEX clusters, there are in fact three different cluster redshifts for the given cluster position \citep{2009A&A...499..357G}: (i) $\zs=0.08025$, the redshift obtained from a low-S/N spectrum; (ii) $\zs=0.138$ is an average of five galaxy redshifts; (iii) $\zs=0.285$ is an average of six galaxy redshifts. The cluster is matched to \object{SPT-CL\,J2032-5627} at the robust redshift $\zs\,=\,0.284$ obtained from 32 cluster member spectra, as reported by \citet{2012ApJ...761...22S}. The authors describe it as a superposition of structures along the same line of sight. More details can be found in the paper's appendix.   
Given the above, we have adopted  $\zs\ =\ 0.284$ from \citet{2012ApJ...761...22S}.\\

\noindent{\bf \object{MCXC\,J2135.2+0125}:} 
This is a REFLEX cluster cross-identified with  \object{Abell\,2355}, located at 2.7\arcmin\ separation. The original redshift $z\,=\,0.1244$,  with reference  \citet{1999ApJS..125...35S}, is based on two galaxies.  This is one of the cases of mismatch identified by \citet{2014ApJ...783...80R}, between the MCXC redshift and redMaPPer photometric redshift ($\zp=0.234$).  The RM cluster, \object{RM\,J213518.8+012527.0}, corresponds to \object{WHL\,J101135.7+545005}, both located at 0.6\arcmin from the X-ray position, at $\zs=0.2301$. This value is consistent with the redshift of the BCG of the RM cluster ($\zs=0.2306$).  \object{MCXC\,J2135.2+0125} is also in the \planck\ and  ACT catalogues and matches RXGCC\,853 (z\ =\ 0.2324).  The redshift histogram of \object{RXGCC\,853} shows two clear peaks,   one at $z\,=\,0.1187$ (the galaxies are mostly in the south, but one galaxy is very close to the Abell position) and the other at $z\,=\,0.2324$ with galaxies centred on the X-ray cluster. 

The REFLEX redshift is likely that of foreground galaxies.  The WHL redshift is based on two galaxies. More redshifts are now available from SDSS-DR17, and we, therefore, re-estimated the cluster redshift to be $ z\ =\ 0.229\pm0.004$ from ten spectroscopic galaxies. \\

\noindent{\bf \object{MCXC\,J2306.5-1319}:} 
This is a REFLEX cluster with an original redshift $z=0.0659$ based on two galaxies, estimated from the dedicated ESO Key Project follow-up. \object{MCXC\,J2306.5-1319} also matches  \object{RXGCC\,913} at a consistent $z=0.0683$. We identified a potential counterpart,  \object{WHY\,J230621.3-131552}, at  $4.9\arcmin$ but at a discrepant redshift of $z\,=\,0.1095$. \object{WHY\,J230621.3-131552} matches \object{Abell\,2529}, at $z=0.11$  from \citet{1999ApJS..125...35S}.   
\object{MCXC\,J2306.5-1319} is also close to the SGP cluster, \object{MCXC\,J2306.8-1324}, south-east of  \object{MCXC\,J2306.5-1319}, at  $z=0.0659$. This redshift is based on two galaxies with a reference given as {\it H. Boehringer \& L. Guzzo, 1999, ESO Key Project, private communication}, i.e. likely the same data as that used for \object{MCXC\,J2306.5-1319}.

The \xmm\ observation, together with the information in the RXGCC database allows us to disentangle this case. There are two clusters, clearly separated in the \xmm\ image,  \object{WHY\,J230621.3-131552} (A2529)  in the north at  $z\,=\,0.1095$, and the SGP cluster in the south-east.  The \xmm\ emission of the latter is much fainter than that of the former. The position of the REFLEX cluster is in between, closer to  \object{WHY\,J230621.3-131552}, with an extraction aperture encompassing both clusters.  The \object{RXGCC\,913} ROSAT image does not resolve the two clusters either, but the maximum is at the WHY cluster location.  The redshift histogram shows two peaks. A main peak is at $z\sim0.0683$ with galaxies located in the south, consistent with the SGP redshift. There is a second peak at $z=0.1077$, with galaxies concentrated in the north, around the WHY cluster position. 
In conclusion, this confirms the position and redshift of the SGP cluster,  \object{MCXC\,J2306.8-1324}, which may be however contaminated by the more massive cluster in the North. The REFLEX (and RXGCC) detection is a confusion between the two objects at different redshifts but is dominated by the brighter component in the north.  We thus changed the redshift to  $z\,=\,0.1095$.\\

\noindent{\bf \object{MCXC\,J2334.0+0704}:} 
This is a NORAS cluster initially associated with \object{Abell\,2620}. The redshift is $z\,=\,0.099$ from \citet{1999ApJS..125...35S} with reference  \citet{1993AstL...19..198F}. 
We note that the Abell position is about $9.7\arcmin$ away from the NORAS position, so the initial association is probably not correct. We found 20 galaxy redshifts from SDSS at $z \sim 0.29$ matching with the X-ray contours within 6\arcmin\ and more than half of these within 3\arcmin of the X-ray peak. The average $z\,=\,0.295$\ (0.2952$\pm$0.0022) based on these 20 galaxy redshifts was adopted for \object{MCXC\,J2334.0+0704}.\\

\noindent{\bf \object{MCXC\,J2341.1+0018}:}
This is a NORAS cluster with $z=0.11$ from dedicated follow-up, with reference \citet{1998A&AS..129..399K}. The number of galaxies is not specified. We cross-identified the cluster with  \object{WHL\,J234106.9+001833} at 0.36\arcmin separation, lying at $z=0.2768$ from four galaxies ($ R_L= 41$). This object also corresponds to a rich cluster in the RM catalogue,  \object{RM\,J234105.4+001815.2} at consistent $ \zp=0.2928$.  From the latest SDSS data, there are 18 spectroscopic redshifts including the BCG within the X-ray contours, with an average $z=0.2757\pm0.0031$, in agreement with the WHL value. The NORAS z estimate was likely based on foreground objects.

\section{$K$, $\KL$  and aperture corrections}\label{ap:kcor}

\subsection{$K$-correction factors}

Table~\ref{tab:K} Details the $K$-correction as a function of redshift and temperature for the $[0.1-2.4]$ and $[0.5-2]$~keV energy bands. Table~\ref{tab:KL} gives the $\KL$-correction factor,  the ratio of the luminosity in $\bando$ energy band to that in the $\bandi$ band, as a function of the temperature.

 \begin{table}[!ht]
    \caption{\label{tab:K}  $K$-correction as a function of redshift for four different temperatures ($1, 2, 4$ and $10 $ \keV) and for two energy bands of the ROSAT input flux, $\bandi$ and $\bando$.}

 \resizebox{\columnwidth}{!} {
\begin{tabular}{ ccccccccccc}
\toprule
\toprule
        \multicolumn{1}{c}{{$z$}} &
        \multicolumn{1}{c}{} &
    \multicolumn{4}{c}{{  $[0.1$--$2.4]$ \keV  }} & 
       \multicolumn{1}{c}{} &
    \multicolumn{4}{c}{{  $[0.5$--$2.0]$ \keV  }}  \\
\cmidrule{3-6} 
 \cmidrule{8-11} 
        \multicolumn{1}{c}{{  }} & 
        \multicolumn{1}{c}{{  }} & 
   \multicolumn{1}{c}{{1.0 }} &  
   \multicolumn{1}{c}{{2.0}} &  
        \multicolumn{1}{c}{{ 4.0 }} & 
        \multicolumn{1}{c}{{ 10.0 }} & 
         \multicolumn{1}{c}{{  }} & 
  \multicolumn{1}{c}{{1.0}} &  
  \multicolumn{1}{c}{{2.0}} &  
        \multicolumn{1}{c}{{ 4.0 }} & 
        \multicolumn{1}{c}{{ 10.0 }} \\
\midrule 
 0.00 &  & 1.000 & 1.000 & 1.000 & 1.000 &  & 1.000 & 1.000 & 1.000 & 1.000\\
 0.01 &  & 1.000 & 0.997 & 0.995 & 0.994 &  & 1.000 & 0.995 & 0.995 & 0.993\\
 0.02 &  & 1.005 & 0.994 & 0.991 & 0.987 &  & 0.998 & 0.990 & 0.989 & 0.987\\
 0.05 &  & 1.007 & 0.984 & 0.978 & 0.970 &  & 1.000 & 0.983 & 0.977 & 0.969\\
 0.10 &  & 1.025 & 0.973 & 0.955 & 0.941 &  & 1.004 & 0.978 & 0.959 & 0.942\\
 0.15 &  & 1.036 & 0.963 & 0.936 & 0.915 &  & 1.011 & 0.976 & 0.943 & 0.917\\
 0.20 &  & 1.039 & 0.956 & 0.920 & 0.890 &  & 1.019 & 0.973 & 0.929 & 0.894\\
 0.30 &  & 1.051 & 0.940 & 0.890 & 0.849 &  & 1.028 & 0.968 & 0.905 & 0.853\\
 0.40 &  & 1.061 & 0.929 & 0.864 & 0.812 &  & 1.062 & 0.970 & 0.884 & 0.818\\
 0.50 &  & 1.067 & 0.921 & 0.842 & 0.779 &  & 1.096 & 0.972 & 0.867 & 0.787\\
 0.60 &  & 1.072 & 0.914 & 0.823 & 0.751 &  & 1.156 & 0.978 & 0.853 & 0.760\\
 0.70 &  & 1.078 & 0.909 & 0.806 & 0.726 &  & 1.236 & 0.978 & 0.838 & 0.735\\
 0.80 &  & 1.084 & 0.905 & 0.792 & 0.703 &  & 1.386 & 0.992 & 0.830 & 0.715\\
 0.90 &  & 1.091 & 0.904 & 0.780 & 0.684 &  & 1.607 & 1.002 & 0.820 & 0.696\\
 1.10 &  & 1.106 & 0.903 & 0.760 & 0.649 &  & 2.338 & 1.051 & 0.810 & 0.663\\
 1.27 &  & 1.119 & 0.902 & 0.743 & 0.621 &  & 3.830 & 1.098 & 0.802 & 0.637\\
\bottomrule
\end{tabular}
}

\end{table}

 \begin{table}[!h]
    \caption{ \label{tab:KL}  $\KL$-correction factor,  the ratio of the luminosity in $\bando$ energy band to that in the $\bandi$ band, as a function of the temperature, $T$,  in $\keV$.  }
\centering
\begin{tabular}{ cc}
\toprule
\toprule
 $\log(T)$  & $\KL$ \\
\midrule
-1.00 & 5.4916 \\
-0.95 & 4.6962 \\
-0.90 & 4.0376 \\
-0.85 & 3.4923 \\
-0.80 & 3.0408 \\
-0.75 & 2.6670 \\
-0.70 & 2.3576 \\
-0.65 & 2.1018 \\
-0.60 & 1.8904 \\
-0.55 & 1.7159 \\
-0.50 & 1.5720 \\
-0.45 & 1.4534 \\
-0.40 & 1.3559 \\
-0.35 & 1.2757 \\
-0.30 & 1.2098 \\
-0.40 & 1.3559 \\
-0.30 & 1.2098 \\
-0.20 & 1.1113 \\
-0.10 & 1.0449 \\
-0.00 & 1.0000 \\
 0.10 & 0.9696 \\
 0.20 & 0.9491 \\
 0.30 & 0.9351 \\
 0.40 & 0.9257 \\
 0.50 & 0.9195 \\
 0.60 & 0.9156 \\
 0.70 & 0.9138 \\
 0.80 & 0.9138 \\
 0.90 & 0.9138 \\
 1.00 & 0.9138 \\
 \bottomrule
\end{tabular}
 \end{table}
\newpage
\subsection{Aperture correction}\label{ap:apcor}

 The luminosity within a given radius $R$ scales as 
\begin{eqnarray}
 \label{eq:ycyl}
L(<R) &\equiv&\int_{0}^{R} 2\pi r dr \int_{r}^{R_{\rm max}} \frac{2\,\rho_{\rm gas}^2(r')r'dr'}{\sqrt{r'^2 -r^2}}  \\
& = &  \int_{0}^{R_{\rm max} }4\pi \rho_{\rm gas}^2r^2dr-\int_{R}^{R_{\rm max}} 4\pi\,\rho_{\rm gas}^2\sqrt{r^2 -R^2}rdr  \nonumber \\
 \end{eqnarray}
\noindent where $ \rho_{\rm gas}$ is the gas density profile and  $R_{\rm max}$ is the cluster radial extent. We  adopted $R_{\rm max}=5\Rv$. For a universal gas profile:
\begin{equation}
\rho_{\rm gas}^2(r) = A\,F_{\rm g}(x) ~~~{\rm with } ~~~~ x   =  r/\Rv 
\end{equation}
and defining :
\begin{eqnarray}
I(x)\!&=&\!\int_{0}^{x}F_{\rm g}(u)\,u^2\,du   \\
J(x)\!&=&\int_{x}^{5} F_{\rm g}(u)\,\sqrt{u^2-x^2}\,u\,du  
\end{eqnarray}
The aperture correction is:
\begin{eqnarray}
\frac{\Lv}{\Lap} &=& f_{\rm ap}(x_{\rm ap}=\Rap/\Rv)  \\
f_{\rm ap}(x) & = & \frac{I(5)-J(1)}{I(5)-J(x)}
 \end{eqnarray}
The numerical values for the \rexcess\ profile:
\begin{equation}
\rho_{gas} \propto  \left( \frac{x}{x_c} \right)^{-\alpha} \times \left[1+ \left( \frac{x}{x_c} \right)^{2} \right]^{-3 \beta /2  + \alpha/2} ,
\label{nmean:eq} 
\end{equation}
with $x_{\rm c} = 0.303$, $\alpha =  0.525$, and  $\beta=  0.768$ are given in the Table~\ref{tab:aperture}. 

 \begin{table}[!ht]
    \caption{ \label{tab:aperture}  Aperture correction factor,  the ratio of the luminosity within the aperture radius, $\Rap$, to that within $\Rv$, as a function of the ratio $\Rap/\Rv$.  }
\centering
\begin{tabular}{ cc}
\toprule
\toprule
 $\log(\Rap/\Rv)$  & $\Lv/\Lap$ \\
\midrule
-1.00 & 5.4916 \\
-0.95 & 4.6962 \\
-0.90 & 4.0376 \\
-0.85 & 3.4923 \\
-0.80 & 3.0408 \\
-0.75 & 2.6670 \\
-0.70 & 2.3576 \\
-0.65 & 2.1018 \\
-0.60 & 1.8904 \\
-0.55 & 1.7159 \\
-0.50 & 1.5720 \\
-0.45 & 1.4534 \\
-0.40 & 1.3559 \\
-0.35 & 1.2757 \\
-0.30 & 1.2098 \\
-0.40 & 1.3559 \\
-0.30 & 1.2098 \\
-0.20 & 1.1113 \\
-0.10 & 1.0449 \\
-0.00 & 1.0000 \\
 0.10 & 0.9696 \\
 0.20 & 0.9491 \\
 0.30 & 0.9351 \\
 0.40 & 0.9257 \\
 0.50 & 0.9195 \\
 0.60 & 0.9156 \\
 0.70 & 0.9138 \\
 0.80 & 0.9138 \\
 0.90 & 0.9138 \\
 1.00 & 0.9138 \\
\bottomrule
\end{tabular}
\end{table}

\subsection{Iteration convergence}\label{ap:iter}
 \begin{figure}
 \includegraphics[width=\columnwidth]{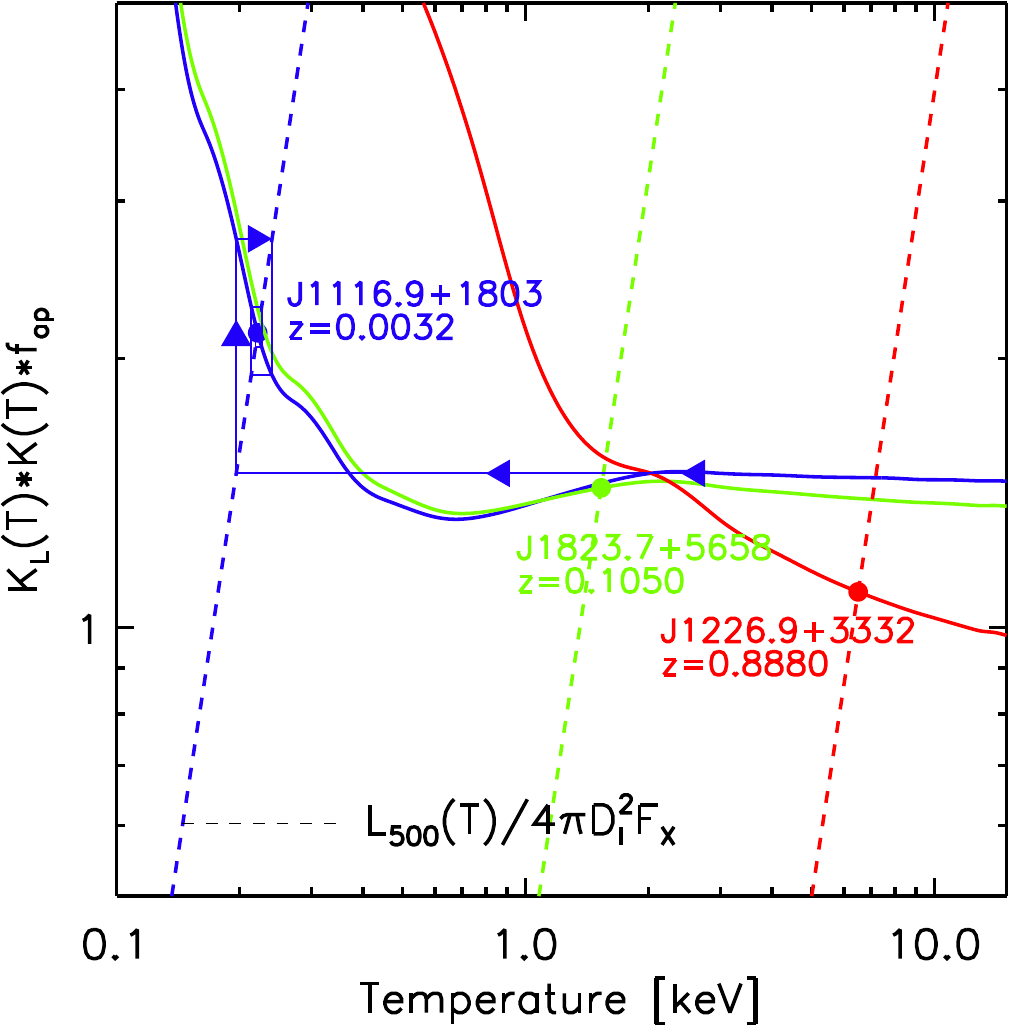}
 \caption{Illustration of the determination of the temperature for three 400SD clusters. The dashed line is the  $f_1(T)$ function, the luminosity corresponding to the temperature, $T$,  given the \LT\ relation, divided by the luminosity ($4\, \pi D_{\rm l}^{2}(z)\  \Fap$) estimated from the input aperture flux without correction (Eq.~\ref{eq:Timplicit1}).
 The solid line  is the correction function $f_2(T)$ defined in Eq.~\ref{eq:Timplicit2}, the product of the K--correction, the $\KL$ correction factor used to convert the luminosity from the $\bandi$ input band to  $\bando$ band and $\Fap$ is the aperture correction (fixed in that case). Both functions depend on cluster redshift. The  curves are  color-coded in blue, green and red for the \object{MCXC\,J1116.9+1803} ($z=0.0032$), \object{MCXC\,J1823.7+5658} ($z=0.1050$), and \object{MCXC\,J1226.9+3332} ($z=0.888$),  respectively.  The temperature is the intersection of the two curves (points). Starting from $T=3$ \keV, the iterative trajectory  is shown in a thin blue line with arrows for the cluster \object{MCXC\,J1116.9+1803} ($z=0.0032$). We note the rapid convergence.}
 \label{fig:iter}
 \end{figure}

We discuss here the convergence of the iterative process described in Sect.~\ref{sec:method}, which is used to determine the temperature, mass and luminosity. The convergence stems from the properties of the implicit equation that must verify the temperature. Combining the Eqs. \ref{eq:KK}, \ref{eq:KL}, \ref{eq:lapl500}, which provide the estimated luminosity from the input flux, and the \LT\ relation (Eq.~\ref{eq:lt}), the implicit equation on the temperature is of the form $f_1 (T)= f_2(T)$ with:
\begin{eqnarray}
\label{eq:Timplicit1}
f_1 (T) & = &A \left(\frac{T}{5\ \keV}\right)^{\alpha_{\rm T}} \\
\label{eq:Timplicit2}
 f_2(T) & =  & K_{\rm L,[E_1-E_2]}(T)\    K_{\rm [E_1-E_2]}\ (T,z) f_{\rm ap}(R_{\rm ap}/\Rv) 
  \end{eqnarray}
where $K_{\rm L,[E_1-E_2]}(T)$ can be omitted if the input flux is given in the $\bando$ energy band.  The normalisation factor  $A$ of $f_1 (T)$ is a constant for a given cluster, determined by its flux and redshift: $A= A_{\rm T}  E(z)^{-\beta_{\rm T} }/(4\, \pi D_{\rm l}^{2}(z)\  F_{\rm ap, [E_1-E_2]})$).

The left-hand side of the equation, $f_1 (T)$,  is simply a power law with a slope of $\alpha_{\rm T}=3$, which is a very steeply increasing function of the temperature. The right hand function is slightly more complex:
 \begin{itemize}
 \item  
For a fixed  aperture correction, $f_2 (T)$ is 
essentially a decreasing function of $T$, except via the $K_{\rm L}$ factor in the $\sim[0.5$--$2]\ \keV$ band. Here  $K_{\rm L}$ increases mildly with $T$ but much less steeply than the $f_1(T)$ function (see Fig.~\ref{fig:iter}).  In the typical cluster temperature range, $0.1$--$15\,\keV$,  there is only one intersection between the steeply increasing $f_1(T)$ and the mostly decreasing $f_2(T)$ function, spanning a smaller range of variation. In other words, the $T$ solution is unique. Furthermore, the convergence is rapid and the starting $T_0$ point is not critical. After the first iteration, the temperature, $T_1$ is where $f_1 (T) \propto T^3$ is equal to $f_2(T_0)$, already close to the final value. The next iterations simply refine the $T$ value,  due to the milder variations of $f_2 (T)$ as compared to $f_1 (T)$. This is illustrated in Fig.~\ref{fig:iter}, where we plot the  $f_1 (T)$ and $f_2 (T)$ functions for three 400SD clusters, at very low redshift, median redshift and high redshift. For this catalogue,  the aperture correction is fixed but we have to apply a $K_{\rm L}$ correction.   
\item  
For cases of available aperture flux, the aperture correction, $f_{\rm ap}(R_{\rm ap}/\Rv) =  \Lv /\Lap$ in $ f_2(T)$  is itself determined iteratively and depends on the temperature of the iteration step in question. This dependence is via the computation of the current $\Lap$, which appears in the corresponding implicit equation that has to be solved for $ \Rv$. Combining Eq.~\ref{eq:KK} and Eq.~\ref{eq:KL}, which determines $\Lap$ from the input flux,  and  the scaling relation in Eqs. \ref{eq:lm} and  \ref{eq:r500}, we must solve: 
\begin{eqnarray}
f_{\rm ap}(R_{\rm ap}/\Rv)  =   B\  \Rv^{3*1.675} / (K_{\rm L}(T)\    K (T,z))
\end{eqnarray}
where $B$ is a constant for a given cluster.   The solution always exists due to the restricted range of the aperture variation  (see Table~\ref{tab:aperture}) as compared to that of the right-hand power law. It is unique as the left hand and right hand functions in the equation increase monotonically with increasing $\Rv$.   Generally, their intersection moves to the right in the log-log plane (that is, to higher $\Rv$ and aperture correction) for a higher iteration step temperature, as the normalisation of the power law increases (lower $K$ corrections).  The convergence is again rapid and is only weakly dependent on the starting point due to the steepness of the right hand power law.  Note that this is this second iteration on $\Rv$ which in practice fixes the next iteration step temperature via the \LM\ and \LT\ scaling relations with the corresponding $\Mv$. 
 \end{itemize}

\newpage

\section{OBSID of \xmm\ observations}\label{app:xmmobsid}

Table~\ref{tab:OBSID} gives the \xmm\ OBSIDs that were used in catalogue validation.

\begin{table}[!h]
    \caption{ \label{tab:OBSID}  List of \xmm\ observations used in this work. Col. 1: MCXC-II cluster; Col. 2: Name of the target of the observations; Col. 3: OBSID.  }
\centering
\begin{tabular}{lrl}
\toprule
\toprule
MCXC-II name  & Target name   & OBSID \\
\midrule
      J0019.6+2517 &    PSZ2G113.91-37.01 &  0827021001 \\ 
      J0034.2-0204 &      RXCJ0034.6-0208 &  0720250401 \\ 
      J0040.0+0649 &              Abell 76 &  0405550101 \\
      J0056.0-3732 &  NGC 300              &  0305860301 \\
      J0125.4+0145 &             A 189    &  0109860101 \\ 
      J0152.9-1345 &             NGC 720  &  0602010101 \\ 
      J0159.0-3412 &    PSZ2G243.15-73.84 &  0827011301 \\ 
      J0210.4-3929 &       RXJ0210.4-3929 &  0401130101 \\ 
      J0236.0-5225 &             WW Hor   &  0098810101 \\
      J0338.4-3526 &  NGC 1399             &  0400620101 \\ 
      J0345.7-4112 &      RXCJ0345.7-4112 &  0201900801 \\ 
      J0416.7-5525 &  NGC1549              &  0205090201 \\
      J0553.4-3342 &    PSZ2G239.27-26.01 &  0827010401 \\ 
      J0748.1+1832 &  RXCJ0748.1+1832      &  0651780201 \\
      J0823.1+0421 &              ZwCl1665 &  0741580501 \\
      J0857.7+2747 &             3C 210   &  0210280101 \\
      J0900.6+2054 &                Z2089 &  0402250701 \\
      J0909.1+1059 &    PSZ2G218.81+35.51 &  0827351301 \\ 
      J0926.7+1234 &          PG 0923+129 &  0783270401 \\ 
      J0956.4-1004 &  Abell 901            &  0148170101 \\
      J1022.0+3830 &  RXCJ1022.0+3830      &  0503601301 \\
      J1032.2+4015 &             A1035    &  0653810501 \\ 
      J1058.1+0135 &          cl1058+0137 &  0601930101 \\ 
      J1253.2-1522 &             NGC 4756 &  0551600101 \\ 
      J1310.4+2151 &      RXCJ1310.9+2157 &  0841900201 \\ 
      J1311.7+2201 &             Z5768    &  0402250301 \\ 
      J1314.4-2515 &     RXCJ 1314.4-2515 &  0501730101 \\ 
      J1329.4+1143 &       NGC 5171 Group &  0041180801 \\ 
      J1330.8-0152 &  A1750                &  0112240301 \\
      J1359.2+2758 &  PSZ2G040.03+74.95    &  0827031901 \\
      J1411.4+5212 &             XMM-RM19 &  0804271501 \\ 
      J1414.2+7115 &  PLCKESZG113.82+44.35 &  0692933601 \\
      J1415.2-0030 &  MCXC J1415.2-0030    &  0762870501 \\
      J1419.3+0638 &  RXJ1419.3/RXJ1419.9 &  0303670101 \\ 
      J1419.8+0634 &  RXJ1419.3/RXJ1419.9 &  0303670101 \\ 
      J1501.1+0141 &             NGC 5813 &  0554680301 \\ 
      J1506.4+0136 &  NGC 5846             &  0723800201 \\
      J1544.0+5346 &         SBS 1542+541 &  0060370901 \\ 
      J1700.7+6412 &  HS 1700+6416         &  0723700201 \\
      J1705.1-8210 &  S0792                &  0761111801 \\
      J1730.4+7422 &             RXJ 1730 &  0014150401 \\ 
      J1736.3+6803 &  RXCJ1736.3+6803      &  0203610401 \\
      J1755.7+6752 &  PLCKESZ G098.12+30.3 &  0841950201 \\
      J1847.3-6320 &  2XMMi J184725-63172 &  0694610101 \\ 
      J1925.4-4256 &             A3638    &  0765020101 \\ 
      J2004.8-5603 &         CL 2003-5556 &  0673180401 \\ 
      J2032.1-5627 &     SPT-CLJ2032-5627 &  0674490401 \\
      J2034.3-3429 &  Abell 3693 &  0404520201 \\
      J2127.1-1209 &  Abell 2345           &  0604740101 \\ 
      J2218.2-0350 &        1E2216/1E2215 &  0800380101 \\ 
      J2256.9+0532 &  MCXC J2256.9+0532    &  0762871101 \\
      J2306.5-1319 &      RXCJ2306.6-1319 &  0765030201 \\ 
      J2306.8-1324 &      RXCJ2306.6-1319 &  0765030201 \\ 
      J2311.5+0338 &                A2552 &  0693010201 \\
      J2318.5+1842 &          ABELL 2572b &  0741580601 \\ 
      J2325.6-5443 &           XBCSM2-19b &  0677820144 \\ 
      J2359.3-6042 &      RXCJ2359.3-6042 &  0677180601 \\    
\bottomrule 
\end{tabular}
\end{table}

\section{Catalogue content}\label{ap:mcxcfield}

Table~\ref{mcxc_field} gives an overview of the 46 fields contained in the MCXC-II catalogue.

We added an individual formatted note for 59 clusters that are displayed in the column {\tt NOTES}. Each note comprises three components: 1) a simplified category code; 2) if the object is discussed in the paper, the corresponding Section location; and 3) a text note providing information concerning the cluster category code.

The category code covers eight different cases: ({\tt PLc}) -- Point-Like contamination: cluster emission contaminated by AGN (no AGN redshift available); ({\tt PLc\_intra}) -- Point-Like contamination intra: cluster emission contaminated by AGN inside the cluster; ({\tt PLc\_extra}) -- Point-Like contamination extra: cluster emission contaminated by foreground/background AGN; 
({\tt complexStr}) -- Complex X-ray structure: cluster with sub-structures and/or close-by clusters on the plane of the sky; ({\tt Dupl})	-- Duplicate: two different objects in MCXC-I identified as one single object in MCXC-II; ({\tt losStr}) -- line-of-sight structure: structure on the line of sight; ({\tt XOoff}) -- Offset between the X-ray and Optical-counterparts as discussed in the paper; and ({\tt Other}) --	Any other useful information identified during the construction of MCXC-II.

 \begin{table*}[!ht]
  \caption{Summary overview of catalogue fields.}
\begin{tabular}{llll}
\toprule
\toprule
     \multicolumn{1}{c}{{\bf Field Name}} &
    \multicolumn{1}{l}{{\bf FORMAT }} &  
    \multicolumn{1}{l}{{\bf UNIT}} &
    \multicolumn{1}{l}{{\bf DESCRIPTION}} \\
\midrule 
{\tt INDEX}  			& SHORT  &                         	&	Cluster index  \\
{\tt NAME\_MCXC}	 	& STRING &                         	&	Name in MCXC   \\
{\tt NAME} 			& STRING &                         	&	Name in the input catalogue   \\
{\tt NAME\_ALT}	 	& STRING &                         	&  	Other names   \\
{\tt \_RAJ2000} 		& STRING & h:m:s       		&	Right Ascension (J2000)   \\
{\tt \_DEJ2000} 		& STRING & d:m:s 			&	Declination (J2000)   \\
{\tt RAJ2000}	 		&FLOAT  	& 	deg 			&	Right Ascension (J2000)   \\
{\tt DEJ2000}	 		& FLOAT 	& 	deg 			&	Declination (J2000)   \\
{\tt GLON}	 		& FLOAT 	& 	deg 			&	Galactic longitude  \\
{\tt GLAT} 			& FLOAT 	& 	deg 			&	Galactic latitude  \\
{\tt CATALOGUE}	 	& STRING &                         	& Catalogue  name \\
{\tt SUB\_CATALOGUE} 	& STRING &                         	& Sub-catalogue name   \\
{\tt Z} 				& FLOAT 	&                         	& Redshift  \\
{\tt Z\_TYPE} 			& STRING &                         	& Redshift type (1) \\
{\tt Z\_REF} 			& STRING &                         	& Redshift Reference (BIBCODE and/or present work (PW)) \\
{\tt Z\_FLAG} 			& STRING &                         	& Redshift Flag (2)   \\
{\tt SCALE} 			& FLOAT 	&  ${\rm kpc\ arcsec^{-1}}$		&Scale factor   \\
{\tt ERANGE}    		& FLOAT(2) & keV 			& Energy range of the flux  \\
{\tt R\_AP\_ARCMIN}	 & DOUBLE & 	arcmin 		& Radius of aperture of flux extraction (-1. for input total flux) \\
{\tt FX}				& DOUBLE &$10^{-12}$ ${\rm erg\ s^{-1}\ cm^{-2}}$ & Flux \\
{\tt ERRFX}    	 		& DOUBLE &	 $10^{-12}$ ${\rm erg\ s^{-1}\ cm^{-2}}$ 	&Error on the flux\\
{\tt LX2L500}   		& STRING &                         	& Method to convert  L\_AP  to L500 (3) \\
{\tt L\_AP} 			& DOUBLE & $10^{44}$ ${\rm erg\ s^{-1}}$ 	& Luminosity within aperture  (-1. for input  total flux)  \\
{\tt L500} 			& DOUBLE & $10^{44}$ ${\rm erg\ s^{-1}}$ 	&Luminosity within R500   \\
{\tt ERRML500} 	& DOUBLE & $10^{44}$ ${\rm erg\ s^{-1}}$ 	&	Lower error on L500   \\
{\tt ERRPL500} 	& DOUBLE & $10^{44}$ ${\rm erg\ s^{-1}}$ 	&Upper error on L500   \\
{\tt R500} 			& DOUBLE & 	Mpc 	&Radius corresponding to a density contrast of 500   \\
{\tt M500} 			& DOUBLE & 	$10^{14}$ solar mass 	&Mass corresponding to a density contrast of 500   \\
{\tt ERRMM500} 	& DOUBLE & 	$10^{14}$ solar mass 	&Lower error on M500   \\
{\tt ERRPM500} 	& DOUBLE & 	$10^{14}$ solar mass 	&Upper error on M500   \\
{\tt OVLP} 			& STRING &                         	&Overlap in different catalogues   \\
{\tt CAT1}			 &STRING&                         	&First overlapped catalogue \\
{\tt CAT2}			 &STRING&                         	&Second overlapped catalogue   \\
{\tt CAT3}			 &STRING&                         	&Third overlapped catalogue   \\
{\tt CAT4}			 &STRING&                         	&Fourth overlapped catalogue   \\
{\tt L5001}			& DOUBLE& $10^{44}$ ${\rm erg\ s^{-1}}$ 	& Luminosity within R500  for first catalogue overlapped   \\
{\tt ERRML5001}			& DOUBLE& $10^{44}$ ${\rm erg\ s^{-1}}$ 	& Lower error on L5001\\
{\tt ERRPL5001}			& DOUBLE& $10^{44}$ ${\rm erg\ s^{-1}}$ 	& Upper error on L5001\\
{\tt L5002} 			& DOUBLE&  $10^{44}$ ${\rm erg\ s^{-1}}$ 	& Luminosity within R500 for second catalogue overlapped   \\
{\tt ERRML5002}			& DOUBLE& $10^{44}$ ${\rm erg\ s^{-1}}$ 	& Lower error on L5002\\
{\tt ERRPL5002}			& DOUBLE& $10^{44}$ ${\rm erg\ s^{-1}}$ 	& Upper error on L5002\\
{\tt L5003}			& DOUBLE &  $10^{44}$ ${\rm erg\ s^{-1}}$ 	&Luminosity within R500  for third catalogue overlapped    \\
{\tt ERRML5003}			& DOUBLE& $10^{44}$ ${\rm erg\ s^{-1}}$ 	& Lower error on L5003\\
{\tt ERRPL5003}			& DOUBLE& $10^{44}$ ${\rm erg\ s^{-1}}$ 	& Upper error on L5003\\
{\tt L5004}			& DOUBLE & $10^{44}$ ${\rm erg\ s^{-1}}$ 	& Luminosity within R500  for fourth catalogue overlapped   \\
{\tt ERRML5004}			& DOUBLE& $10^{44}$ ${\rm erg\ s^{-1}}$ 	& Lower error on L5004\\
{\tt ERRPL5004}			& DOUBLE& $10^{44}$ ${\rm erg\ s^{-1}}$ 	& Upper error on L5004\\
{\tt NOTES\_ORIG} 	& STRING &                         	& Note (expanded) on the cluster as given in the source catalogue  (4) \\
{\tt NOTES}			& STRING &                         	& Note on the cluster (5)  \\
{\tt INDEX\_MCXC\_P2011} & SHORT &                        	&Index in MCXC published by  Piffaretti et al. 2011   \\
\bottomrule
\end{tabular}

\tablefoot{  

(1) redshift type: spectroscopic (S), photometric (P),  a combination of photometric and spectroscopic values (SP), estimated (E) or unknown (U)

(2) Redshift flag:
\begin{itemize}[noitemsep,topsep=0pt,label=$-$]
\item 'Catalogue': original value from catalogue.
\item 'New': No redshift was available in the source catalogue; the new  redshift origin is given in Z\_REF
\item 'Revised': redshift in the catalog has been updated
\item 'Consolidated': the source of the redshift in the catalogue and/or its type was unclear. We propose another reference (given in {\tt Z\_ref}), with  the corresponding redshift for consistency. The difference with the catalogue value  is negligible. 
\end{itemize}

(3) Method to convert aperture luminosity to L500: by  iteration ('ITER' ) or  using a fixed size correction  ('SC')

(4) Collated notes from original input source catalogues. In some cases, these have been expanded from the original abbreviated form, or combined from different tables in the input catalogue paper.

(5) Individual formatted notes on the cluster (Code | Section | Description). More details are in the text, see Appendix \ref{ap:mcxcfield}.

}

   \label{mcxc_field} 
\end{table*}

\end{appendix}

\end{document}